\documentclass[%
 reprint, prab,
 amsmath, amssymb,
 aps,
 superscriptaddress,
 floatfix,
]{revtex4-2}

\usepackage{graphicx}
\usepackage{longtable}
\usepackage[colorlinks=true,allcolors=blue]{hyperref}

\begin{document}
\raggedbottom

\preprint{FERMILAB-PUB-26-0388-PIP2}

\title{Higher-order space-charge stability in anisotropic beams: Vlasov--Poisson derivation, refined dispersion relations, and stability charts}

\author{Abhishek Pathak}
\thanks{Contact author: abhishek@fnal.gov}
\affiliation{Fermi National Accelerator Laboratory, Batavia, IL 60510, USA}

\date{September 6, 2026}

\begin{abstract}
The Hofmann stability chart is widely used for working-point screening in
space-charge-dominated linacs, including through a design-code diagnostic
that implements its higher-order ($\ell\!=\!3,4$) relations. We identify two
errors in the published equations: the $\ell\!=\!3$ second-order
space-charge ($S^{4}$) $\alpha$-coupling residues omit factors
$(1\!\mp\!2\hat\eta^{2}/\alpha)$, and the stated isotropic reduction of the
$\ell\!=\!4$ relation contains a sign error. Both corrections follow from
Hofmann's Vlasov--Poisson equations without fitted parameters. They also
satisfy external consistency checks: the corrected relations reproduce
coherent tune-shift coefficients tabulated in the author's later monograph,
which the printed forms miss by $24\%$ and $127\%$. Mode-resolved figures
from a principal published application agree with the corrected relations
and reject the printed forms, indicating that the equations published in
1998 are inconsistent with the calculations underlying those tested figures.
We quantify the effect on the non-oscillatory stability chart. Inside the
adopted $S^{2}\!\le\!10$ comparison domain, the printed and corrected forms
disagree on $0.73$--$2.11\%$ of the cells, with no preferred direction.
Among the excluded cells, the disagreement reaches $22\%$, and the printed
relation over-predicts instability at every sampled anisotropy. This
concentration outside the comparison domain may help explain why the errors
persisted, although the calculation does not establish their historical
cause. For PIP-II, the corrected chart flags four of thirty-two evaluable
periods, including one on an $\ell\!=\!3$ odd branch missed by an
$\ell\!=\!2$ screen. This count covers non-oscillatory modes only and remains
conditional on an unresolved factor-five disagreement between two codes on
transverse emittance growth.
\end{abstract}

\maketitle

\section{Introduction}
\label{sec:intro}

The Hofmann stability chart~\cite{Hofmann1998} has been the standard
working-point screen of
high-intensity linac design for nearly three decades, mapping the
$(\nu_{z}/\nu_{x},\,\nu_{x}/\nu_{0x})$ plane into stable and unstable
regions for a Kapchinskij--Vladimirskij~\cite{KV1959}
beam under continuous focusing.  In the form considered here it is
specifically a \emph{non-oscillatory coherent-mode} chart: the
boundaries we draw isolate the purely growing
($\mathrm{Re}\,\omega\!=\!0$, $\mathrm{Im}\,\omega\!>\!0$) roots of the
$\ell\!=\!2,3,4_{e}$ KV-dispersion polynomial, which are the
branches targeted by the present working-point screen and are often
among the operationally relevant ones in long high-intensity linacs.
Hofmann's published charts additionally annotate the oscillatory branches.
This convention follows Ref.~\cite{HofmannFranchetti2003}, which states
that ``oscillatory instabilities should be discarded as KV artifact'' and
draws its own charts over non-oscillatory modes only, on the ground that such
modes are specific to the KV distribution's delta-function Hamiltonian
dependence.  We discuss the treatment as a scope choice in Sec.~\ref{sec:track1validation}.  The chart builds on the periodic-system KV-stability analysis of Hofmann,
Laslett, Smith, and Haber~\cite{HLSH1983} and the broader rms-envelope
framework reviewed in Refs.~\cite{Wangler2008,Reiser2008,Chao1993}.
Experimental tests include the GSI~UNILAC stop-band
campaigns~\cite{Groening2009,Groening2009b} and subsequent observations of
space-charge-driven resonances of orders four and
six~\cite{Cheon2020FourthOrder,Jeon2016ExpResonances,Jeon2015Sixth}.
The two-dimensional KV coherent-mode spectrum was derived by
Gluckstern~\cite{Gluckstern1970Modes} and generalized to anisotropy by
Hofmann~\cite{Hofmann1998}. Extensions treat anisotropic two- and
three-dimensional beams~\cite{HofmannFranchetti2003} and, at equal
emittances, three-dimensional periodically focused
beams~\cite{HofmannBF2017}. The charts provide a direct assessment of an
operating point's distance from parametric resonance bands.

The PIP-II 800-MeV H$^{-}$ linac at
Fermilab~\cite{PIP2CDR,PIP2FDR2024,PIP2LinacOpt} adopts a
non-equipartitioned design whose geometric emittance ratio
nonetheless remains mildly anisotropic: $\varepsilon_{z}/\varepsilon_{x}$
is largest ($\sim\!1.7$--$1.8$) in the low-energy HWR/SSR1 front end and
falls to $\sim\!1.2$ through the LB650/HB650 sections (mean
$\approx\!1.48$). The operating point traces a path that crosses or skirts several resonance bands in this plane.  Hofmann's 1998 analysis~\cite{Hofmann1998}
covers modes up to $\ell\!=\!4$ in both parities (his Sec.~IV.A--C); in
operational chart-based screening, however, the dispersion relations
are routinely used only in their $\ell\!=\!2$ form and isotropic-limit
reductions, under restrictive assumptions: continuous focusing, a KV
phase-space distribution, the $\ell\!=\!2$ collective mode, zero
incoherent tune spread, and a binary stable/unstable verdict.

We determine where the published and corrected $\ell\!=\!3$ relations give
different stability classifications, first across chart space and then along
a linac trajectory. On the same grid, we also quantify the regions where
higher-order modes change the classification obtained from an
$\ell\!=\!2$ screen.

A search of the publisher index, INSPIRE-HEP and Crossref
records found no published correction addressing the two errors discussed
here, as of 5~September~2026. Appendix~\ref{app:correction-search} records
the search method and its limitations; this negative result does not
establish that no correction exists.

The anisotropic $\ell\!=\!2$
relations, in both parities, are established --- Hofmann's own monograph prints
them, and they have been re-derived from moment equations and shown equivalent
to the linearized Vlasov--Poisson system~\cite{Yuan2017}.  We use the established $\ell\!=\!2$ even relation to
fix the two normalization constants. The present derivations concern
$\ell\!=\!3$ and $\ell\!=\!4$. The chart's internal structure has also
been studied: Li and
Jameson~\cite{LiJameson2018} give an independent anisotropic eigenmode
treatment to fourth order in periodic focusing, and Jameson~\cite{Jameson2022}
examines the published charts and records an exchange with Hofmann about their asymmetric stop bands.  Neither prints the anisotropic
$\ell\!\ge\!3$ relations in closed form, and neither reports an error in them.

A preliminary analysis appeared as a
three-page conference contribution~\cite{PathakIPAC2026}.  The present paper supersedes that contribution in the following respects.  That contribution identified the missing $(1\!\mp\!2\hat\eta^{2}/
\alpha)$ factors from the isotropic-limit inconsistency alone; it did not
derive them, did not identify the Eq.~(41)/(42) sign conflict, and contains no
chart-wide census.  Its PIP-II numbers are superseded: it paired
$\alpha\!=\!1/R_{\rm eff}$ where the correct pairing is
$\alpha\!=\!R_{\rm eff}$ (Sec.~\ref{sec:coords}), and consequently reported seven flags among $38$ periods with rates to $0.24$ whereas the corrected map gives four flags among the $32$ physically
evaluable in-domain periods (Sec.~\ref{sec:pip2}), with a maximum in-domain
rate of $0.049$.  It also
described the $S^{2}/S^{4}/S^{6}$ expression as a non-convergent partial sum,
whereas these are exact finite polynomials in $S^{2}$ and the domain
restriction adopted here rests on model adequacy rather than on convergence.
Finally, it deferred the out-of-domain periods to a lattice-periodic Floquet
treatment in this paper; that treatment was attempted, failed validation, and
is reported as such in the Supplemental Material rather than presented as a
result.  Every number in the present paper is recomputed from the corrected
map.

Two preliminaries are required before the chart-impact question can be asked.
The first is an equation audit, and it yields both corrections examined
here.  The $S^{4}$ $\alpha$-coupling residues of
Hofmann's $\ell\!=\!3$ relation cannot reproduce his own stated isotropic
limit as printed.  The
missing factors follow from the determinant structure implied by his
two-term potential basis
(App.~\ref{app:l3e-derivation}), and from direct
evaluation of his Vlasov--Poisson equations, Eqs.~(19) and~(23). The same
evaluation reproduces the
$\ell\!=\!4$ relation exactly as printed --- there at twelve independent
anisotropy pairs rather than symbolically in $(\alpha,\hat\eta)$
(Sec.~\ref{sec:vlasov-full}).  The companion defect is a sign: the
$\sigma_{p}^{4}$ term of Eq.~(42) conflicts with Eq.~(41), and the five
closed-form roots Hofmann prints in his own Eq.~(43) settle the conflict
against Eq.~(42) (App.~\ref{app:l4e}).  The second preliminary concerns the coordinate mapping: Hofmann's envelope ratio and frequency ratio
must be taken from the same plane assignment, and the tune ratio built from
depressed rather than zero-current tunes.  This rule is stated by Jameson~\cite{Jameson2022}. The earlier coordinate
mapping used in this work violated it, producing an unphysical discontinuity
at $\hat\eta\!=\!1$.  The two defects have different signatures at isotropy.  The Eq.~(36) defect is \emph{detected} at isotropy --- the
printed equation fails to reduce to its own Eq.~(37) --- but its off-isotropic
\emph{form} is invisible there, since the disputed factors degenerate to
constants at $\alpha\!=\!\hat\eta\!=\!1$.  The coordinate defect is
invisible at isotropy outright: both faulty pairings coincide with the correct
one when $\alpha\!=\!\hat\eta\!=\!1$.  In neither case can an isotropic-limit regression suite determine the anisotropic form.

Section~\ref{sec:census} quantifies the
resulting changes in stability classification.  Inside the adopted $S^{2}\!\lesssim\!10$ comparison domain for our
$\ell\!\le\!4$ KV chart, the printed and corrected forms
of Eq.~(36) disagree on the verdict of $0.73$--$2.11\%$ of the cells that
domain admits, and which of the two flags more changes sign with anisotropy, at the ratios
where that direction survives grid refinement.
Among the cells it excludes the disagreement affects $6.65$--$22.27\%$ and
acquires a direction: the printed relation over-predicts higher-order
instability at every ratio we sampled.  Two comparisons with the published literature are relevant.  A reader implementing Eq.~(36) as printed on the axes of
Ref.~\cite{HofmannFranchetti2003} would change the verdict of a comparable
fraction of cells.  Hofmann's mode-resolved Fig.~6 matches the corrected relation and rejects the
printed one, providing strong evidence that the tested curves were computed
with the corrected form (Sec.~\ref{sec:fig6}).  A comparison drawn from inside
$S^{2}\!\le\!10$ is meanwhile sampling a population in which the two forms
almost always agree cell for cell, so unless it lands on one of the few that
discriminate, it tests a prediction they share.  Separating them needs a point
chosen for the purpose, which Sec.~\ref{sec:census} identifies, or the
excluded region.

The same census quantifies the contribution of higher-order modes.  The
fraction of operating points flagged by $\ell\!=\!3,4_{e}$ \emph{alone} is
$21.6$--$22.4\%$ across $\varepsilon_{z}/\varepsilon_{x}\!\in\![1.2,8]$ ---
varying by $3.9\%$ on that grid, and about $11\%$ after refinement --- and, more usefully for design, occupies bands spatially
separated from the $\ell\!=\!2$ region, so it cannot be avoided by adding
margin to a classical screen.

That figure is a property of this $(R,\eta)$ rectangle under uniform cell
weight, not of the relations alone --- Sec.~\ref{sec:census} reports how far it
moves with the window --- and the operational statement is narrower still.  We report these quantities separately.
Restricting the $\ell\!\le\!4$ KV chart to the adopted
$S^{2}\!\lesssim\!10$ comparison domain, the same fraction reads $6.3\%$ at
$\varepsilon_{z}/\varepsilon_{x}\!=\!1.2$ rising to $20.7\%$ at $5.0$.  The
difference is quantitative and informative: the restriction admits $19\%$ of
the higher-order region at $1.2$ and $77\%$ at $5.0$, so the region does not
grow with anisotropy --- it migrates into the range an $\ell\!\le\!4$
treatment can address.  All \emph{operating-point} claims in this paper,
including every PIP-II number, are made inside the restriction.  The gated
number is essentially independent of the $S^{4}$ correction, because there the
printed $S^{2}$ block sets where the channels are active; the ungated number
is more exposed to it, and we quantify both in Sec.~\ref{sec:pip2} rather
than assert robustness for either.

PIP-II provides a worked example of a trajectory through this chart space.  It is a mildly anisotropic machine
($\varepsilon_{z}/\varepsilon_{x}\!\approx\!1.2$--$1.8$) and therefore sits in
the corner of the census where the \emph{gated} higher-order-only fraction is
smallest ($6$--$7\%$, against $21.6\%$ ungated).  The trajectory returns four weak
flags among the $32$ in-domain periods whose depressed tunes are
physically evaluable --- three classical $\ell\!=\!2$
modes and one $\ell\!=\!3$ odd-branch flag that an $\ell\!=\!2$ screen does
not see --- and integrating the rates over the actual per-period phase
advance leaves a small local exponent budget, the largest single period
contributing $N_{e}\!=\!0.072$.  Whether any of that is transported into net
growth across the linac is a question a frozen-per-period analysis cannot
answer, and we do not claim it.  PIP-II thus illustrates the framework in a regime of weak predicted growth.

The solver is restricted to non-oscillatory roots; on a $2000$-point sweep of the
census domain it recovers the exact real-root set at every sampled point
inside and outside the adopted $S^{2}\!\le\!10$ domain on the current
sampled set; the earlier uniform-only scan failed once outside it for a
grid-resolution reason we identify and retain as a historical regression case; and we quantify the oscillatory
channel it excludes, which is not negligible.  An attempt at an
independent 21-dimensional second-moment ODE closure (analogous to Sacherer's
envelope formalism~\cite{Sacherer1971}) did not reproduce the $\ell\!=\!2$
growth rate at a known unstable chart point; the diagnosis is reported in
the Supplemental Material~\cite{SupplementalMaterial} rather than omitted, and confirmation outside the adopted model domain, by lattice-periodic Floquet
or multiparticle analysis, remains
future work.  A probabilistic overlay under an engineering jitter budget
(the Supplemental Material~\cite{SupplementalMaterial}) and a solver-labeled surrogate
(App.~\ref{app:surrogate}) are reported as auxiliary tools; the surrogate in
particular reaches near-unity in-distribution ranking yet recovers only half
the flags on the PIP-II subset, which limits its usefulness for this trajectory.

The paper is organized as follows.  Section~\ref{sec:coords} fixes notation,
the chart parametrization, and the corrected coordinate map.
Section~\ref{sec:track1} presents the anisotropic $\ell\!=\!2,3,4_{e}$ solver,
its validation, and a sweep-based audit of its root-set completeness.
Section~\ref{sec:vlasov-full} derives the corrected $\ell\!=\!3_{e}$ block
from Hofmann's Eqs.~(19) and~(23) directly, surface and volume terms, and
excludes the alternatives away from isotropy.
Section~\ref{sec:census} locates where the errors change the stability
verdict, and gives the chart census it
rests on.  Section~\ref{sec:pip2} applies the stack to PIP-II as a worked
example.  Section~\ref{sec:limits} states limitations and
Section~\ref{sec:concl} concludes.  App.~\ref{app:anisotropic} documents the
corrected dispersion blocks, their isotropic reduction, and the
determinant-structure route to the $S^{4}$ factors;
App.~\ref{app:surrogate} the machine-learning surrogate;
App.~\ref{app:perperiod} the per-period PIP-II table.  The Supplemental
Material~\cite{SupplementalMaterial} carries the probabilistic-margins
analysis, the moment-ODE failed-validation record, the section-level
worst-case growth table, and the full 41-period dataset.

\section{Coordinate mapping and data provenance}
\label{sec:coords}

We work in the standard Hofmann coordinates:
\begin{align}
R     &= \nu_{z} / \nu_{x}     \quad \text{(tune ratio)} \\
\eta  &= \nu_{x} / \nu_{0x}    \quad \text{(transverse tune depression)} \\
\varepsilon_{z}/\varepsilon_{x} &\quad \text{(emittance ratio, geometric)}
\end{align}
Following Ref.~\cite{Hofmann1998} we introduce the auxiliary envelope ratio
\begin{equation}
  \hat\eta_{0} = \sqrt{R/(\varepsilon_{z}/\varepsilon_{x})},
  \label{eq:etahat_raw}
\end{equation}
which is the pre-flip value.  Hofmann's variables require $\hat\eta\!\ge\!1$
(his Eq.~(24)), so when $\hat\eta_{0}\!<\!1$ we exchange the two planes.
Writing $F$ for that flip, the variables actually fed to the dispersion
polynomial are
\begin{equation}
  (\hat\eta,\,R_{\rm eff}) =
  \begin{cases}
    (\hat\eta_{0},\ R), & \text{no flip},\\[3pt]
    (1/\hat\eta_{0},\ 1/R), & F,
  \end{cases}
  \qquad \alpha = R_{\rm eff}.
  \label{eq:coordmap}
\end{equation}
Hofmann's Eq.~(24) requires
$\alpha=R_{\rm eff}$ rather than $1/R_{\rm eff}$: it is the choice for which
$\varepsilon_{x}/\varepsilon_{y}=\hat\eta^{2}/\alpha$ holds, and adopting the
reciprocal violates that identity by exactly $R_{\rm eff}^{2}$.  This is the
first of the two coordinate defects noted in Sec.~\ref{sec:intro}, and it is
invisible at $\alpha\!=\!\hat\eta\!=\!1$.

The dimensionless space-charge strength follows from Hofmann's Eq.~(5), and
because the flip exchanges which plane is the reference it too has two
branches:
\begin{equation}
  S^{2} =
  \begin{cases}
    \dfrac{(1+\hat\eta)(1-\eta^{2})}{\eta^{2}}, & \text{no flip},\\[10pt]
    \dfrac{(1+\hat\eta)(1-\eta^{2})}{\hat\eta\,R^{2}\,\eta^{2}}, & F,
  \end{cases}
  \qquad S^{2} = 0 \text{ for } \eta \ge 1,
  \label{eq:S2_def}
\end{equation}
The convention $S^{2}=0$ for $\eta\!\ge\!1$ captures the fact that
space-charge cannot \emph{raise} the tune above its zero-current
value.

We restrict all
quantitative operating-point claims to $S^{2}\!\lesssim\!10$ and call points inside that restriction
\emph{in-domain} and points outside it \emph{out-of-domain}.
These labels follow the literature and specify membership of the adopted domain.  They do \emph{not} mean that the dispersion
relations are perturbative expansions in $S^{2}$: they are exact polynomials
in $S^{2}$, and Sec.~\ref{sec:limits} explains what the restriction actually
rests on.  Seven PIP-II periods (Sec.~\ref{sec:pip2},
App.~\ref{app:perperiod}) return matched-envelope phase advances at
or above $\nu_{0x}$; these are lattice locations where the matched
envelope is either at the no-current transition (zero-current
sampling points, or a section-boundary period such as
SSR1$\to$SSR2 at idx~16, $\eta\!=\!1.05$) or where the matched-envelope
solver returns a super-zero-current value that is a tune-extraction
artifact rather than a physical depressed tune (idx~9, with
$\eta\!=\!1.50$, i.e.\ $50\%$ above $\nu_{0x}$, is the strongest such
case, in the early SSR1 cells just past the HWR$\to$SSR1 transition;
a depressed tune cannot physically exceed $\nu_{0x}$, so this reflects
energy gain and matched-envelope overshoot, not a real effect).  We treat all such periods as space-charge-free for the chart evaluation.  Setting
$S^{2}\!=\!0$ makes the Hofmann dispersion trivially stable, so these seven
periods are stable \emph{by convention} rather than by computation; we therefore quote the flagged count against the $32$ periods that were
actually evaluated, ``4 of 32'', rather than against all $39$ inside the
domain.  Including the seven conventionally stable periods gives 4 of 39; the denominator 32 counts only periods with a physically evaluated depressed tune.  Clamping all seven
to $\eta\!=\!0.99$ --- a chosen near-zero-space-charge value
within the physically admissible interval --- leaves every one sub-threshold and the flagged
set unchanged at the same four periods (idx~10, 18, 21, 26).  That test is favourable by
construction, however, and does not bound the answer: $\eta\!=\!0.99$ is a
\emph{weak} space-charge case, with still weaker cases as $\eta\!\to\!1^{-}$.  Scanning each of
the seven across $\eta\!\in\![0.20,0.99]$ inside the gate, five never flag at
any depression, but two would --- SSR1 idx~9 for
$\eta\!\in\![0.28,0.315]$ and SSR2 idx~16 for $\eta\!\in\![0.36,0.56]$.
Their true depressed tunes are not available from the export, so we cannot
say whether either lies there.  The flagged set is unchanged under this chosen weak-space-charge correction. Its sensitivity to the underlying tune-extraction problem remains unresolved and requires corrected matched-envelope tunes.  The growth rate in units of the zero-current transverse tune is
$\gamma/\nu_{0x} = \sqrt{-s}\,G_{\rm scale}$, where
$s=(\omega/\nu_{\rm ref})^{2}$ is the normalized squared coherent
eigenfrequency defined in Sec.~\ref{sec:track1} (unstable modes have
$s<0$), with
\begin{equation}
  G_{\rm scale} =
  \begin{cases}
    \eta,     & \text{no flip},\\
    R\,\eta,  & F,
  \end{cases}
  \label{eq:gscale}
\end{equation}
since without the flip $s$ is normalized to $\nu_{x}$ and with it to
$\nu_{z}$.  Throughout the rest of the paper $S^{2}$ denotes a single
dimensionless quantity (the space-charge strength), while $S^{4}$
and $S^{6}$ denote the second and third powers of $S^{2}$ (not fourth or
sixth powers of a separate $S$).  As shown in Sec.~\ref{sec:vlasov-full}
these arise as minors of a determinant whose entries are linear in $S^{2}$,
so they are exact terms of a polynomial rather than orders of an expansion.
The ``even-parity fourth-order'' mode $\ell\!=\!4_{e}$, used
throughout, refers to the polynomial-perturbation order
$\ell\!=\!4$ branch with the angular harmonic
$\cos(4\varphi)$~\cite{Hofmann1998}; the odd branch
$\ell\!=\!4_{o}$ is treated separately in Sec.~\ref{sec:limits}.

\textit{Data provenance.}\quad The numerical implementation used in
Secs.~\ref{sec:track1}--\ref{sec:pip2} was evaluated with a 516-test regression suite
covering the corrected anisotropic
dispersion solver (Sec.~\ref{sec:track1}, App.~\ref{app:anisotropic}),
the probabilistic chart with ensemble Monte~Carlo uncertainty
(the Supplemental Material~\cite{SupplementalMaterial}), and the machine-learning surrogate (App.~\ref{app:surrogate}).
A frozen legacy implementation was retained for the comparisons described below.
The PIP-II lattice follows the design basis of the Conceptual Design
Report~\cite{PIP2CDR} with the post-2023 physics-design
revision~\cite{PIP2FDR2024}; the per-period tune ratio $R$ and tune
depression $\eta$ are taken from the \textsc{TraceWin}~\cite{Uriot}
transfer-matrix output of that lattice, while the geometric emittance
ratio $\varepsilon_{z}/\varepsilon_{x}$ is taken from the HELIX
beam-dynamics emittance export (see below).
The per-period chart coordinates and growth rates are tabulated in the
Supplemental Material~\cite{SupplementalMaterial}; the extraction conventions
are specified here.  Zero-current betatron tunes
$\nu_{0x,y,z}$ are extracted as the diagonal phase advances of the
period transfer matrix at vanishing space-charge potential;
depressed tunes $\nu_{x,y,z}$ are extracted from the matched
self-consistent envelope at the period's nominal current
($5$~mA peak at the RFQ exit, $2$~mA average through the SRF linac
after the upstream chopper);
the geometric emittance ratio is taken from the HELIX beam-dynamics
export of the normalized rms emittances $\varepsilon_{n,x}$ and
$\varepsilon_{n,z}$ (both in $\pi$\,mm\,mrad in the common
$(x,x')$/$(z,z')$ slope-coordinate convention); each plane is
de-normalized by $\beta\gamma$, the standard relativistic normalization
for a slope coordinate, so that
$\varepsilon_{z}/\varepsilon_{x}\!=\!\varepsilon_{n,z}/\varepsilon_{n,x}$
exactly ($\beta\gamma$ cancels).  That the longitudinal column is a
$(z,z')$ emittance and not a $(z,\Delta p/p)$ quantity --- the
distinction that sets the anisotropy scale --- is corroborated by the
\textsc{TraceWin} $\varepsilon_{zz'}$ chart output, which carries the
same magnitude ($\approx\!0.35\,\pi$\,mm\,mrad) as the HELIX
$\varepsilon_{n,z}$ column throughout the linac; a $(z,\Delta p/p)$
column would differ from it by orders of magnitude.  We note that the
two codes agree closely in the longitudinal plane
($\varepsilon_{n,z}$ falls by $9\%$ from injection to
HB650-2 exit while \textsc{TraceWin}'s $\varepsilon_{zz'}$ rises by
$1\%$) but differ in the transverse plane, where HELIX reports $27\%$
normalized-emittance growth against \textsc{TraceWin}'s $5\%$.  All four
are endpoint-to-endpoint over the $175$~m the two exports share.  The
resulting spread in $\varepsilon_{z}/\varepsilon_{x}$ at the
high-energy end is quantified as a sensitivity in
Sec.~\ref{sec:limits}.

\section{The higher-order anisotropic dispersion solver}
\label{sec:track1}

\subsection{Dispersion relations}

Second-order collective screening contains two distinct parities.  The
$\ell\!=\!2_e$ coherent envelope modes follow from linearizing the
matched-envelope equations; their quadrupole oscillation is relevant to
parametric stopbands in periodically focused lattices, including the familiar
$2:1$ resonance description.  Hofmann's $\ell\!=\!2_o$ branch is instead an
odd tilting/coupling mode, whose low-frequency instability is driven by the
internal space-charge difference resonance.  Our continuous-focusing chart
includes both parities and does not calculate periodic-lattice excitation;
the second-order PIP-II flags below all come from $\ell\!=\!2_o$.  Hofmann's 1998 paper~\cite{Hofmann1998} also
gives the $\ell\!=\!3$ and $\ell\!=\!4$ (even and odd) dispersion
relations explicitly (his Eqs.~(36), (41), (45)), but operationally these higher-order branches are used less often in practice than the
$\ell\!=\!2$ one, although they are available in design software: the \textsc{TraceWin}
chart diagnostic from which we take the PIP-II lattice generates its charts
from this same theory, taking maxima over the unstable modes to fourth
order~\cite{Uriot,HofmannBook2017}.  For a non-equipartitioned beam the
$\ell\!=\!3$ and $\ell\!=\!4$ modes open additional stopbands that the
$\ell\!=\!2$ analysis does not see; these become physically relevant when
$\varepsilon_{z}/\varepsilon_{x}$ departs strongly from unity.  Use of these relations in a deployed chart diagnostic motivates measuring
the effect of the printed errors. We have not established which algebraic
form that code evaluates, and
our own Sec.~\ref{sec:fig6} shows that Hofmann's published curves used the
corrected relation while citing the printed equation --- so an implementation
may well carry an internal correction too.  The discriminating point of
Sec.~\ref{sec:census} would settle it for any given code in a single
evaluation, providing a direct implementation check.
We therefore implement Hofmann's $\ell\!=\!3$ and $\ell\!=\!4_{e}$
(even-parity) equations directly, retaining their full anisotropic
$S^{4}/S^{6}$ blocks rather than the iso-limit reductions, so that the
chart can screen for higher-order anisotropic growth across the full
emittance-ratio range and quantify the margin of a given trajectory
--- such as PIP-II's (Sec.~\ref{sec:pip2}) --- to the onset of those
modes.

For each mode order $\ell\ge 3$ the coherent eigenfrequency
$s=(\omega/\nu_{\rm ref})^{2}$, normalized to the active-frame depressed
reference tune $\nu_{\rm ref}$ ($\nu_{\rm ref}=\nu_{x}$ without the
symmetry flip of Sec.~\ref{sec:coords} and $\nu_{z}$ with it, matching the
$G_{\rm scale}$ branches of Eq.~\eqref{eq:gscale}), is a root
of the dispersion function
\begin{multline}
  D_{\ell}(s) =
  (1 + \hat\eta)^{\ell}
  + S^{2} \!\! \sum_{k \in \mathcal{P}_{\ell}}
    \frac{N^{(1)}_{\ell k}(\alpha,\hat\eta)}{d_{\ell k} - s} \\
  + S^{4} \!\!  \sum_{(i,j) \in \mathcal{Q}_{\ell}}
    \frac{N^{(2)}_{\ell ij}}{(d_{\ell i} - s)(d_{\ell j} - s)}
  + \cdots = 0,
  \label{eq:disp}
\end{multline}
where the poles $d_{\ell k}$ are the space-charge-free mode
frequencies and the residues $N^{(1)}_{\ell k},\, N^{(2)}_{\ell ij}$,
together with the pole sets
$\mathcal{P}_{\ell}, \mathcal{Q}_{\ell}$, are taken from Hofmann's
equations~(36), (41), and (45).  We implement the full anisotropic
$S^{2}$ block for $\ell\!=\!2,3,4_{e},4_{o}$, together with the full
anisotropic higher-order blocks of Eqs.~(36) and (41) --- $S^{4}$ for
$\ell\!=\!3$, which is the highest power Eq.~(36) contains, and
$S^{4}$ and $S^{6}$ for $\ell\!=\!4_{e}$; the
$\ell\!=\!4_{o}$ $S^{4}$ block was formerly retained in the
isotropic-limit closed form Eq.~(46).  The current solver instead evaluates
the full anisotropic block derived in Sec.~\ref{sec:vlasov-full}, with
Eq.~(46) recovered at isotropy.  The $\ell\!=\!4_{o}$ branch remains
disabled by default in the principal aggregator (see Sec.~\ref{sec:limits}).
Consequently, no PIP-II period in Sec.~\ref{sec:pip2} carries a
contribution from the $\ell\!=\!4_{o}$ branch.  On the corrected
PIP-II trajectory (Sec.~\ref{sec:pip2}) three of the four
in-domain threshold-crossing flags in
Table~\ref{tab:pip2_flagged} originate in the classical $\ell\!=\!2_o$
tilting/coupling mode, and the fourth (SSR2 idx~18) in the $\ell\!=\!3$ odd
branch.  The $\ell\!=\!3$ even and $\ell\!=\!4_{e}$ blocks return growth
below threshold at every in-domain period for the design
emittance ratio $\varepsilon_{z}/\varepsilon_{x}\!\approx\!1.2$--$1.8$;
the only nonzero $\ell\!=\!3$ even contribution (SSR1 idx~12) occurs at
$S^{2}\!=\!33$, outside the adopted model domain, and is not propagated.
The higher-order channels therefore do add an in-domain flag here, and additionally serve to quantify the margin to
anisotropy-induced onset (Sec.~\ref{sec:pip2}).
The $\ell\!=\!3_{o}$ (odd) branch is evaluated by the Hofmann
interchange $\alpha\!\to\!1/\alpha$, $\hat\eta\!\to\!1/\hat\eta$ of
his~p.~4719 applied to Eq.~(36); the $S^{4}$ correction documented in
App.~\ref{app:anisotropic} for the even branch propagates
identically to the odd branch under this variable substitution, so
the same iso-limit reduction Eq.~\eqref{eq:appA-iso-target}
constrains both parities.  Section~\ref{sec:vlasov-full} independently derives the odd block directly from Eqs.~(19)
and~(23) on the odd-parity basis and recovers exactly this relation, so the
substitution is confirmed rather than assumed.  Hofmann specifies the interchange as acting
on his Eqs.~(24) and~(25), which \emph{define} the dimensionless
intensity and eigenfrequency by normalizing to $\nu_{x}$:
$S^{2}\!=\!\omega_{p}^{2}/\nu_{x}^{2}$ and $s\!=\!(\omega/\nu_{x})^{2}$.
Exchanging $\nu_{x}\!\leftrightarrow\!\nu_{y}$ therefore rescales both
alongside the inversions,
\begin{equation}
  \alpha \to \frac{1}{\alpha}, \quad
  \hat\eta \to \frac{1}{\hat\eta}, \quad
  S^{2} \to \frac{S^{2}}{\alpha^{2}}, \quad
  s \to \frac{s}{\alpha^{2}} .
  \label{eq:odd-interchange}
\end{equation}
The plasma frequency $\omega_{p}^{2}$ --- the physical beam-intensity
scalar --- is unchanged by relabeling the mode planes; it
is the \emph{normalizer} $\nu_{x}^{2}$ that changes, so the
dimensionless $S^{2}$ does move.  Operationally, the odd branch is
evaluated from the even-branch function
under the full substitution~\eqref{eq:odd-interchange}; because the
result is still written as a function of the original $s$, the root
scan requires no further rescaling.
An
internal-consistency audit identifies that the $\alpha$-coupling
terms of the $S^{4}$ block in Eq.~(36) require the
$(1\!\mp\!2\hat\eta^{2}/\alpha)$ factors, with the same
structural form as Hofmann's $S^{2}$ block, for the equation to
reduce to its own stated isotropic-limit form Eq.~(37); the
algebraic identity
$-(9\!-\!s)^{2}+6(1\!-\!s)(9\!-\!s)+27(1\!-\!s)^{2}=-32s(3\!-\!s)$
that closes the reduction is shown in App.~\ref{app:anisotropic}.  We
adopt that form throughout; the $S^{4}/S^{6}$ blocks of Eq.~(41) are
transcribed term-by-term, with the iso-limit reduction checked against
Eq.~(42) \emph{in the sign-corrected form}
Eq.~\eqref{eq:appA-eq42-corrected} that Eq.~(43)'s closed-form roots
require (App.~\ref{app:l4e}; Sec.~\ref{sec:track1validation}).
Eq.~(45) for the
$\ell\!=\!4_{o}$ branch supplies the $S^{2}$ anisotropic block; its
$S^{4}$ block is now derived and evaluated at full anisotropy.  The former
implementation used Eq.~(46) for that block at all anisotropies, a limitation
whose measured consequences are retained in Sec.~\ref{sec:limits}.
Hofmann's published isotropic closed forms Eqs.~(37), (42), (46) therefore
serve as \emph{validation anchors}; Eq.~(46) is the exact isotropic limit of
the current $\ell\!=\!4_{o}$ block, rather than its running off-isotropic
form (the odd-branch implementation carries no $S^{6}$ term).  One qualification on that
status: because we correct Eq.~(42)'s $S^{4}$ signs, it is no longer an
\emph{independent} anchor for $\ell\!=\!4_{e}$.  The independent anchor
there is Eq.~(43), whose closed-form roots we do not modify and which
forces the correction; Eq.~(42) then serves as a derived consistency
check rather than an external one.
A negative real root, $s<0$, signals an exponentially growing
(purely non-oscillatory) instability with growth rate
$\gamma/\nu_{0x}=\sqrt{-s}\,G_{\rm scale}$.  Roots are located by a hybrid scan in $s\in[s_{\min},0)$: the
original $2000$ uniform points are supplemented by $1200$ logarithmically
spaced points near zero, followed by $50$ bisection iterations per detected
sign change.  This resolves the paired-root failure of the earlier uniform
scan in the sampled audits below.  Finite precision and proximity to zero
preclude a uniform relative-error guarantee; we report measured matched-root
errors and the scan settings rather than inferring $10^{-15}$ relative
accuracy from the iteration count alone.

\subsection{Validation}
\label{sec:track1validation}

\begin{figure*}[t]
  \centering
  \includegraphics[width=0.98\textwidth]{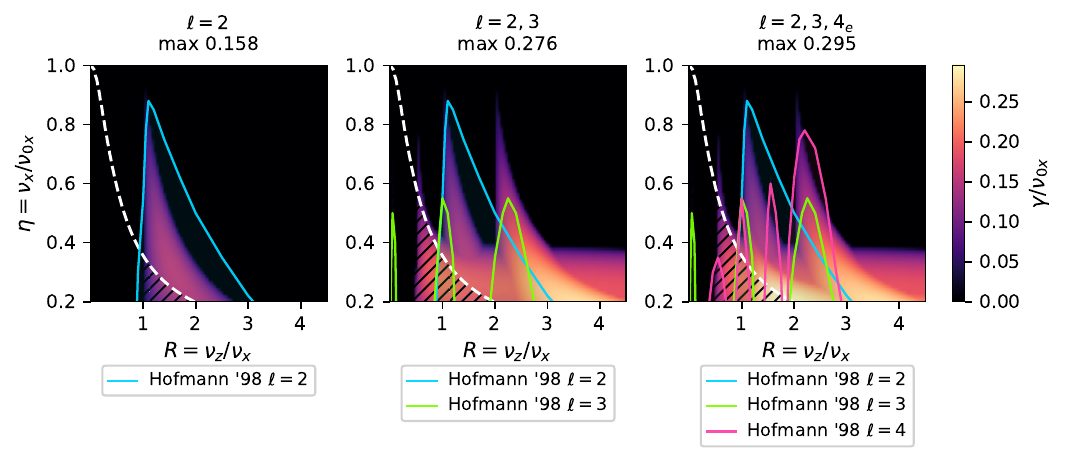}
  \caption{Higher-order dispersion-solver chart at
  $\varepsilon_{z}/\varepsilon_{x}=5.0$ (matching Hofmann 1998
  Fig.~10).  Panels show the solver growth rate $\gamma/\nu_{0x}$ for
  the cumulative mode sets $\ell=2$, $\ell=2,3$, and
  $\ell=2,3,4_{e}$; hand-digitized boundaries of Hofmann's
  corresponding instability regions are overlaid in cyan
  ($\ell=2$), green ($\ell=3$), and magenta ($\ell=4$).  All three panels
  share one colour scale so that the cumulative mode sets can be read against
  each other; the largest growth rate on each is $\gamma/\nu_{0x}=0.158$,
  $0.276$ and $0.295$ respectively, and no panel requires colour clipping.  Each cumulative mode set fills in
  under its corresponding digitized boundary --- the $\ell=2$ lobe
  under the cyan curve, the $\ell=3$ lobes under green, the
  $\ell=4_{e}$ lobes under magenta.  That agreement is qualitative as
  drawn; quantified as the minimum distance from each of the $68$ digitized
  boundary points to our $\gamma/\nu_{0x}\!=\!10^{-2}$ contour, it is
  median $0.14$, RMS $0.23$, maximum $0.52$ in $(R,\eta)$ --- against
  $0.27/0.43/0.98$ under the previous coordinate map (see text).  The out-of-domain corner lies at
  \emph{low} $R$ and low $\eta$, where the space-charge strength grows
  rapidly ($S^{2}\!=\!140$ at $R\!=\!0.3$, $\eta\!=\!0.3$ against
  $S^{2}\!=\!2.4$ at $R\!=\!2.7$, $\eta\!=\!0.3$); growth rates shown
  inside the hatching are formal only (see Sec.~\ref{sec:limits}).  The dashed white contour and
  the diagonally hatched region in each panel mark the
  $S^{2}\!=\!10$ in-domain boundary and the
  out-of-domain $S^{2}\!>\!10$ corner respectively
  (Sec.~\ref{sec:limits}, App.~\ref{app:convergence}).  No PIP-II
  period falls inside this specific $(R,\eta)$ corner; two periods
  in the low-energy SSR1 section sit at $S^{2}\!=\!14$ and $33$
  and are flagged as outside the adopted model domain
  (Sec.~\ref{sec:pip2}, Sec.~\ref{sec:limits}).}
  \label{fig:fig2}
\end{figure*}

At isotropy, the
$(1\!\mp\!2\hat\eta^{2}/\alpha)$ factors of
App.~\ref{app:anisotropic} give the reduction of Hofmann's Eq.~(36) to his
stated isotropic-limit form Eq.~(37).  Without them the residue at the $(9-s)^{-2}$
pole is $9/8$ rather than the $27/8$ demanded by Eq.~(37), and the
closing algebraic identity Eq.~\eqref{eq:appA-key-id}
(symbolically verified) fails.  The asymmetry with Hofmann's
Eq.~(41), whose printed $S^{4}$ block does carry the
$(1\!\mp\!\hat\eta^{2}/\alpha)$ factors on its $\alpha$-coupling
terms, indicates that the omission in Eq.~(36) is a localized print
artifact rather than a convention difference.  The solver is checked by two quantitative comparisons and one qualitative comparison.  \emph{First}, on
the isotropic plane ($\alpha=\hat\eta=1$) each anisotropic
$S^{2}/S^{4}/S^{6}$ block of our implementation reduces analytically
to Hofmann's published isotropic forms Eqs.~(37), (42), (46) to
machine precision (max relative error $1.7\!\times\!10^{-16}$ for
$\ell=3_{e}$, $4.4\!\times\!10^{-16}$ for $\ell=4_{e}$ across a dense
$(s,\,S^{2})$ test grid).  For $\ell=4_{e}$ the anchor is Eq.~(42) in
the sign-corrected form Eq.~\eqref{eq:appA-eq42-corrected}; against the
form as printed the reduction fails, which is how the misprint was
found (App.~\ref{app:l4e}).  \emph{Second}, away from isotropy our
corrected $\ell=3_{e}$ and $\ell=4_{e}$ branches diverge from the
legacy isotropic-limit prototype by a controlled, term-by-term amount
that is independently verifiable from Eq.~(36)/(41) structure; an internal regression test verifies this difference from the legacy implementation.
\emph{Third}, Fig.~\ref{fig:fig2} overlays the solver's cumulative
$\ell=2$, $\ell=2,3$, and $\ell=2,3,4_{e}$ charts at
$\varepsilon_{z}/\varepsilon_{x}=5.0$ with the hand-digitized
instability-region boundaries of Hofmann (1998) Fig.~10.  The solver
recovers the location of the dominant lobes; the agreement is
qualitative rather than quantitative because our root-finder isolates
the non-oscillatory ($\mathrm{Re}\,\omega=0$) modes only, while
Hofmann's published chart additionally marks oscillatory
($\mathrm{Re}\,\omega>0$) modes, and because the digitized boundaries
are coarse ($\pm0.1$ in $R,\eta$).

Reduction of the corrected anisotropic dispersion relations to the
isotropic-limit forms at $\alpha\!=\!\hat\eta\!=\!1$ holds to machine
precision.  For $\ell\!=\!3_{e}$, $\ell\!=\!3_{o}$ and $\ell\!=\!4_{o}$ the
observed maximum relative error is $\sim\!1.7\!\times\!10^{-16}$, verified by
comparison against the frozen isotropic-limit
prototype, to a tolerance of $10^{-12}$ over a randomized
$S^{2}\!\in\![0.1,5.0]$ and a dense $s$ grid.

$\ell\!=\!4_{e}$ is deliberately excluded from that comparison.  The prototype
encodes Eq.~(42) \emph{as printed}, and we adopt the sign-corrected form
Eq.~\eqref{eq:appA-eq42-corrected}, so the two differ at isotropy by
construction --- by exactly $2S^{4}\sum_{i}M_{i}$, which is verified by a regression test.  Against the
corrected Eq.~(42) the reduction holds to $4.4\!\times\!10^{-16}$ over
$8{,}000$ $(s,S^{2})$ pairs, and the sign resolution is verified symbolically
by substituting Eq.~(43)'s closed-form
roots.  Eq.~(43) determines the sign; comparison with the corrected Eq.~(42) verifies the implemented reduction.

\emph{Root-set completeness and the oscillatory channel.}\quad The
grid-scan-plus-bisection root finder locates sign changes of $D_{\ell}$ on
the real negative-$s$ axis.  That procedure is structurally blind to roots of
even multiplicity, to pairs of roots inside one grid interval, and to every
complex root.  For the ten-point PIP-II audit we retain the higher-order
branches $\ell=3_e,3_o,4_e$ used in the principal screen.  To bound what the
scan misses in that scope we cleared the denominators of each
$D_{\ell}$ symbolically, formed the exact numerator polynomial, and solved it
by companion matrix, so that the complete root set is obtained at once.
The polynomial coefficients and eigenvalue solve use double precision:
``exact-root'' below denotes this full-polynomial reference, not exact
arithmetic for its numerical roots.

For the sampled PIP-II points, the scan recovers the full-polynomial non-oscillatory result: across the PIP-II flagged set and a sample of quiet
periods --- ten points in all --- the largest discrepancy between the scanned
growth rate and the exact real-root growth rate is
$5.2\!\times\!10^{-17}$.  The reported values at those points are therefore
not limited by the root finder.

That comparison is of the largest growth per branch, which by itself would
not register a \emph{subdominant} root the scan missed, so we also compare
the root \emph{sets} directly, and over the census domain rather than at ten
points.  For $2000$ sampled
$(R,\eta,\varepsilon_{z}/\varepsilon_{x})$ and all four scan-based branches
we take the exact real roots lying inside the scan's own search window and
ask which of them the scan locates, and whether it returns anything that is
not a root at all.  Inside the adopted domain, at the
$1524$ sampled points with $S^{2}\!\le\!10$, all $678$ exact roots are
recovered, no spurious root is produced, and the largest matched-pair
discrepancy is $4.6\!\times\!10^{-14}$.

Of the $476$ sampled points with $S^{2}\!>\!10$, the current hybrid
scan recovers all $523$ exact roots, with $0$ missed and $0$ spurious roots;
the maximum matched-pair discrepancy is $2.0\!\times\!10^{-13}$.
The earlier uniform-only scan missed a close root pair.  We retain the example as a historical resolution test: one point --- at
$S^{2}\!=\!66.4$, on $\ell\!=\!3_{o}$ --- misses two roots, at
$s\!=\!-0.512$ and $s\!=\!-0.056$.  The search window scales as
$S^{2}(1+\hat\eta)^{2}$, so at that point it spans $[-3.3\!\times\!10^{4},0)$
on a fixed $2000$-point grid; both roots fall inside the single final cell,
where $D$ has the same sign at each end and no sign change exists to bisect.
At this point, the exact solve returns
$\gamma/\nu_{0x}\!=\!0.095$ on that branch where the scan returns zero, and
$\ell\!=\!2$ is quiet there, so the cell is one the ungated column would
otherwise have counted as higher-order-only.
This is the paired-root mode named above, and it is a statement about grid
resolution at large $S^{2}$, not about the dispersion relations.

The historical uniform-scan comparison above was measured on the
\emph{census} box ($R\!\in\![0.15,3]$, $\eta\!\in\![0.22,0.98]$),
and its missed-root rate does not transfer to other axes.  The earlier audit on the axes
of Table~\ref{tab:h2003}, which
reach $\eta\!=\!0.05$, found that the uniform scan missed $3.79\%$ of exact roots outside
the gate --- $48$ of $1266$ on the same four branches --- against $0.50\%$ on
the census box, a factor $7.6$.  Inside the gate that earlier scan missed nothing, $0$
of $374$.  Those historical counts also used the former fourth-order odd
closure.  The current audit recovers all $477$ in-domain and $1601$
out-of-domain roots of the four corrected branches, with no misses or
spurious roots.  Including the two printed third-order branches as separate
comparisons gives $749$ in-domain and $2675$ out-of-domain roots; both the
production hybrid scan and its supplementary refinement recover all of them.  The uniform-grid resolution explains the historical failure: at
$(R,\eta,\varepsilon_{z}/\varepsilon_{x})\!=\!(0.10,0.05,5.0)$ the window
$s_{\min}=-\max(200,\,20S^{2}(1+\hat\eta)^{2})$ reaches
$5.9\!\times\!10^{7}$ while the entire real-root support is
$|s|\!\le\!21.5$, so a $2000$-point uniform grid has spacing
$3.0\!\times\!10^{4}$ and every root of a branch can fall in one cell.

Every higher-order branch entering Table~\ref{tab:h2003} ---
including the shared $\ell\!=\!4_{e}$ --- is evaluated through the log-refined
near-zero scan.  Measured
as a root set, that refinement recovers the exact real roots at all $2000$
audited points, in and out of gate: $0$ missed, $0$ spurious, largest
matched-pair error $2.0\!\times\!10^{-13}$.  The current production
scanner already contains that logarithmic refinement; the supplementary pass
is retained as a regression cross-check.  The earlier refined audit had a
maximum matched-pair error of $4\!\times\!10^{-12}$, which we retain as
historical context rather than the current precision measurement.

The failure-mode lower-bound argument that follows is specific to the census
count; it describes the direction a missed root would induce, not a measured
correction still required by the current sampled audit.  The $\ell\!=\!2$ branch is solved in closed form --- its
roots come from a quadratic, not from the grid scan --- so it cannot lose a
root this way; only the higher-order branches can.  A miss can therefore only
\emph{remove} a higher-order flag from a cell, never create one, so the ungated
higher-order-only fraction is a lower bound with respect to this failure mode
rather than being inflated by it.  That reasoning does \emph{not} carry to a
flip count between two forms that share $\ell\!=\!4_{e}$, and its direction
reverses there: a cell where the shared branch exceeds threshold is unstable
under both forms and does not flip, so losing that branch \emph{manufactures} a
flip.  For Table~\ref{tab:h2003} the corresponding statement would be an upper
bound, which is one more reason that table is computed on the refined scan.  At the historical rate
--- one point in $476$ sampled outside the gate, two roots of $397$ --- the
implied correction was far below the spread reported for the factor
structure.  With the current scanner the corresponding audit has $0$ missed
roots of $523$ outside the gate, so that historical rate is no longer its
measured failure rate.  The sweep establishes completeness on the sampled set inside the adopted domain, extending the ten-point audit without proving completeness over the continuous domain.

The retained higher-order blocks also carry \emph{oscillatory} roots ---
complex-conjugate pairs $s$, $s^{*}$, giving $\mathrm{Re}\,\omega\!\neq\!0$ ---
at $5$ of the $10$ PIP-II periods we audited, and they appear at periods for
which the non-oscillatory analysis of the same branch returns exactly zero.
They are carried by $\ell\!=\!4_{e}$ (at $5$ periods) and by $\ell\!=\!3_{e}$
(at $3$); $\ell\!=\!3_{o}$ shows none.  The formal rates we extract span
$\gamma/\nu_{0x}\!=\!0.014$--$0.061$, comparable to the non-oscillatory rates
of Table~\ref{tab:pip2_flagged} rather than dominating them.  These formal rates require comparison with the published bound: Hofmann states that the
oscillatory instabilities left of the equipartitioning line have normalized
growth rates ``limited to $0.05$'' (his~p.~4723).  Three of our eight nonzero
oscillatory rates exceed that number, but the comparison must respect the
domain of his statement, which is the region left of equipartition,
$T\!<\!1$ with $T=a^{2}\nu_{x}^{2}/b^{2}\nu_{y}^{2}=\hat\eta^{2}/\alpha^{2}$
(his Eq.~(8)).  Evaluating $T$ from the corrected map --- note that it is
flip-dependent, equal to $R\,\varepsilon_{z}/\varepsilon_{x}$ with the
interchange and to its reciprocal without --- only two of the eight audited
rates lie at $T\!<\!1$: SSR1 idx~12 ($T\!=\!0.21$, $0.0605$) and SSR2 idx~18
($T\!=\!0.31$, $0.021$).  \emph{One} of those exceeds $0.05$, by $21\%$.  The
other two exceedances sit at $T\!=\!1.9$ and $1.4$, outside the region his
bound describes.  The discrepancy within the stated region is therefore confined to one rate,
which exceeds the bound by $21\%$.
We record it as an unresolved discrepancy and \emph{not} as a third erratum:
our solver discards oscillatory roots by construction, so it is not the
instrument with which to adjudicate a bound on them, and the two corrections
this paper does advance are independent of it.

Our caution about these magnitudes does \emph{not} rest on a truncation
argument.  Eqs.~(36) and~(41) are exact polynomials in $S^{2}$
(Sec.~\ref{sec:limits}), so there is no expansion to be least-controlled here.
Nor is the degeneracy one between different $\ell$: each $D_{\ell}$ is solved
independently, so the collisions are \emph{within} a single branch's own pole
set.  The colliding pair varies with the lattice period: at SSR1 idx~10 the $\ell\!=\!3_{e}$ pair
emerges between $(1\!-\!2\alpha)^{2}\!=\!0.72$ and $1$, and its
$\ell\!=\!4_{e}$ pair between $4\alpha^{2}\!=\!3.42$ and $4$ (as at idx~21
and~26); at SSR1 idx~12 the $\ell\!=\!4_{e}$ pair sits between $4$ and
$4(1\!-\!\alpha)^{2}\!=\!12.99$, and at SSR2 idx~18 between $16$ and
$4\alpha^{2}\!=\!18.93$.  These are genuine coupled-mode
instabilities, arising where two space-charge-free resonances of the same
order nearly coincide and space charge couples them.  Our caution is that the
$\delta$-function KV spectrum is precisely where such near-degeneracies are
most model-dependent.  We therefore regard the magnitudes as indicating a
non-negligible excluded channel rather than as quantitative predictions, and
we do not propagate them.
Their existence and onset are numerically resolved: they are
well-resolved conjugate pairs --- $|\mathrm{Im}\,s|/|\mathrm{Re}\,s|$ ranges
from $0.014$ to $0.251$ across the audited points, against a
$4.6\!\times\!10^{-13}$ imaginary floor on the genuinely real roots ---
eleven orders of magnitude of separation, so they are not numerical noise
--- they appear only above a finite onset in $S^{2}$ and vanish below it, as a
coupled-mode instability must.  Above the onset, the dependence is nonmonotonic: at SSR1 idx~10 the
$\ell\!=\!3_{e}$ rate is zero at $S^{2}\!=\!0.61$, first appears at $0.758$,
and rises to $0.056$ at the nominal $3.04$, but at SSR2 idx~18 the
$\ell\!=\!4_{e}$ rate peaks at $0.024$ near $S^{2}\!=\!2.4$ and has fallen to
$0.021$ by its nominal $2.66$, while SSR1 idx~12 shows two disjoint windows in
$S^{2}$ entirely.  The onset is robust; the shape above it is not a simple
scaling;
in the $\ell\!=\!4_{e}$ cases at idx~10, 21 and~26 they sit between the
$s\!=\!4$ and $s\!=\!4\alpha^{2}$ poles, i.e.\ on the near-degenerate $2\nu_{x}$/$2\nu_{y}$
pair that space charge splits into the complex plane.  We verified separately
that the $\ell\!=\!2$ quadratics have non-negative discriminant at every
PIP-II period, so this is additional structure in the higher-order relations
and not $\ell\!=\!2$ physics re-entering.  We note that the
$\ell\!=\!3_{e}$ oscillatory roots are not a consequence of the $S^{4}$
correction or of the Eq.~(42) sign: $D_{3,e}$ is untouched by the latter, and
the roots persist under all three factor structures of
Sec.~\ref{sec:pip2}.

The scope restriction to non-oscillatory modes is therefore a much stronger
restriction on this trajectory than the phrase ``often among the
operationally relevant ones'' conveys, and the flag counts in
Sec.~\ref{sec:pip2} should be read strictly as
\emph{non-oscillatory} counts.  Extending the solver to report the
oscillatory branch, and reconciling its magnitudes with Hofmann's stated
$0.05$ bound, is the most immediate extension of this work.

\emph{Quantitative check at the operating anisotropy.}\quad The
digitized comparison below is drawn at
$\varepsilon_{z}/\varepsilon_{x}\!=\!5$, because that is where Hofmann's
Fig.~10 is plotted --- a factor ${\approx}3.3$ from the ratio at which the
PIP-II conclusions of Sec.~\ref{sec:pip2} live, and carrying
hand-digitization noise of $\pm0.1$ per axis.  Hofmann's Fig.~13, however, is
drawn at $\varepsilon_{x}/\varepsilon_{y}\!=\!1.5$, essentially the PIP-II
mid-linac ratio, and in the accompanying text (his p.~4723) he states several
\emph{numerical} conclusions about it.  Comparison with the printed numerical statements avoids figure-digitization
uncertainty.

Writing his anisotropy parameter as $T=(\nu_{x}/\nu_{y})(\varepsilon_{x}/
\varepsilon_{y})$ --- consistent with his marking $T\!=\!1$ at
$\nu_{x}/\nu_{y}\!=\!0.2$ for $\varepsilon\!=\!5$ and at $0.667$ for
$\varepsilon\!=\!1.5$ --- two of his statements are directly testable.

First, he states that at $T\!=\!1/3$ the transverse tune depression ``must be
below $0.6$'' to enter the third-order non-oscillatory even-mode unstable
region, ``and even lower'' for the fourth order.  Our solver places the
$\ell\!=\!3_{e}$ onset at $\eta\!=\!0.577$ and the $\ell\!=\!4_{e}$ onset at
$\eta\!=\!0.425$: the $\ell\!=\!3_{e}$ value sits $3.8\%$ below his stated
bound, and the two are correctly ordered.
This numerical agreement is obtained at the operating anisotropy, rather than at $\varepsilon\!=\!5$.

Second, he suggests that ``the region of transverse tune depression between
$0.7$ and $1$ should be safe from a practical point of view.''  We find small
residual growth there: over $R\!\in\![0.05,1.2]$ our solver flags 34 grid
points with $\eta\!\in\![0.7,1)$, 13 of them in the practical band
$T\!\in\![0.5,2]$, with a maximum rate of $0.030$.  We read this as consistent
with his wording --- he claims practical safety, not vanishing growth, and
$\gamma/\nu_{0x}\!=\!0.03$ gives a local amplitude factor
$\exp(0.03\sigma_{0x})$ over one focusing period, with the zero-current
phase advance $\sigma_{0x}$ in radians.  Its excess is approximately $3\%$
only when $\sigma_{0x}\!\approx\!1$, using the measure of Sec.~\ref{sec:pip2} --- but we record it as a
residual discrepancy rather than a clean confirmation.

\emph{External consistency check against Hofmann's published
chart.}\quad We hand-digitized the boundaries of the unstable
lobes of Hofmann's Fig.~10 at $\varepsilon_{z}/\varepsilon_{x}\!=\!5$
($N\!=\!68$ boundary points across $\ell\!=\!2$ (1 lobe),
$\ell\!=\!3$ (3 lobes), and $\ell\!=\!4_{e}$ (4 lobes)) and
computed, for each digitized point, the minimum geometric distance
to the threshold contour $\gamma/\nu_{0x}\!=\!10^{-2}$ of our
corrected anisotropic solver in $(R,\eta)$ coordinates.  The
distribution of distances has median $0.14$, RMS $0.23$, and
maximum $0.52$ over the 68 boundary points.  (Under the previous
coordinate map the same statistic was $0.27/0.43/0.98$; the roughly
two-fold improvement is an independent, if coarse, corroboration of the
map correction of Sec.~\ref{sec:coords}, since the digitized boundaries
and the metric are unchanged.)  These distances are
still larger than the operational chart margin
$\Delta\eta\!\approx\!0.05\text{--}0.10$ relevant in
the Supplemental Material~\cite{SupplementalMaterial},
so this comparison cannot bound off-iso
solver errors at the level that matters for operational claims;
it is intended as a qualitative external consistency check, not as a
precision benchmark, and is not used either to calibrate the solver
or to support the quantitative PIP-II conclusions of
Sec.~\ref{sec:pip2}; the latter rest on the analytic
isotropic-limit reduction (Sec.~\ref{sec:track1validation}) and on
internal regression tests of the 516-test suite.  The comparison is
intentionally conservative because the present solver is restricted
to non-oscillatory roots; disagreement in regions dominated by
oscillatory branches is therefore \emph{expected} rather than
indicative of a solver defect.  Three known sources limit the
agreement at this resolution: (i)~hand-digitization noise on
the published chart at $\sim\!10$~px resolution ($\sim\!0.05\!-\!0.1$
per axis); (ii)~our root-finder isolating only the non-oscillatory
($\mathrm{Re}\,\omega\!=\!0$) modes while Hofmann's chart
additionally marks the oscillatory branches; (iii)~display of the
cumulative growth-rate envelope in Fig.~\ref{fig:fig2} versus
individual lobes in Hofmann's original.  A branch-resolved
re-scan of the published charts at $\ge\!300$~dpi, with
oscillatory-mode coverage added to the solver, is the natural
next step but is beyond the present scope.

\section{Derivation of the corrected $\ell=3_{e}$ block from
  Vlasov--Poisson}
\label{sec:vlasov-full}

Section~\ref{sec:track1} stated the corrected $\ell=3_{e}$ block without
justifying it.  That correction can be reached by several arguments from
\emph{internal consistency} --- the determinant structure of Hofmann's
Eq.~(35), the isotropic-limit reduction to his Eq.~(37), the residue pattern,
and the surface term --- and App.~\ref{app:anisotropic} gives each in full.
None of them is sufficient on its own: the surface term does not supply the
residues' common $\hat\eta$ weight, and the determinant argument must assume
Eq.~(36) is complete in which pole pairs it contains.  This section closes both
gaps by evaluating Hofmann's Eqs.~(19) and~(23) directly, volume term included.  This calculation derives the factors without imposing
Eq.~(37) and reproduces Hofmann's $\ell=2$ relations as
controls and, for $\ell=4_{e}$, his printed relation exactly in
$(\sigma,S^{2})$ at twelve independent anisotropy pairs.  These control calculations distinguish the $\ell=3$ equation discrepancy from an implementation error.

\paragraph*{The system.}  Eq.~(19) gives the perturbed Poisson equation with
two pieces: a $\delta$-function surface charge on the beam boundary, and a
volume charge.  For a polynomial interior ansatz these separate cleanly.  The
volume piece is the interior equation, and Eq.~(23) --- built from the same
surface piece --- is the boundary jump:
\begin{equation}
  \nabla^{2}\Phi = \frac{C_{V}S^{2}}{2\pi}\,\mathcal{V}[\Phi], \qquad
  \Bigl[\tfrac{\partial\Phi}{\partial\xi}\Bigr]_{\xi_{0}-0}^{\xi_{0}+0}
    = C_{S}S^{2}\,\mathcal{S}[\Phi],
  \label{eq:appA-vlasov-system}
\end{equation}
where the intensity enters only through the explicit $S^{2}$, so both sides are
\emph{linear} in $S^{2}$ --- a fact we return to below.  Writing
$\mathcal{K}[\,\cdot\,] \equiv (e^{-i\sigma L}-1)^{-1}\!\int_{0}^{L}
e^{-i\sigma u}\,[\,\cdot\,]\,du$ for Hofmann's periodic-orbit average, with
$L=2\pi m$ and
$\mathcal{G}\equiv p_{x}'\,\partial\Phi/\partial x'
 + T p_{y}'\,\partial\Phi/\partial y'$, the two operators are
\begin{align}
  \mathcal{S}[\Phi] &= \mathcal{K}\bigl[\,\mathcal{G}\,\bigr]_{P^{2}=0},
  \label{eq:appA-surface-op}\\
  \mathcal{V}[\Phi] &= \mathcal{K}\Bigl[\int_{0}^{2\pi}\!
    \frac{d\mathcal{G}}{dP^{2}}\Bigl|_{P^{2}=P_{s}^{2}}\,d\Theta\Bigr],
  \label{eq:appA-volume}
\end{align}
evaluated on the KV shell
$P_{s}^{2}=m^{2}\gamma^{2}\bigl[\nu_{x}^{2}(a^{2}-x^{2})-T\nu_{y}^{2}y^{2}\bigr]$.
Throughout this subsection we set $\nu_{x}\!=\!a\!=\!m\gamma\!=\!1$, which
fixes the units of $\sigma$, of length, and of momentum respectively; no
generality is lost because the dispersion relation depends only on the
dimensionless $(\sigma,S^{2},\alpha,\hat\eta)$.
Only even powers of $P$ survive the $\Theta$ average, so the shell substitution
introduces no square roots and $\mathcal{V}$ is again polynomial.  Matching
Fourier harmonics in the elliptic angle $\psi$ turns
Eq.~\eqref{eq:appA-vlasov-system} into a homogeneous linear system in the
expansion coefficients; the dispersion relation is its vanishing determinant,
exactly as Hofmann states below his Eq.~(23).

\paragraph*{Selection of the independent equations.}
Matching Eq.~\eqref{eq:appA-vlasov-system} produces more conditions than
unknowns.  We use the interior conditions at the
\emph{leading} monomial order together with the \emph{leading} boundary
harmonic $k\!=\!\ell$.  For $\ell\!\le\!3$ that is every interior condition;
for $\ell\!=\!4_{e}$ it is not, and the surplus is accounted for below:
\begin{center}
\begin{tabular}{lccc}
\hline\hline
mode & coeffs & residual monomials & harmonics \\
\hline
$\ell\!=\!2_{o}$ & 1 & --- (none) & $\{2\}$ \\
$\ell\!=\!2_{e}$ & 2 & $1$ & $\{0,2\}$ \\
$\ell\!=\!3_{e}$ & 2 & $x$ & $\{1,3\}$ \\
$\ell\!=\!4_{e}$ & 3 & $1,\,x^{2},\,y^{2}$ & $\{0,2,4\}$ \\
\hline\hline
\end{tabular}
\end{center}
The residual $\mathcal{R}[\Phi]=\nabla^{2}\Phi-(C_{V}S^{2}/2\pi)\,
\mathcal{V}[\Phi]$ is not $\nabla^{2}\Phi$: because $\mathcal{V}$ is evaluated
on the KV shell $P_{s}^{2}=1-x^{2}-\hat\eta^{2}y^{2}$ it contributes a constant
that the Laplacian does not.  It is a polynomial of degree $\ell-2$ carrying
the parity of $\Phi$, so for $\ell\!\le\!3$ its monomial count plus one equals
the number of coefficients and the system is square.  For $\ell\!=\!4_{e}$ the
count is three ($1$, $x^{2}$, $y^{2}$), so with the three available harmonics
$\{0,2,4\}$ the pool is six conditions for three coefficients and the system
is over-determined.  For $\ell\!=\!2_{o}$ the count is $0+1=1$: there is a single
coefficient and hence no determinant to take, which is why Eq.~(32) is an
expression rather than a $2\!\times\!2$ condition.

The lower harmonics $k\!<\!\ell$ are not discarded arbitrarily.  They fix the
lower-order terms of $\Phi$ --- the ones Hofmann explicitly sets aside when he
writes that ``only the leading terms in the $x,y$ expansion of $\Phi$ are
needed to determine the eigenfrequency'' (his Sec.~IV) --- and those terms do
not enter the leading-order determinant.  Using them instead of the leading
harmonic does not reproduce the printed relations, providing a check on the equation selection: of the
$\binom{6}{3}\!=\!20$ possible three-equation subsets for $\ell\!=\!4_{e}$,
only $\{x^{2},\,y^{2},\,k\!=\!4\}$ returns Eq.~(41), as
$\det=-12\,\hat\eta^{-4}D_{4,e}$; the other nineteen give ratios that depend
on $\sigma$, as verified by symbolic regression tests.

The surplus condition is not a defect of the selection.  Restoring the
lower-order even terms $b_{2}x^{2}+b_{4}y^{2}$ that Hofmann sets aside leaves
the $x^{2}$, $y^{2}$ and $k\!=\!4$ rows free of $b_{2}$ and $b_{4}$, so the
system is block triangular in the leading coefficients and the surplus row is
a consequence of the three retained ones rather than an independent
constraint.  The over-determination is therefore apparent, not real.

\paragraph*{Normalization.}  The two constants $C_{V},C_{S}$ carry the
prefactors of Eqs.~(19)/(23).  Rather than track them through the
$\delta$-function normalization we fix them \emph{once}, by requiring that
$\ell\!=\!2$ even reproduce the undisputed printed Eq.~(28).  That fit is
unique --- a rational function in $\sigma$ and $S^{2}$ has many independent
coefficients and all must match simultaneously --- and returns
\begin{equation}
  C_{V} \to 2, \qquad C_{S} \to 1/\hat\eta .
  \label{eq:appA-vlasov-consts}
\end{equation}
That $C_{S}=1/\hat\eta=b/a$ is the metric factor an elliptic-coordinate jump
condition must carry is a check on the fit rather than an input to it.

\paragraph*{Results.}  With Eq.~\eqref{eq:appA-vlasov-consts} held fixed, every
other relation is a parameter-free prediction.  Writing $\ell$ for the mode
order, the derived determinants satisfy
\begin{equation}
  \det\mathbf{M}_{\ell} = c_{\ell}\,\hat\eta^{-\ell}\,D_{\ell},
  \qquad c_{2_{e}}=2,\ c_{3_{e}}=\tfrac{3}{2},\ c_{4_{e}}=-12,
  \label{eq:appA-vlasov-scales}
\end{equation}
for the even modes.  $\ell\!=\!2$ odd has a single expansion coefficient and
so no determinant to take; its single jump equation reproduces $D_{2,o}$ with scale
$-1/(2\hat\eta^{2})$.  In each case the system is the interior Poisson
equation(s) together with the \emph{leading} boundary harmonic
($k\!=\!\ell$).  The verification has two levels.  For $\ell\!=\!2$ and $\ell\!=\!3$ --- both parities in each
case --- and for $\ell\!=\!4_{o}$, the identity is established
\emph{symbolically}, holding identically in all four of $\sigma$, $S^{2}$,
$\alpha$, $\hat\eta$.  For $\ell\!=\!4_{e}$ the fully
symbolic $3\!\times\!3$ determinant is impractically large, so we substitute
rational $(\alpha,\hat\eta)$ before forming it and keep $\sigma$ and $S^{2}$
symbolic; the identity then holds exactly in $(\sigma,S^{2})$ at each of
twelve independent $(\alpha,\hat\eta)$ spanning $\alpha\!\in\![1/3,11/6]$
and $\hat\eta\!\in\![6/5,7/2]$, with the same constant $-12\hat\eta^{-4}$
at every one.  The twelve are chosen on three grounds: they are rational, so
the determinant is formed and compared in exact arithmetic with no
floating-point tolerance anywhere; they are balanced six below and six above
isotropy in $\alpha$, so the sample is not one-sided in the variable the
disputed factors depend on; and none lies on the
isotropic plane $\alpha\!=\!1$ itself, where the disputed factors degenerate
to constants and the test would carry no information.  This establishes the identity at the tested rational points, but is not a symbolic proof in $(\alpha,\hat\eta)$.  Specifically:

\begin{itemize}
\item $\ell\!=\!2$ odd: $\Phi=a_{1}xy$ is harmonic, so the volume term
  vanishes identically and only the jump condition survives.  The raw surface integrand reproduces
  Eq.~(32)'s $S^{2}$ bracket exactly, including the factor $\tfrac12$ Hofmann
  prints, using no fitted constant at all.  Assembling the actual $k\!=\!2$
  jump equation --- which does use $C_{S}$ --- reproduces the whole of
  $D_{2,o}$ with scale $-1/(2\hat\eta^{2})$.  The assembled equation provides a cross-check: $C_{S}$ was fixed on
  $\ell\!=\!2$ \emph{even} and is not refitted here, so its working
  unchanged on the odd mode is a prediction.
\item $\ell\!=\!3$ even: the derived determinant equals
  $\tfrac{3}{2}\hat\eta^{-3}D_{3,e}$ with $D_{3,e}$ the \emph{corrected}
  Eq.~(36), i.e.\ including the $(1\!\mp\!2\hat\eta^{2}/\alpha)$ factors on the
  $S^{4}$ $\alpha$-coupling residues, and including the $(3+\hat\eta)$ weight
  that the surface term alone did not supply.  Repeating the comparison against
  the printed (unfactored) form and against the $\ell\!=\!4$-style
  $(1\!\mp\!\hat\eta^{2}/\alpha)$ form fails in both cases: neither gives a
  ratio independent of $\sigma$.
\item $\ell\!=\!3$ odd: the same calculation run on the odd-parity basis
  $\Phi=b_{0}x^{2}y+b_{2}y^{3}$, with the boundary harmonic taken in the sine
  series, yields a determinant equal to $\tfrac{3}{2}D_{3,o}$, where $D_{3,o}$
  is the \emph{corrected} Eq.~(36) under Hofmann's interchange --- exactly, and
  symbolically in all four of $\sigma$, $S^{2}$, $\alpha$, $\hat\eta$.  This independently derives the odd branch carrying the higher-order PIP-II flag
(Sec.~\ref{sec:pip2}). The interchange rule --- the $\alpha\!\to\!1/\alpha$,
  $\hat\eta\!\to\!1/\hat\eta$ interchange of his p.~4719, with the
  $S^{2}\!\to\!S^{2}/\alpha^{2}$, $s\!\to\!s/\alpha^{2}$ rescaling that
  Eqs.~(24)/(25) force --- is recovered by the calculation.  Repeating the comparison against the printed and the
  $\ell\!=\!4$-style structures fails on this parity too: neither gives a
  ratio independent of $\sigma$.

\item $\ell\!=\!4$ odd: the odd quartic basis $\Phi=c_{1}x^{3}y+c_{3}xy^{3}$
  has two coefficients rather than three, so unlike $\ell\!=\!4_{e}$ its
  determinant is tractable \emph{fully symbolically}.  It equals
  $\tfrac{3}{4}\hat\eta^{-4}D_{4,o}$ with $D_{4,o}$ a relation whose $S^{2}$
  block is Eq.~(45) exactly as printed and whose $S^{4}$ block reduces at
  $\alpha\!=\!\hat\eta\!=\!1$ to Eq.~(46) exactly --- so the isotropic closure
  is \emph{confirmed} as the limit of the anisotropic block rather than
  assumed to stand in for it.  The anisotropic $S^{4}$ block itself, which
  Hofmann does not print, has poles at the six Eq.~(45) families with the two
  sum and difference pairs $(1\!\mp\!\alpha)^{2}$ doubled, as a
  $2\!\times\!2$ determinant over those residues requires, and its residues
  each carry the coupling factor of their pole family --- $(1\!\mp\!3\alpha)
  (1\!\mp\!3\hat\eta^{2}/\alpha)$, $(3\!\mp\!\alpha)(3\!\mp\!\hat\eta^{2}/\alpha)$,
  $(1\!\mp\!\alpha)(1\!\mp\!\hat\eta^{2}/\alpha)$ --- exactly the pattern
  App.~\ref{app:anisotropic} infers from the printed relations.  Its closed
  form is implemented in the solver.  One symmetry checks the whole block: the
  odd quartic basis is invariant under $x\!\leftrightarrow\!y$, so
  $D_{4,o}$ must be invariant under the interchange of Sec.~\ref{sec:coords}
  up to normalization, and it is --- symbolically, the ratio is
  $\hat\eta^{4}$ with no $\sigma$ or $S^{2}$ dependence.

\item $\ell\!=\!4$ even: the derived determinant equals $-12\hat\eta^{-4}
  D_{4,e}$ with $D_{4,e}$ \emph{exactly as printed} --- the
  $(1\!\mp\!\hat\eta^{2}/\alpha)$ factors and the $-S^{4}$ sign --- at each of
  the twelve $(\alpha,\hat\eta)$, this being exact verification at the sampled rational points rather than the full symbolic identity distinguished above.  The
  $+S^{4}$ form is rejected, independently confirming the Eq.~(42) resolution
  of App.~\ref{app:l4e} by a route that does not use Eq.~(43).
\end{itemize}

The kinetic derivation establishes the $S^{4}$ correction and excludes both
alternative factor structures at every $(\alpha,\hat\eta)$, extending the
isotropic reduction argument.  Within Hofmann's leading-polynomial ansatz, the derivation also resolves the completeness assumption --- the pole set is an output of
Eqs.~\eqref{eq:appA-vlasov-system}, not an input, so the question of whether
Eq.~(36) omits pole pairs no longer arises at that order.  Because the $\ell\!=\!4$ ratio is also
$\sigma$-independent, Eq.~(41)'s pole set is reproduced
exactly, which settles the four-pair question raised above.

The derivation is verified by symbolic tests of the determinant identities.
These computationally demanding checks provide the evidence for the claims
in this subsection.

\subsection{Comparison with the 2017 monograph}
\label{sec:book-confirmation}

Hofmann's 2017 monograph~\cite{HofmannBook2017} provides
a later treatment of these modes and a comparison for both corrections.

The monograph does not print an
anisotropic dispersion relation of any order above the second anywhere:
the only dispersion equations it states are the second-order even
and odd forms (its Eqs.~(5.11) and~(5.17)).  For every higher order it refers
the reader back to the 1998 paper --- ``For the full dispersion relation
expression we refer to [2]'' in third order, and ``the fourth order even mode
dispersion relation in [2]'' in fourth, where the monograph's own Ref.~[2] is Ref.~\cite{Hofmann1998} here.  Neither Eq.~(36) nor Eq.~(42) is restated, corrected,
or remarked upon.  Both errors therefore stand uncorrected in the author's own
2017 treatment, and the 1998 paper remains the reference text for these
relations.

The monograph prints
round-beam closed-form roots --- its Eq.~(5.22) in third order and Eq.~(5.26)
in fourth --- and these are satisfied by our corrected relations and not by the
printed ones.  These expressions are character-for-character reproductions of Hofmann's own Eqs.~(38) and~(43), reprinted with the
idiosyncratic unexpanded groupings $4(64-2\sigma_{p}^{2})$ and
$8(128+10\sigma_{p}^{2})$ intact.  Testing our relations against them is
therefore the same test as testing them against Eqs.~(38) and~(43), which
Sec.~\ref{sec:track1} and App.~\ref{app:anisotropic} already do.  In the
$\ell\!=\!3$ case it is weaker still: Eq.~(38) gives the roots of Eq.~(37) by
construction, and our primary argument is precisely that the corrected Eq.~(36)
reduces to Eq.~(37).

An internal consistency check using Eq.~(43) cannot exclude a misprint in
that equation's root set, and its reprinting by the same author provides
limited additional evidence. An independent comparison is available from
Gluckstern~\cite{Gluckstern1970Modes},
who derived the round-beam two-dimensional spectrum twenty-eight years
earlier using coupled integral equations and a hypergeometric eigenfunction
basis on a round cross section. His Eq.~(30) is exactly the $\sigma^{2}_{2,3}$ pair of Eq.~(43).  Over $\sigma_{p}^{2}\!\in\!\{0.5,1,2,4,8,16,31\}$ that pair satisfies Eq.~(41)
with its printed $-\sigma_{p}^{4}$ to $2.3\!\times\!10^{-13}$ and rejects the
flipped sign by $11.5$--$51.7$ --- fourteen orders of separation.  So two
of the five roots are corroborated by an independent derivation in a different
formalism; the remaining three, and the entire anisotropic structure, still rest on the
Vlasov--Poisson calculation of this section, which excludes the rejected sign
by a route that uses neither the monograph nor Eq.~(43).  

The monograph also gives a useful external consistency check in a different
representation, while retaining its dependence on those closed-form roots.  Its Table~5.1 lists
first-order coherent tune-shift coefficients $F_{k}$, defined by
$\omega_{k}=k(\nu_{x}+F_{k}\Delta\nu_{x})$ for the highest-frequency branch of
each even mode.  These are pure numbers rather than root sets, but the book obtains them by
expanding its own root formulas, Eqs.~(5.22) and~(5.26), at small space charge.
They therefore provide a consistency check against that source, not an
additional theoretical derivation independent of those roots.  Its equal-tune row gives
$F_{3}\!=\!1/4$ and $F_{4}\!=\!3/16$ (its Eqs.~(5.23) and~(5.27)), and the
1998 relations must reproduce them.  At the round beam
$\alpha\!=\!\hat\eta\!=\!1$ they do --- but only in corrected form:

\begin{center}
\begin{tabular}{lccc}
\hline\hline
mode & Table~5.1 & this paper & as printed \\
\hline
$\ell\!=\!3$ & $1/4=0.250000$    & $0.2500000$ & $0.3110042$ \\
$\ell\!=\!4$ & $3/16=0.187500$   & $0.1874999$ & $0.4257733$ \\
\hline\hline
\end{tabular}
\end{center}

The corrected Eq.~(36) and Eq.~(41)'s printed $-\sigma_{p}^{4}$ reproduce both
entries to seven decimals; the printed Eq.~(36) misses by $24\%$ and the
rejected $+\sigma_{p}^{4}$ by $127\%$.  So Hofmann published, in 2017, two
numbers that his own 1998 Eqs.~(36) and~(42) as printed do not reproduce.

The monograph's other coefficients do not distinguish the two forms.  Its split-tune row and its
anisotropic Eq.~(5.20) are reproduced by both forms, at every $\hat\eta$ we
tested.  This follows from the pole structure: away from $\alpha\!=\!1$ the
$\alpha$-coupling poles $(1\!\mp\!2\alpha)^{2}$ sit off resonance and the
disputed factors contribute nothing at first order in $\Delta\nu$.  They become
resonant only as $\alpha\!\to\!1$ --- which is exactly where the printed
Eq.~(36) fails to reduce to Eq.~(37).  The two observations are the same
observation, reached from a later book rather than from the 1998 paper's own
isotropic form.

\section{Effect of the corrections on stability classification}
\label{sec:census}

Hofmann established that the $\ell\!=\!3$ and $\ell\!=\!4$ channels go unstable
over extended regions of the chart, and mapped them: his Fig.~10 charts
second-, third- and fourth-order modes at
$\varepsilon_{x}/\varepsilon_{y}\!=\!5$, and his Fig.~13 the third- and
fourth-order modes at $1.5$.  Using the correction derived in
App.~\ref{app:l3e-derivation}, we determine where evaluating the published
$\ell\!=\!3$ relation changes a stable/unstable classification relative to
the corrected Eq.~(36).

We classify every cell of a $60\!\times\!60$ grid over $R\!\in\![0.15,3]$,
$\eta\!\in\![0.22,0.98]$ twice --- once with Eq.~(36) exactly as printed, once
with the corrected $S^{4}$ factors --- and count the cells whose verdict
\emph{differs}.  We count the symmetric difference rather than the net change
in the flagged fraction, because those are not the same quantity: at
$\varepsilon_{z}/\varepsilon_{x}\!=\!5.0$ the net inside the gate is $5$ cells
while $39$ actually flip, the remainder cancelling in pairs.  Only the
$\ell\!=\!3$ branches carry the disputed factors, so $\ell\!=\!2$ and
$\ell\!=\!4_{e}$ are evaluated once per cell and shared between the two
variants.

\begin{table*}[t]
\caption{Cells whose stable/unstable verdict differs between Hofmann's printed
Eq.~(36) and the corrected form, $\gamma_{\rm th}\!=\!10^{-2}$, on a
$60\!\times\!60$ grid ($3600$ cells).  ``printed'' counts cells the printed relation flags as higher-order-only and
the corrected one does not; ``corrected'' the reverse; ``differ'' is their sum,
i.e.\ the symmetric difference.  Counts are split by the gate and percentages
taken against the corresponding population, as in Table~\ref{tab:h2003}; a
single whole-grid figure would mix the two and is not quoted.  Inside the $S^{2}\!\le\!10$ domain the disagreement is small and
changes sign with anisotropy; outside it the printed relation flags more cells
at every ratio.}
\label{tab:misprint}
\begin{ruledtabular}
\begin{tabular}{lrrrrrrr}
 & & \multicolumn{3}{c}{inside $S^{2}\!\le\!10$} & \multicolumn{3}{c}{outside} \\
\cline{3-5}\cline{6-8}
$\varepsilon_{z}/\varepsilon_{x}$ & in gate & printed & corrected & differ & printed & corrected & differ \\
\hline
1.2 & 2332 & 0 & 18 & 18 \ (0.77\%) & 238 & 41 & 279 \ (22.00\%) \\
1.5 & 2419 & 17 & 34 & 51 \ (2.11\%) & 224 & 39 & 263 \ (22.27\%) \\
2.5 & 2787 & 30 & 17 & 47 \ (1.69\%) & 147 & 6 & 153 \ (18.82\%) \\
5.0 & 2999 & 22 & 17 & 39 \ (1.30\%) & 80 & 0 & 80 \ (13.31\%) \\
8.0 & 3029 & 8 & 14 & 22 \ (0.73\%) & 38 & 0 & 38 \ (6.65\%) \\
\end{tabular}
\end{ruledtabular}
\end{table*}

Inside the adopted reporting
domain for the $\ell\!\le\!4$ KV chart the misprint changes relatively few classifications and has no preferred direction: it moves $0.73$--$2.11\%$ of the
in-gate cells, and which relation flags more changes sign across the ratios where that
direction is stable under grid refinement.  That figure understates what a benchmark would encounter, because it
is diluted by the large majority of in-gate cells where the higher-order
channels are silent and the two forms agree trivially.  Conditioned on the
cells whose verdict those channels alone decide, the printed and corrected
$S^{4}$ blocks disagree on $12.2$, $26.3$, $9.8$, $6.1$ and $3.6\%$ at the
five ratios --- up to one cell in four.  A measurement is in any case sited in
a stop band and reports a rate rather than a verdict, and on that same
conditioning set the two forms differ in growth rate by more than
$\gamma_{\rm th}$ on $50$--$66\%$ of cells.  We tested the per-ratio direction
against grid resolution rather
than asserting it from the production grid alone, because the margins
are thin --- $22$ against $17$ cells at
$\varepsilon_{z}/\varepsilon_{x}\!=\!5.0$.  Across $40\!\times\!40$,
$60\!\times\!60$ and $90\!\times\!90$ the direction is stable at four of the
five ratios: the corrected form flags more at $1.2$ and $8.0$, the printed
form at $2.5$ and $5.0$.  At $1.5$ it is not stable --- printed, corrected,
printed as the grid refines --- so we do not assign a direction there, and no
claim in this paper rests on one.  Among the cells the gate \emph{excludes}, the disagreement affects
$6.65$--$22.27\%$ of cells --- an order of magnitude more --- and it acquires
a direction: the printed relation flags more at every ratio, by at least
$5$:$1$, and at $\varepsilon_{z}/\varepsilon_{x}\!\ge\!5$ the corrected form
flags nothing outside the gate that the printed one does not.  For Eq.~(36), where the $S^{4}$ terms are small the two forms are
nearly interchangeable, and where they are not, the printed block
over-predicts.  The two percentages have different denominators:
in-gate flips are counted against the cells the gate admits, excluded-cell
flips against the cells it rejects, and the whole-grid figure --- which mixes
them --- is $1.67$--$8.72\%$ and belongs to neither statement.

That has a consequence for the published literature, because these relations
have been used to draw charts.  Hofmann \emph{et al.}~\cite{HofmannFranchetti2003}
present anisotropic stability charts in the same plane we use --- growth rate
over $(\nu_{z}/\nu_{x},\,\nu_{x}/\nu_{0x})$ at
$\varepsilon_{z}/\varepsilon_{x}\!=\!0.6$, $1.2$, $2$, $3$ and $5$ (their
Figs.~11 and~12) --- and state that their analytical basis is ``the dispersion
relations in Ref.~[7],'' which is Hofmann's 1998 paper.  They also take the
maximum over \emph{non-oscillatory} modes up to fourth order, which is the
convention adopted here.  Those charts are therefore computable under both
forms of Eq.~(36).

We evaluate both relations on the published axes. The original code, grid and contour levels are
unavailable, so this comparison measures the effect of the equation choice
on those axes; it does not reproduce the published figures or establish an
error in a particular published contour.  Their comparisons extend beyond the domain where the misprint has a small effect: the tightest quantitative
agreement they report, the KV stop-band width of their Fig.~8 at
$\nu_{x}/\nu_{0x}\!=\!0.8$, sits at $S^{2}\!\approx\!1$, but their broader
scans at $\nu_{x}/\nu_{0x}\!=\!0.5$ reach $S^{2}$ of order $10^{2}$.

Table~\ref{tab:h2003} reports the comparison with the same population-specific denominators as Table~\ref{tab:misprint}: separately for the
cells their axes place inside $S^{2}\!\le\!10$ and for those they place
outside.  Inside the gate the two forms
disagree on $1.19$--$2.88\%$ of cells --- comparable to the
$0.73$--$2.11\%$ found on our own census grid, providing a consistency check across the two chart windows --- and the direction is \emph{not} uniform: at
$\varepsilon_{z}/\varepsilon_{x}\!=\!1.2$ the corrected form flags more, not
the printed one ($113$ against $97$, strictly one-sided over $16$ flips).  At
$0.6$ the two are within a single cell of each other ($96$ against $95$ over
$21$ flips), which is a net rather than a direction, and we assign none.  Outside the gate the disagreement is $10.89$--$21.04\%$ and the direction is
uniform, the printed relation declaring $14$--$24\%$ more of that region
unstable at every ratio.  Their published panels reach
$\nu_{x}/\nu_{0x}\!=\!0.05$, so on average $46\%$ of the area we evaluate
lies outside the gate, where on the census box only $24.6\%$ of cells lie
beyond it; $88\%$ of all flips fall there.  A single whole-grid percentage over
these axes would mix the two populations and is not quoted.  The largest single growth-rate difference on those axes is $0.2296$ in units
of $\nu_{0x}$, but it sits at $S^{2}\!=\!43.7$ and so outside the domain; the
largest in-gate difference is $0.1480$.  Both values are reported with their domain membership. Comparison with the published grey scale also requires its normalization:
Ref.~\cite{HofmannFranchetti2003} calibrates growth per betatron period, i.e.\
$2\pi\,\mathrm{Im}\,\omega/\nu_{0x}$ --- their own p.~024202-5 converts a rate
of $0.03$ to $(2\pi\!\times\!0.03)^{-1}\!\approx\!5.3$ betatron wavelengths ---
so the same difference reads $1.44$ on their axes, against a scale maximum of
about $1.2$.

The mode aggregation differs slightly from the published one: Ref.~\cite{HofmannFranchetti2003} takes maxima over
``all nonoscillatory modes including second (odd), third, or fourth order,''
whereas our principal census omits $\ell\!=\!4$ odd and includes
$\ell\!=\!2$ even.  These differences do not compensate: for physical
positive parameters and $s<0$, every term in Eq.~(28) is positive, so the even
second-order branch contributes no non-oscillatory instability.  Omitting
$\ell\!=\!4_o$ can only leave the combined growth unchanged or reduce it.
A shared unstable fourth-order odd branch can also mask a printed/corrected
third-order disagreement.  We retain the restricted comparison and quantify
this scope sensitivity on the census grid in Sec.~\ref{sec:limits}, rather
than presenting it as an exact reproduction of their aggregation.

\begin{table*}[t]
\caption{Effect of the Eq.~(36) misprint evaluated on the published axes of
Ref.~\cite{HofmannFranchetti2003} (their Figs.~11 and~12).  Measured from a
$400$~dpi render, their panels span
$\nu_{z}/\nu_{x}\!\in\![0.000,2.484]$ and
$\nu_{x}/\nu_{0x}\!\in\![0.024,0.987]$; we evaluate
$\nu_{z}/\nu_{x}\!\in\![0.10,2.50]$, $\nu_{x}/\nu_{0x}\!\in\![0.05,0.90]$ on
$48\!\times\!48$ cells, which truncates the low-$R$ and low-$\eta$ corners
where $S^{2}$ diverges, truncates the top of the ordinate at $0.90$, and
over-runs the abscissa by $0.016$.  Also $\gamma_{\rm th}\!=\!10^{-2}$, maximum over
non-oscillatory modes to fourth order (the aggregation differs slightly from theirs; see text).  Counts are split by the
$S^{2}\!\le\!10$ gate, and percentages are taken against the corresponding
population.  ``corr.''/``print.'' count the cells each form declares unstable;
``differ'' is the symmetric difference.  This is our relations evaluated on
their axes, not a reproduction of their figures; Sec.~\ref{sec:fig6}
finds that the tested published curves agree with the corrected form and
reject the printed one, supporting that inference about their computation.}
\label{tab:h2003}
\begin{ruledtabular}
\begin{tabular}{lrrrrrr}
 & \multicolumn{3}{c}{inside $S^{2}\!\le\!10$} & \multicolumn{3}{c}{outside} \\
\cline{2-4}\cline{5-7}
$\varepsilon_{z}/\varepsilon_{x}$ & corr. & print. & differ & corr. & print. & differ \\
\hline
0.6 & 96 & 95 & 21 \ (2.04\%) & 855 & 1061 & 268 \ (21.04\%) \\
1.2 & 113 & 97 & 16 \ (1.48\%) & 859 & 1008 & 191 \ (15.64\%) \\
2.0 & 307 & 323 & 36 \ (2.88\%) & 788 & 926 & 138 \ (13.11\%) \\
3.0 & 549 & 562 & 23 \ (1.65\%) & 672 & 769 & 99 \ (10.89\%) \\
5.0 & 672 & 683 & 17 \ (1.19\%) & 593 & 699 & 106 \ (12.09\%) \\
\end{tabular}
\end{ruledtabular}
\end{table*}

\subsection{Comparison with the mode-resolved curves of
Ref.~\cite{HofmannFranchetti2003}}
\label{sec:fig6}

Table~\ref{tab:h2003} quantifies the effect of implementing Eq.~(36) as
printed. To compare with the published calculations, we use Fig.~6 of
Ref.~\cite{HofmannFranchetti2003}. Unlike the contour maps in its Figs.~11
and~12, Fig.~6 provides mode-resolved quantitative data:
separate labelled curves for each order at a stated
$\nu_{x}/\nu_{0x}\!=\!0.5$, with ordinate $\mathrm{Im}\,\omega/\nu_{0x}$.
Individual $\ell\!=\!3$ features can therefore be read off and tested
against both forms.

The parity labels require a change of convention. Our ``even'' and ``odd'' are defined operationally, by the
$\alpha\!\to\!1/\alpha$ interchange of Sec.~\ref{sec:coords}, relative to the
plane assignment \emph{after} the flip that map applies.  Comparing against
Hofmann's own legend, the two namings agree exactly when the map flips ---
that is, when $R\!<\!\varepsilon_{z}/\varepsilon_{x}$ --- and interchange when
$R\!>\!\varepsilon_{z}/\varepsilon_{x}$.  All three quantitative features
matched below are drawn by Hofmann as third-order \emph{even} modes and are
reproduced here by the branch our code calls odd; the correspondence is a
relabelling, not a difference in the physics.

They match the corrected form and reject the printed one.  The third-order
bands at $\varepsilon_{z}/\varepsilon_{x}\!=\!1.2$ and $2.0$ are published at
$0.087$ and $0.120$; the corrected relation gives $0.0873$ and $0.1199$, the
printed relation $0.0708$ and $0.1063$.  Two further features are decided by
the \emph{support} of a curve rather than its height, and so do not depend on
reading a value off a plot.  At $\varepsilon_{z}/\varepsilon_{x}\!=\!0.6$ an
isolated third-order island is published near $R\!\approx\!2.05$; the corrected
relation produces it ($0.0223$ against a published $\approx\!0.023$) and the
printed relation gives \emph{identically zero} there.  At the same ratio the
published band on the small-$R$ side closes near $R\!\approx\!0.58$: the
corrected relation closes it at $R\!=\!0.577$, the printed relation not until
$R\!=\!0.705$.  Over $R\!\in\![0.62,0.70]$, therefore, the corrected relation
gives identically zero, matching the published curve, while the printed
relation would still place a band reaching $0.0997$.  These two support comparisons favor the corrected form independently of the uncertainty in reading curve heights.

The comparison strongly indicates that the misprint did not propagate into
the computations behind the tested Fig.~6 features.  This is an inference
from published curves, not an inspection of the original calculation code.
We have tested Fig.~6 only --- one figure, one tune depression, third order,
three of five ratios.  It suggests the same for Figs.~11 and~12, but their
use of the corrected relation remains untested here.  Another published application permits a similar check: Ref.~\cite{HofmannFranchetti2006} states that its charts were computed
from the 1998 dispersion polynomials, so the same question could be put to it.  The tested curves therefore make Ref.~\cite{HofmannFranchetti2003} a quantitative corroboration
predating this work by more than two decades and independent of the algebra in
dispute --- of Eqs.~(37), (38) and~(43), of the monograph, and of the
derivation in Sec.~\ref{sec:vlasov-full}.  It is not independent of the
\emph{author}: Hofmann is its first author and the sole author of
Ref.~\cite{Hofmann1998}.  The correction reconciles the printed equation with these tested figures.

The branches without disputed factors provide control comparisons.  Across the three panels the published $\ell\!=\!2$ odd and
$\ell\!=\!4_{e}$ curves are reproduced to $0.6$, $1.5$, $4.0$ and $1.2\%$ ---
an external check on the coordinate map and on the absolute growth-rate
normalization, neither of which had one before.  The published heights are read from a printed plot, so the
quantitative agreements are good to a few per cent rather than exactly, and we
have not attempted to reproduce the figure itself.  The two support-based
features carry the argument, because they do not depend on that reading.

On the sampled grid, the largest retained-channel non-oscillatory
discrepancy occurs at one chart point.  At
$(R,\eta,\varepsilon_{z}/\varepsilon_{x})\!=\!(1.454237,0.297288,1.5)$,
where $S^{2}\!=\!9.68$ and the point is therefore inside the domain, the
printed
relation gives a non-oscillatory $\gamma/\nu_{0x}\!=\!0.1503$ where the
corrected one gives exactly zero.  The coordinates are quoted to six
decimals deliberately: the point sits at $97\%$ of the gate boundary, and
rounding it to three gives $S^{2}\!=\!9.70$, so a reader reproducing it
from fewer digits would not recover the value printed here.  The corrected relation still has oscillatory instability at this point; the zero-growth result applies only to its \emph{non-oscillatory} branch.  Among the originally retained channels, its oscillatory maximum is
$0.0997$, from $\ell\!=\!4_e$ (the $\ell\!=\!3_e$ value is $0.0861$).
The full implemented spectrum also includes $\ell\!=\!4_o$, whose
oscillatory growth is $0.1160$.  The corrected and printed spectral maxima
are therefore $0.1160$ and $0.1503$, a ratio of $1.30$.  The original
zero-versus-$0.1503$ discriminator remains a restricted non-oscillatory
third-order code test.  Isolating that contrast in a simulation requires
seeding and projecting the odd $\ell\!=\!3$ harmonic; an experimental
measurement additionally depends on excitation and the diagnostic observable.

A second target improves the domain margin and the spectral contrast, although
it does not establish a mode-independent experimental observable.  Restricting the same search to
$S^{2}\!\le\!7$ --- comfortably inside the gate rather than at $97\%$ of it ---
the sharpest disagreement lies at
$(R,\eta,\varepsilon_{z}/\varepsilon_{x})\!=\!(0.971186,0.426102,5.0)$, where
$S^{2}\!=\!6.89$.  There the printed relation gives
$\gamma/\nu_{0x}\!=\!0.1176$ and the corrected one again exactly zero, but now
the originally retained channels have a maximum of $0.0090$ from
oscillatory $\ell\!=\!4_e$ growth.  The full implemented spectrum, however,
contains non-oscillatory $\ell\!=\!4_o$ growth of $0.0472$.  Thus its
corrected-versus-printed spectral contrast is $0.0472$ against $0.1176$, a
factor of $2.49$, rather than the factor of thirteen inferred from the
restricted channels.  This remains larger than the first point's full-mode
ratio of $1.30$, and it has more room against the domain boundary.  Both
points are retained as reproducible numerical tests.  The search maximizes
the stated non-oscillatory discrepancy on its sampled grid; it does not
optimize a full-spectrum experimental observable or prove a continuous-domain
global maximum.  The second point is a useful candidate for a mode-resolved
simulation, whose seeding, projection and diagnostic response must be specified
before recommending a parity-unresolved measurement.

The same grid answers a second question, about the gate rather than about the
misprint.  Classifying each cell as stable, flagged by $\ell\!=\!2$ only,
flagged by both, or flagged by the higher-order channels \emph{alone} gives
Table~\ref{tab:census}.  The domain dependence of these fractions is examined below.

\begin{table}[t]
\caption{Verdict census at five emittance ratios,
$\gamma_{\rm th}\!=\!10^{-2}$, on a $60\!\times\!60$ grid ($3600$ cells).
``HO only'' counts points flagged by $\ell\!=\!3,4_{e}$ but not by
$\ell\!=\!2$ --- the operating points an $\ell\!=\!2$ screen misses.  The
gated columns count cells with $S^{2}\!\le\!10$; the final column counts them
over all $3600$ cells.  The ungated fraction is flat
($21.6$--$22.4\%$ on this grid, with a relative spread of about $11\%$
on the $90\!\times\!90$ refinement) while the gated one rises
by a factor $3.3$.  The trend therefore tracks how much of the higher-order
region the gate admits as the bands migrate relative to it, not a change in
the size of that region.}
\label{tab:census}
\begin{ruledtabular}
\begin{tabular}{lrrrrr}
 & \multicolumn{4}{c}{gated, $S^{2}\!\le\!10$} & ungated \\
\cline{2-5}\cline{6-6}
$\varepsilon_{z}/\varepsilon_{x}$ & in gate & $\ell{=}2$ only & both & HO only & HO only \\
\hline
1.2 & 2332 &  21 &  17 & 148 \ (6.3\%) & 779 \ (21.6\%) \\
1.5 & 2419 &  53 &  48 & 177 \ (7.3\%) & 778 \ (21.6\%) \\
2.5 & 2787 & 117 & 162 & 450 \ (16.1\%) & 808 \ (22.4\%) \\
5.0 & 2999 & 170 & 359 & 621 \ (20.7\%) & 808 \ (22.4\%) \\
8.0 & 3029 & 197 & 473 & 609 \ (20.1\%) & 789 \ (21.9\%) \\
\end{tabular}
\end{ruledtabular}
\end{table}

\begin{figure*}[t]
  \centering
  \includegraphics[width=0.98\textwidth]{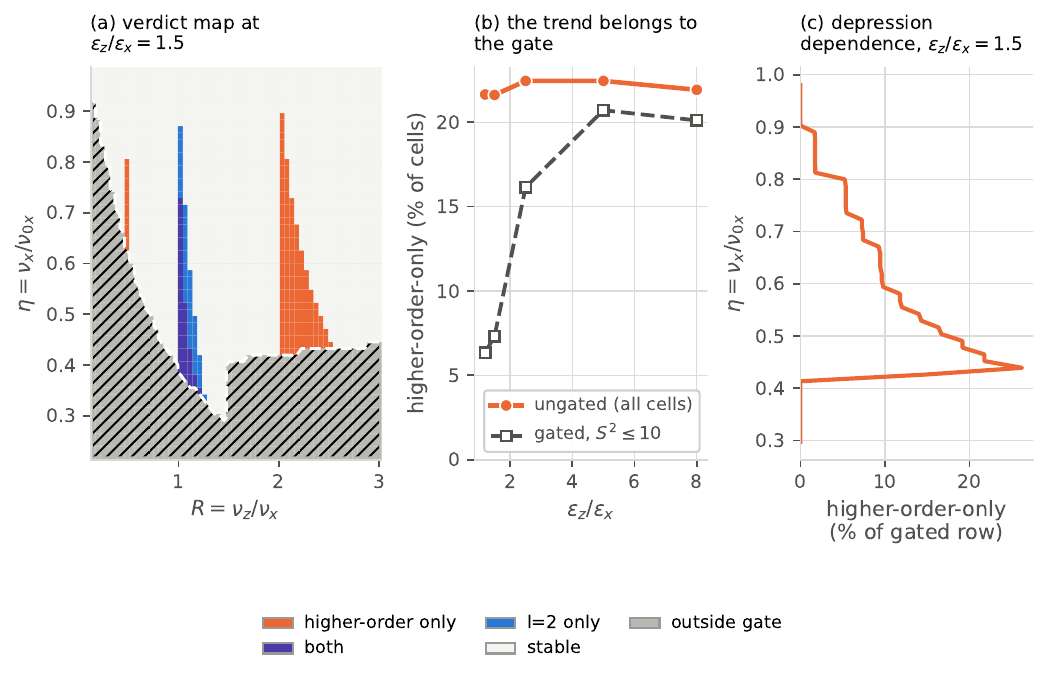}
  \caption{Where carrying $\ell\!=\!3,4_{e}$ changes the verdict.
  (a)~Verdict map at $\varepsilon_{z}/\varepsilon_{x}\!=\!1.5$; the hatched
  region is outside the $S^{2}\!\le\!10$ gate and is not classified.  The
  higher-order-only class forms two bands, near $R\!\approx\!0.5$ and
  $R\!\approx\!2.0$--$2.5$, on either side of the $\ell\!=\!2$ band at
  $R\!\approx\!1$.  (b)~Fraction of cells flagged by higher-order channels
  alone, versus emittance ratio, both ungated (solid, over all $3600$ cells)
  and restricted to $S^{2}\!\le\!10$ (dashed).  The ungated curve is flat;
  the gated one rises by a factor $3.3$, tracking the changing overlap between
  the higher-order bands and the gate.  (c)~The gated fraction resolved by
  tune depression at $\varepsilon_{z}/\varepsilon_{x}\!=\!1.5$, peaking near
  $\eta\!\approx\!0.44$, where the gate boundary and the higher-order bands
  coincide; we read the peak as that coincidence rather than as a feature of
  the channels themselves.}
  \label{fig:census}
\end{figure*}

The higher-order-only fraction is approximately constant across the sampled anisotropy range.  Over
$\varepsilon_{z}/\varepsilon_{x}\!\in\![1.2,8]$ it spans only
$21.6$--$22.4\%$ on the production grid.  The constancy is approximate: refining to $90\!\times\!90$ gives $21.3$--$23.7\%$, so the
maximum-to-minimum spread is $11.2\%$, and the third significant figure is not converged.  Roughly one chart point in
five is flagged
by $\ell\!=\!3,4_{e}$ and not by $\ell\!=\!2$ --- counted over the whole
sampled space, not over the gated subset --- and that is as true of a nearly
equipartitioned machine as of a strongly anisotropic one.

Inside the $S^{2}\!\le\!10$ restriction the same quantity behaves quite
differently: $6.3\%$ at $1.2$, rising to $20.7\%$ at $5.0$ and turning over to
$20.1\%$ at $8.0$ --- a factor $3.3$.  This increase reflects the changing overlap with the adopted domain. In Table~\ref{tab:census}, the higher-order-only region contains $779$
cells at $\varepsilon_{z}/\varepsilon_{x}\!=\!1.2$ and $808$ at $5.0$ --- the
same size to within $3.7\%$ --- but only $148$ of the first ($19\%$) lie inside
$S^{2}\!\le\!10$, against $621$ of the second ($77\%$).  The higher-order region therefore shifts into the
adopted domain as anisotropy increases. The gated fraction measures the
portion covered by the quantitative reporting range of the
$\ell\!\le\!4$ KV chart.

Although the
relations are exact in $S^{2}$, the cell fraction also depends on the
$(R,\eta)$ rectangle, uniform cell weighting and binary classification.
Changing the rectangle changes this fraction.  On the same $3600$ cells
the higher-order-only fraction is $21.6$--$22.4\%$ over the full box, but
$15.6\%$ falling to $12.0\%$ under $R\!\le\!2.0$, $16.4\%$ rising to $25.2\%$
under $R\!\ge\!0.5$, and identically zero at
$\varepsilon_{z}/\varepsilon_{x}\!=\!1.2$ and $1.5$ on the interior window
$R\!\in\![0.5,2]$, $\eta\!\ge\!0.4$ --- the corner nearest PIP-II.  Two
sub-windows of one rectangle carry opposite trends in anisotropy, so neither
the level nor the flatness is a statement about Hofmann's relations alone.

The nearly constant total on this rectangle results from compensating changes between its subregions. From
$\varepsilon_{z}/\varepsilon_{x}\!=\!1.2$ to $8.0$ the flagged population is
almost unchanged in size, $779\!\to\!789$ cells, while its content moves
entirely: the $R\!<\!0.5$ band falls $266\!\to\!3$, the interior
$0.5\!\le\!R\!\le\!2$ rises $100\!\to\!278$, and $R\!>\!2$ rises
$413\!\to\!508$.  What is physical is that migration --- the same movement the
gate-admission figures already report, $19\%$ of the region admitted at $1.2$
against $77\%$ at $5.0$ --- not the constancy of the total, which is an
accident of where this rectangle's two edges fall.  The gated fraction is what
survives our model-adequacy restriction, and it is the one every
operating-point claim in this paper --- including all of Sec.~\ref{sec:pip2}
--- is made against.

The higher-order-only class is also spatially separated from the
$\ell\!=\!2$ class rather than surrounding it
(Fig.~\ref{fig:census}a).  It occupies bands on either side of the
$\ell\!=\!2$ region, so it is not reachable by adding margin to an
$\ell\!=\!2$ screen: a design that keeps a comfortable distance from the
classical band can sit inside a higher-order one.

Both readings are consistent with Table~\ref{tab:misprint}.  Measured as a
net shift rather than as a symmetric difference, the $S^{4}$ factor structure
moves the gated higher-order-only fraction by $0.17$--$0.77$ percentage points
across the five ratios, while the ungated one is more
exposed, $0.89$--$4.97$.  Inside the restriction it is the printed $S^{2}$
block that sets the broad support of these channels and fixes the aggregate
fraction to within a percentage point; outside it the higher-order minors are
no longer small by comparison.  Small changes in the aggregate fraction can coexist with changes in cell membership: the same $S^{4}$ structure that moves the
in-gate fraction by less than a point still changes the individual verdict of
$0.73$--$2.11\%$ of in-gate cells, as Table~\ref{tab:misprint} counts
directly, most of which cancels in the net.  The \emph{size} of the
higher-order region is therefore insensitive to the derivation of
App.~\ref{app:l3e-derivation}; which particular cells are in it is not, and
neither are the growth rates.
\section{Application to PIP-II}
\label{sec:pip2}

We apply the framework of Secs.~\ref{sec:track1} and~\ref{sec:vlasov-full},
with the surrogate of App.~\ref{app:surrogate}, to the PIP-II
trajectory.  At each of the 41 lattice periods we take $(R,\eta)$ from
the physics-design TraceWin output~\cite{Uriot} and the geometric
emittance ratio $\varepsilon_{z}/\varepsilon_{x}$ from the HELIX
emittance export (Sec.~\ref{sec:coords}), and evaluate the three
screening methods with distinct roles: (i) the Hofmann $\ell=2$ closed-form analysis,
(ii) the higher-order $\ell=2,3,4_{e}$ solver of
Sec.~\ref{sec:track1}, and (iii) the calibrated gradient-boosted
surrogate of App.~\ref{app:surrogate}
\emph{trained on no PIP-II-specific data}:
the labels come from uniform random sampling of the
parameter space.

The 41 periods correspond to the cavity-period boundaries of the
physics-design accelerating lattice, sampled at the entrance of each
focusing-period cell from the HWR injection (idx~0, $E\!\approx\!2.3$~MeV) to the last
cavity-period boundary in HB650-2 (idx~40, $E\!\approx\!715$~MeV).
The final sampled energy is therefore below the nominal PIP-II
800~MeV extraction energy by design: the last entry is the last
cavity-period boundary before the HB650-2 exit and the beam-transfer
line, not the extraction point itself.

\begin{figure*}[t]
  \centering
  \includegraphics[width=0.98\textwidth]{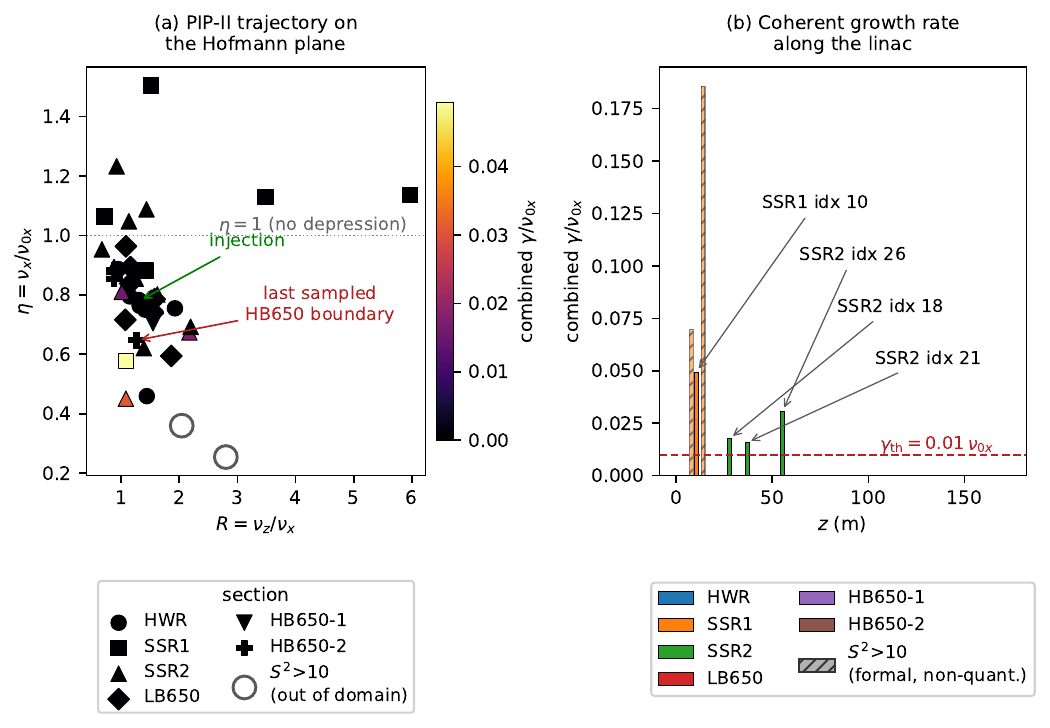}
  \caption{PIP-II operating trajectory through the higher-order
  Hofmann analysis (41 lattice periods, PIP-II
  physics-design trajectory~\cite{PIP2FDR2024,Uriot}).  (a)~The $(R,\eta)$ plane with
  each period colored by its solver combined growth rate
  $\gamma/\nu_{0x}$ and shaped by section; the injection end and the
  last sampled HB650-2 cavity-period boundary ($\sim\!715$~MeV, upstream
  of the 800~MeV extraction point) are flagged, and the $\eta=1$ line marks the no-depression
  boundary above which the solver returns trivial stability.  The
  color axis spans the in-domain range
  ($0$ to $0.049$); the two larger values on the trajectory
  (idx~8 at $0.070$ and idx~12 at $0.186$) are out-of-domain and are
  drawn as open markers with no fill.
  (b)~Combined growth rate versus longitudinal position $z$, with the
  instability threshold $\gamma_{\rm th}=0.01\,\nu_{0x}$ (dashed).
  Open markers / hatched bars denote the two out-of-domain
  $S^{2}\!>\!10$ periods (SSR1 idx~8, idx~12), whose bars
  are formal and not quantitative.  The largest formal bar,
  $\gamma/\nu_{0x}=0.186$ ($\ell=3$ odd, SSR1 idx~12), is one of these
  out-of-domain periods; among the in-domain periods the
  peak is the $\ell=2_o$ tilting/coupling flag at SSR1 idx~10
  ($\gamma/\nu_{0x}=0.049$).  The
  quantitative interpretation of the $S^{2}\!>\!10$ periods requires a treatment valid outside the adopted model domain and is left to
future work (Sec.~\ref{sec:limits}).}
  \label{fig:pip2_traj}
\end{figure*}

Figure~\ref{fig:pip2_traj} overlays the 41 periods on the
$(R,\eta)$ plane and resolves the combined growth rate along the
linac.  Per-period chart coordinates and validity flags are tabulated
in App.~\ref{app:perperiod}; here we summarize the section-level
results subject to the in-domain constraint
$S^{2}\!\lesssim\!10$ (Sec.~\ref{sec:limits}).

\emph{In-domain evaluable subset.}\quad Of the 41 PIP-II records, 39 satisfy
the adopted model domain ($S^{2}\!\le\!10$), and 32 of those 39 are
physically evaluable: the remaining seven have $\eta\!\ge\!1$, which the
anisotropic chart does not admit, and are stable by convention rather than
by computation.  Within the 32 evaluable periods
the corrected anisotropic solver flags 4~periods as showing
chart growth above the threshold $\gamma_{\rm th}\!=\!10^{-2}$.  Three
come from the classical $\ell\!=\!2_o$ tilting/coupling mode --- SSR1 idx~10
($\gamma/\nu_{0x}\!=\!0.049$, $S^{2}\!=\!3.0$), SSR2 idx~26
($0.031$, $S^{2}\!=\!6.4$) and SSR2 idx~21 ($0.016$, $S^{2}\!=\!0.9$)
--- and the fourth, SSR2 idx~18 ($0.018$, $S^{2}\!=\!2.7$), comes from
the $\ell\!=\!3$ \emph{odd} branch and is invisible to the classical
$\ell\!=\!2$ screen.  The $\ell\!=\!3$ even and $\ell\!=\!4_{e}$
channels return growth below threshold at every in-domain
period for the design emittance ratio
($\varepsilon_{z}/\varepsilon_{x}\!\approx\!1.2$--$1.8$).  The
trajectory's in-domain peak is the $\ell\!=\!2_o$ tilting/coupling
rate $\gamma/\nu_{0x}\!=\!0.049$ at SSR1 idx~10.  That the
higher-order channel contributes a flag the $\ell\!=\!2$ analysis does
not see is the operational case for carrying it: on this trajectory the
higher-order screen is not redundant.
The four flags have different sensitivities to the chart coordinates.  Perturbing $R$ at fixed
$(\eta,\varepsilon_{z}/\varepsilon_{x})$, the flags occupy
$R$-bands of $-7.5\%/{+}8.1\%$ (idx~10), $-8.0\%/{+}0.12\%$ (idx~18),
$-1.1\%/{+}2.0\%$ (idx~21) and $-7.6\%/{+}4.8\%$ (idx~26).  Only SSR1 idx~10
survives $\pm5\%$; idx~10 and idx~26 survive $\pm2\%$; idx~18 fails at
$+0.12\%$ and idx~21 at $-1.08\%$.  The flag with the largest rate is also the least sensitive in these scans --- idx~10 flags across the
whole $\pm20\%$ range in $\eta$ as well --- while the $\ell\!=\!3$ odd flag
is the most fragile of the four.  The associated sensitivities are examined below.

We emphasize that ``flagged'' here means \emph{chart points within the
Hofmann model inside its adopted domain show non-zero coherent growth above
$\gamma_{\rm th}$}, not that the corresponding lattice period is operationally
unstable; the chart is a working-point screen, not a beam-loss predictor (see
Sec.~\ref{sec:limits}).

The physical applicability of these coherent modes is limited. Jeon and Jang~\cite{JeonJang2024} argue that parametric instabilities of this family
are unlikely to be observed in real linacs unless waterbag or KV beams are
generated, and Hofmann's own later work notes that Landau damping suppresses
coherent parametric resonances above second order in Gaussian-like
distributions.  These observations limit experimental interpretation. The equations nevertheless require correction because the chart is used as a
design screen regardless, its higher-order branches are carried by a deployed
design code, and the printed relations do not reproduce their author's published coefficients. The flags identify candidate regions for further study; they do not predict the beam response.

The Hofmann chart is a \emph{matched}-equilibrium
KV analysis, and the lattice export carries its own mismatch column: at the
four flagged periods the transverse mismatch factors are $M_{x}\!=\!0.51$,
$0.32$, $0.07$ and $0.67$, and at the out-of-domain SSR1 idx~12 they reach
$M_{x}\!=\!2.86$, $M_{y}\!=\!3.11$.  A matched-beam chart evaluated at $M_{x}\!=\!0.07$--$0.67$ is being used
outside the state it describes.  We do not attempt to correct for this: no anisotropic mismatched-KV
dispersion relation exists to correct \emph{to}, and constructing one is a
larger problem than the errata this paper reports.  The same limitation applies whenever the screened beam is not launched matched,
including the SNS, SPL and ESS studies of
Ref.~\cite{HofmannFranchetti2003} and the UNILAC and J-PARC examples of
Ref.~\cite{HofmannBook2017}, and we are not aware of a published chart
application that reports its mismatch state.  We report the mismatch state alongside the four flags to make this limitation explicit.

\emph{Anisotropy margin.}\quad The design question is how much
additional anisotropy each period could absorb before higher-order
growth turns on --- and, where a period is already past its onset, that
absorption is negative rather than absent.  Holding each period's $(R,\eta)$ fixed and raising
$\varepsilon_{z}/\varepsilon_{x}$ within the adopted model domain, only $6$ of the $32$ evaluable
in-domain periods reach a higher-order onset at all below
$\varepsilon_{z}/\varepsilon_{x}\!=\!9$.  The scan skips periods with
$\eta\!\ge\!1$, so of the $39$ in-domain records $32$ were tested --- $6$ with
an onset and $26$ without --- and $7$ were not.  We do not claim the untested seven never reach one --- two of them
(idx~9 and~16) would flag at admissible depressions.  In
increasing order of headroom above the design ratio, they are SSR2
idx~18 (onset $\varepsilon_{z}/\varepsilon_{x}\!=\!1.47$,
${\times}0.98$), SSR2 idx~26 ($1.38$, ${\times}1.03$), SSR1 idx~10
($2.05$, ${\times}1.16$), SSR2 idx~21 ($1.75$, ${\times}1.20$), SSR2
idx~22 ($2.19$, ${\times}1.51$) and LB650 idx~28 ($6.82$,
${\times}5.39$).

SSR2 idx~18 has its
$\ell\!=\!3$ odd onset at $\varepsilon_{z}/\varepsilon_{x}\!=\!1.47$,
just below its design ratio of $1.503$, so its margin is
${\times}0.98$ --- it sits \emph{past} the higher-order onset rather than
below it, which is the same statement as its appearing in the flagged
set above.  For that period the margin framing does not apply: the
quantity the scan returns is the onset location, not headroom above it.

The trajectory therefore does \emph{not} carry a uniform positive
higher-order margin.  At the nominal working point the six split unevenly:
idx~18 is already \emph{past} its onset, three more (idx~10, 21, 26) would be
brought into the channel by a $20\%$ excursion in longitudinal-to-transverse
emittance ratio, and only idx~22 and~28 carry more than $20\%$ of headroom.
LB650 idx~28 (${\times}5.39$) is the sole onset outside the SSR sections, and
the HWR and HB650 sections reach none anywhere in the scanned range.

The two-decimal values describe the nominal scan; they do not imply corresponding physical precision.  The scan freezes $(R,\eta)$ and varies only the emittance ratio,
but the margin's derivative with respect to the frozen $R$ exceeds its
derivative with respect to the scanned quantity: a $1\%$ change in
$R$ moves the idx~21 and idx~22 margins by $46\%$ and $35\%$ --- more than the
entire $20\%$ emittance excursion the paragraph is built on.  Over
$R\!\pm\!5\%$ the six margins become bands (idx~10 ${\times}0.71$--$2.51$;
idx~18 ${\times}0.70$--$4.40$; idx~21 ${\times}1.00$--$3.06$; idx~22
${\times}0.96$--$2.98$; idx~26 ${\times}0.84$--$1.32$; idx~28
${\times}1.15$--$5.39$), and three of the six --- idx~21, 22 and~28 --- lose
the onset entirely at one or more multipliers inside that band.  The count
``four of the six within ${\times}1.20$ of an onset'' is itself a function of
$R$: it runs $5,4,5,4,2,1,0$ across
$R\!\times\!\{0.95,0.98,0.99,1.00,1.01,1.02,1.05\}$.  We therefore report the
ordered list as a description of the nominal point and the bands as the
design-relevant quantity.  This is in any case a sensitivity statement about
the chart, not a prediction of loss: the local per-period exponents remain
small (the budget below), and no physical excursion varies the emittance ratio
alone.  No onset here
is a coordinate artifact --- an earlier version of this analysis had to
exclude one (SSR1 idx~15) by hand as a discontinuity of the $\hat\eta\!=\!1$
variable interchange, where $S^{2}$ jumped by ${\approx}37$ across a single
scan step, and under the map of Sec.~\ref{sec:coords} that seam does not
occur: all six onsets are smooth and show no interchange discontinuity.

\emph{Out-of-domain subset.}\quad Two periods (SSR1 idx~8 with
$S^{2}\!=\!14$ and SSR1 idx~12 with $S^{2}\!=\!33$) sit outside the
adopted model domain.  Both yield nonzero formal solver growth rates
($\gamma/\nu_{0x}\!=\!0.070$ and $0.186$, in each case set by the
$\ell\!=\!3$ odd branch) that are \emph{not} reliable quantitative
predictions.  These periods are flagged with
$\dagger$ in Table~\ref{tab:pip2_flagged} and require
out-of-domain treatment for a defensible growth-rate value,
which we identify as future work.  We retain them in the table for completeness
and do not propagate them into the section-level claims of this
work.

In summary, inside the adopted model domain ($S^{2}\!\le\!10$) the corrected
Hofmann chart flags \emph{4 of the 32 physically evaluable periods}:
three from the
classical $\ell\!=\!2_o$ tilting/coupling mode (SSR1 idx~10, SSR2 idx~21 and
idx~26) and one from the $\ell\!=\!3$ odd branch (SSR2 idx~18).  The flagged rates are
all weak --- the largest is $\gamma/\nu_{0x}\!=\!0.049$ --- so the
trajectory shows no strong chart-predicted growth anywhere in the
adopted model domain.  But the higher-order channel is \emph{not}
quiescent: it contributes one flag that the classical $\ell\!=\!2$
screen does not see, and the corresponding period sits at its
$\ell\!=\!3$ onset rather than below it.  Higher-order anisotropic screening therefore adds one candidate period for further investigation on this trajectory.

The $(1\!\mp\!2\hat\eta^{2}/\alpha)$ correction of
App.~\ref{app:anisotropic} is \emph{required} for iso-limit consistency
of the $\ell\!=\!3$ block --- a structural property of the dispersion
relation, independent of any PIP-II outcome.  Its effect on this
trajectory runs through the odd branch: the $\ell\!=\!3$ \emph{even}
branch, for which the factors are derived directly
(Sec.~\ref{sec:vlasov-full}),
returns zero growth at every in-domain PIP-II period (its only
nonzero value, SSR1 idx~12 at $0.121$, lies outside the gate), while the
odd branch --- which inherits the same factor structure through the
variable interchange of Sec.~\ref{sec:track1}, and whose separate
Vlasov--Poisson derivation is now given in Sec.~\ref{sec:vlasov-full} --- carries
the idx~18 flag and both out-of-domain rates.  The flagged set
therefore inherits the off-iso caveat of Sec.~\ref{sec:limits}.  Of the
six margin-scan onsets, four are set by $\ell\!=\!4_{e}$ (idx~10, 21, 26, 28)
and two by $\ell\!=\!3$ odd (idx~18, 22).  The $\ell\!=\!4_{e}$ relation carries $\alpha$-coupling
factors \emph{as printed}, so those four onsets --- including the
${\times}1.03$ and ${\times}1.16$ cases --- do not depend on the
$S^{4}$ correction at all.  They do depend on the
Eq.~(42) sign resolution of App.~\ref{app:l4e}.

\emph{Threshold dependence.}\quad $\gamma_{\rm th}\!=\!10^{-2}$ is a
\emph{reporting} threshold, not a mathematical one: in the idealized
model any $\gamma\!>\!0$ is unstable, and the cut selects what we
consider operationally worth flagging.  The flagged set is unchanged for
every threshold from the mathematical onset $\gamma\!>\!0$ up to
$10^{-2}$ --- at all of them the count is 4 and idx~18 is present --- so
the higher-order flag is not an artifact of choosing $10^{-2}$
specifically.  It is, however, sensitive to raising the cut: at
$\gamma_{\rm th}\!=\!2\!\times\!10^{-2}$ only idx~10 and idx~26 survive
and the $\ell\!=\!3$ flag disappears with them, because its rate is
$0.018$.  At $5\!\times\!10^{-2}$ no period is flagged at all.  Readers
who set the bar at $0.02$ should therefore retain two second-order flags and
no higher-order flag; at $0.05$ the trajectory carries no flags at all; at and below the stated threshold, the $\ell\!=\!3$ channel is not redundant.

\emph{Domain-gate dependence.}\quad The sensitivity to the
adopted $S^{2}\!\le\!10$ restriction (Sec.~\ref{sec:limits}) is evaluated by
varying the upper bound around $10$:
\begin{equation*}
  \begin{array}{llll}
    \text{gate} & \text{records} & \text{evaluable} & \text{flags} \\[2pt]
    S^{2}\!\le\!5  & 38 & 31 & 10,\,18,\,21 \\
    S^{2}\!\le\!8  & 39 & 32 & 10,\,18,\,21,\,26 \\
    S^{2}\!\le\!10 & 39 & 32 & 10,\,18,\,21,\,26 \\
    S^{2}\!\le\!15 & 40 & 33 & 8,\,10,\,18,\,21,\,26 \\
    \text{no gate} & 41 & 34 & 8,\,10,\,12,\,18,\,21,\,26 \\
  \end{array}
\end{equation*}
Here ``records'' counts the periods satisfying the gate and ``evaluable''
those among them with $\eta\!<\!1$; the flagged sets are drawn from the
evaluable column, the seven $\eta\!\ge\!1$ records never being computed at
any gate.
The $\ell\!=\!3$ odd flag at idx~18 ($S^{2}\!=\!2.7$) is present for every
choice, including no gate at all, so the qualitative claim --- that the
higher-order channel contributes a flag the $\ell\!=\!2$ screen misses --- does
not depend on the gate.  What the gate controls is whether SSR2 idx~26
($S^{2}\!=\!6.4$) is admitted, and whether the two strongly space-charge
dominated SSR1 periods are counted; relaxing it to $15$ or beyond admits
idx~8 and idx~12, which carry the largest formal rates on the trajectory and
sit at $S^{2}\!=\!14$ and $33$ --- deep in the regime where the
$\ell\!\le\!4$ truncation and the KV assumptions are least defensible, even
though the relations themselves remain exact there.

The 4-of-32 count is not equally robust to the model-domain
cutoff at each period: three of the four flags sit well inside the domain
($S^{2}\!=\!0.9$, $2.7$ and $3.0$) but idx~26 sits at $S^{2}\!=\!6.4$,
and the next period outside the domain is at $S^{2}\!=\!14$.
Table~\ref{tab:pip2_flagged} lists the flagged periods compactly;
section-level worst-case growth rates by mode order, and the full
41-period table, are given in
App.~\ref{app:perperiod} (see Supplemental Material~\cite{SupplementalMaterial}).

\begin{table}[t]
\caption{Compact list of all PIP-II periods at which the corrected
anisotropic solver returns $\gamma/\nu_{0x}\!>\!\gamma_{\rm th}\!=\!10^{-2}$ inside the adopted
model domain.  Three of the four flags originate in
the classical $\ell\!=\!2_o$ tilting/coupling mode; the fourth (SSR2 idx~18) comes from
the $\ell\!=\!3$ odd branch and is invisible to an $\ell\!=\!2$-only screen.
The $\ell\!=\!3$ even and odd branches are listed separately rather than
maximized over, since the parity is what distinguishes the idx~18 flag.
The two rows marked $\dagger$ sit outside the adopted model domain $S^{2}\!\le\!10$ and are listed for completeness only,
not propagated into quantitative claims; both take their combined value from
the $\ell\!=\!3$ odd branch, though idx~12 also carries a nonzero even rate.
The $(1\!\mp\!2\hat\eta^{2}/\alpha)$ correction required for $\ell\!=\!3$
iso-limit consistency (App.~\ref{app:anisotropic}) reaches this trajectory
through the odd branch, which carries the idx~18 flag.}
\label{tab:pip2_flagged}
\begin{ruledtabular}
\begin{tabular}{rlrrrrrrl}
idx & section & $S^{2}$ & $\gamma_{\ell\!=\!2}$ & $\gamma_{\ell\!=\!3_{e}}$ & $\gamma_{\ell\!=\!3_{o}}$ & $\gamma_{\ell\!=\!4_{e}}$ & comb. & origin \\
\hline
10 & SSR1    &  3.0 & 0.049 & 0.000 & 0.000 & 0.000 & 0.049 & $\ell\!=\!2$ \\
26 & SSR2    &  6.4 & 0.031 & 0.000 & 0.000 & 0.000 & 0.031 & $\ell\!=\!2$ \\
18 & SSR2    &  2.7 & 0.000 & 0.000 & 0.018 & 0.000 & 0.018 & $\ell\!=\!3_{o}$ \\
21 & SSR2    &  0.9 & 0.016 & 0.000 & 0.000 & 0.000 & 0.016 & $\ell\!=\!2$ \\
\hline
 8$^\dagger$ & SSR1 & 13.9 & 0.000 & 0.000 & 0.070 & 0.000 & 0.070 & $\ell\!=\!3_{o}$ \\
12$^\dagger$ & SSR1 & 33.2 & 0.000 & 0.121 & 0.186 & 0.000 & 0.186 & $\ell\!=\!3_{o}$ \\
\end{tabular}
\end{ruledtabular}
\end{table}
The surrogate of App.~\ref{app:surrogate} is evaluated under a hard
two-gate eligibility rule: a period is scored only when (i)~its
chart point satisfies the in-domain bound $S^{2}\!\le\!10$
and (ii)~its raw triple
$(R,\eta,\varepsilon_{z}/\varepsilon_{x})$ lies inside the surrogate's
training hyperrectangle $R\!\in\![0.1,3.0]$, $\eta\!\in\![0.2,0.99]$,
$\varepsilon_{z}/\varepsilon_{x}\!\in\![0.9,9.0]$.  Periods that fail
either gate are reported as out-of-scope rather than classified.
Of the 41 PIP-II periods, $32$ pass both gates; the remaining $9$
comprise two at $S^{2}\!>\!10$ (the out-of-domain subset of
Table~\ref{tab:pip2_flagged}) and seven at $\eta\!>\!0.99$
or $R\!>\!3.0$, which sit in the trivially-stable
no-depression / high-aspect-ratio corner of the chart and would not
be scored by the deployed surrogate.  On the $32$-period gated subset
the surrogate agrees with the solver on $29$ periods ($90.6\%$).  Accuracy alone obscures the class imbalance: the
gated subset holds only four positives, so the confusion matrix is
$\mathrm{TP}\!=\!2$, $\mathrm{FN}\!=\!2$, $\mathrm{FP}\!=\!1$,
$\mathrm{TN}\!=\!27$, giving
\begin{equation}
  \text{recall} = 0.50, \quad
  \text{precision} = 0.67.
  \label{eq:ml-pip2-confusion}
\end{equation}
With four positives we do not quote $F_{1}$ or balanced accuracy: on this
sample they carry a precision the counts do not support, and recall $0.50$
means ``missed two''.  The $90.6\%$ is carried almost
entirely by the 27 correctly-returned stable periods.  The surrogate
misses both of the weakest solver flags --- SSR2 idx~18
($\gamma/\nu_{0x}\!=\!0.018$, the $\ell\!=\!3$ odd period) and SSR2
idx~21 ($0.016$) --- and raises one false positive at LB650 idx~28.
All three sit within a factor of two of the decision threshold
$\gamma_{\rm th}\!=\!10^{-2}$, so the surrogate reproduces the chart
well away from the boundary and degrades on exactly the marginal cases
a screen would most want to catch.  In particular it does \emph{not}
recover the higher-order flag, which is the one result the classical
$\ell\!=\!2$ analysis also misses.  A near-unity in-distribution
ROC--AUC therefore says nothing about boundary-case recall on the
operating trajectory, and the two should not be conflated.  With $n\!=\!32$ (and only four
positives) this cross-check is descriptive rather than a statistical
validation, and it argues for keeping the solver in the loop rather
than deploying the surrogate as a standalone screen.

\emph{Sensitivity to the $S^{4}$ block.}\quad The Vlasov--Poisson
calculation of Sec.~\ref{sec:vlasov-full} derives the corrected factors and
excludes both alternatives at every $(\alpha,\hat\eta)$, so the alternatives serve as sensitivity controls rather than competing hypotheses.  Repeating the census on the same
grid and the same fixed emittance ratios with only the $\ell\!=\!3$ $S^{4}$
factors swapped (reproducing
Table~\ref{tab:census} exactly under the derived variant), the three
structures move the gated higher-order-only fraction by $0.17$--$0.77$
percentage points and the ungated one by $0.89$--$4.97$ points.  Those are
\emph{net} shifts in an aggregate, and Sec.~\ref{sec:census} reconciles them
with the stronger cell-by-cell statement of Table~\ref{tab:misprint}: the small net change in an aggregate fraction does not establish stability of individual classifications.

The single higher-order PIP-II flag is sensitive to the factor structure.  At SSR2 idx~18 the
$\ell\!=\!3$ odd rate is $0.0176$ under the derived factors but identically
zero under both alternatives, so the existence of this flag depends on the $S^{4}$ block.  The derivation fixes the factor structure, but the flag remains sensitive to every tested chart coordinate.  It sits at $\gamma/\nu_{0x}\!=\!0.018$ against a reporting threshold
of $0.010$, it disappears if the threshold is raised to $0.02$, and its rate
roughly doubles if the emittance ratio is taken from \textsc{TraceWin} rather
than HELIX.  More sharply, at fixed $(\eta,\varepsilon_{z}/\varepsilon_{x})$
the $\ell\!=\!3$ odd flag occupies $R\!\in\![2.0011,2.1781]$ and this period
sits at $R\!=\!2.1755$ --- $0.12\%$ below the upper edge and $8.0\%$ above the
lower one.  A $+0.2\%$ change in $R\!=\!\nu_{z}/\nu_{x}$ removes the flag
entirely; $-5\%$ roughly triples it, to $\gamma/\nu_{0x}\!=\!0.059$.  The same
flag lies within $+0.43\%$ of an edge in $\eta$ at fixed $R$, and within
$-0.17\%$ along a pure transverse-tune-depression error.  It is therefore inside half a percent of a band edge in every chart
direction we can test.  We read that as a property of the chart rather than a
deficiency of this working point: at $\varepsilon_{z}/\varepsilon_{x}$ near
unity the $\ell\!=\!3$ odd band is narrow --- here $8\%$ wide in $R$ --- so a
trajectory that enters it at all is likely to enter it near an edge, and a
flag drawn from a band this sharp carries a correspondingly sharp input
requirement.  The flag requires an input tolerance tighter than $0.2\%$ in $R$. The lattice export supplies no tolerance on $R$ or $\eta$, so it cannot establish whether this period meets that requirement.  We
therefore present it as a marginal result --- an indication of where a
higher-order screen would look, at a working point where small changes in the
inputs move it --- and not as an established property of the PIP-II design.

\emph{Independent cross-check against Hofmann's own $\varepsilon\!=\!1.5$
chart.}\quad Hofmann's Fig.~13 is computed at
$\varepsilon_{x}/\varepsilon_{y}\!=\!1.5$, which in the present mapping is the
emittance ratio of the PIP-II mid-linac (valid-set mean $1.46$) --- far closer
to this trajectory than the $\varepsilon\!=\!5$ chart used for the solver
validation of Sec.~\ref{sec:track1validation}.  He draws an explicit
operational conclusion from it: that ``the region of transverse tune depression
between $0.7$ and $1$ should be safe from a practical point of view''
(his~p.~4723).  Three of our four flagged periods fall outside that band ---
SSR1 idx~10 at $\eta\!=\!0.58$, SSR2 idx~18 at $0.67$, and SSR2 idx~26 at
$0.45$ --- and the fourth, SSR2 idx~21 at $\eta\!=\!0.81$, falls inside it and
carries the smallest rate of the four.  Three flags are consistent with the region identified by Hofmann's guidance; the exception has the smallest growth rate.  This is an independent, if qualitative, corroboration
that does not rely on our digitization of his figures.  A quantitative
digitization of Fig.~13 --- which covers $\nu_{x}/\nu_{y}\!\lesssim\!1.2$ and
so contains three of the four flagged working points --- would be a sharper
test than the $\varepsilon\!=\!5$ comparison we report, and we identify it as
the most direct validation still available.

\emph{Local exponent budget.}\quad The growth rate must be integrated over
the time spent in an unstable region to obtain the local e-folding exponent,
\begin{equation}
  N_{e} = \int \gamma \, dt
        \;\simeq\; \sum_{j} \frac{\gamma_{j}}{\nu_{0x,j}}\,\sigma_{0x,j},
  \label{eq:Ne}
\end{equation}
where $\sigma_{0x,j}\!=\!k_{x0,j}L_{j}$ is the zero-current transverse phase
advance across period $j$ in radians, and $e^{N_{e}}$ converts an exponent to
an amplitude scale.  We accumulate Eq.~\eqref{eq:Ne} per branch and only over
\emph{contiguous} runs of unstable periods.  That grouping is a bookkeeping
convention, not a physical claim: a mode that stops growing does not stop
existing, and we make no assertion about what its amplitude does across a
gap.

On this trajectory every unstable period is isolated: there is no contiguous
run of two or more.  The lattice export supplies the zero-current phase
advance \emph{per period} directly, in degrees; across the 41 periods it spans
$13$--$84^{\circ}$ ($0.23$--$1.46$~rad), and the four flagged periods span
$34$--$84^{\circ}$.  The four flags give
\begin{equation*}
  \begin{array}{llll}
    \text{SSR1 idx~10} & (\ell\!=\!2)   & N_{e}=0.072 & e^{N_{e}}=1.075 \\
    \text{SSR2 idx~26} & (\ell\!=\!2)   & N_{e}=0.018 & e^{N_{e}}=1.018 \\
    \text{SSR2 idx~18} & (\ell\!=\!3_{o}) & N_{e}=0.013 & e^{N_{e}}=1.013 \\
    \text{SSR2 idx~21} & (\ell\!=\!2)   & N_{e}=0.013 & e^{N_{e}}=1.013 \\
  \end{array}
\end{equation*}
so the strongest flag corresponds to a $7\%$ local amplitude scale over its
own period and the higher-order flag to $1\%$.

The sum of the four exponents is a local diagnostic rather than a propagated amplitude. These periods constitute the complete in-domain non-oscillatory contribution: of the $41$
periods, $35$ return \emph{exactly} zero growth and none returns a rate in
the interval $(0,\gamma_{\rm th}]$, so the reporting threshold hides no tail
of small positive rates and the flagged periods carry the entire
non-oscillatory local exponent budget of the in-domain trajectory.
That total is $N_{e}\!=\!0.116$, and $e^{N_{e}}\!=\!1.123$ is its conversion to
an amplitude scale.

This sum does not bound the transported amplitude. The $\gamma_{j}$ are spectral rates at frozen per-period parameters,
not logarithmic norms of a common propagator, and spectral stability of each
segment does not bound their product: two individually neutral transfer maps
can compose to a hyperbolic one, the familiar parametric mechanism.  Between
flags a mode acquires no \emph{local} exponent, but the intervening maps can
rotate and mix modes, and an amplitude acquired in one band is not erased on
leaving it.  A genuine whole-linac bound would need ordered, mode-resolved
transfer propagation, which we do not compute.  The resulting interpretation is restricted to the local model:
within the smooth-focused KV model, on the non-oscillatory branch, and in a
frozen-per-period sense, PIP-II's design trajectory \emph{generates} very
little local exponent for a coherent mode to draw on, even at the periods
that cross the reporting threshold --- not that little can be transported.
Whether any of it is transported into net growth is a question this analysis
does not answer; a flag marks a working point that touches a stop band, and
how much a mode there actually grows is beyond what a frozen-per-period chart
can decide.

\emph{Physical interpretation of the chart.}\quad The Hofmann
chart is a smooth-focused, continuous-beam KV stability analysis:
it predicts which $(R,\eta,\varepsilon_{z}/\varepsilon_{x})$
operating points are theoretically unstable against coherent
collective modes under those idealizations.  It is a
\emph{working-point screen}, not a beam-quality predictor.  It does
not predict actual RMS emittance growth, halo fraction, beam loss,
or longitudinal-transverse coupling in a real PIP-II beam, nor does
it incorporate the discrete lattice periodicity, finite Landau
damping from the realistic distribution, or the cumulative effect
of imperfections along the linac.  A solver flag therefore should
be read as ``this period sits in a chart region where the
smooth-focused KV analysis predicts non-zero coherent-mode growth'',
a useful design-time warning, but one that requires
out-of-domain tracking and/or lattice-periodic Floquet analysis
(left to future work) for a quantitative
operational claim.

\section{Limitations}
\label{sec:limits}

\begin{itemize}
\item \textbf{Mode eigenvectors.}  The $(1\!\mp\!2\hat\eta^{2}/\alpha)$ factors
  follow from evaluating Hofmann's Eqs.~(19) and~(23) in full
  (Sec.~\ref{sec:vlasov-full}), with the two normalization constants fixed
  once against his undisputed Eq.~(28); both alternative factor structures are
  excluded at every $(\alpha,\hat\eta)$.  The completeness assumption that the
  earlier determinant argument required --- that Eq.~(36) is complete in which
  pole pairs it contains --- is discharged, because the pole set is an output
  of the calculation rather than an input to it.  The implementation does not yet extract the mode structure.  We solve for the
  vanishing of a scalar determinant and stop there; at a root the null vector
  of the coefficient matrix \emph{is} the relative mode amplitude, so the
  information is present in principle and its absence here is an
  implementation gap rather than a limitation of the method.  Extracting it,
  and using it to seed a particle-in-cell run, would permit a particle-in-cell validation beyond the algebraic checks.  It matters here because the $\ell\!=\!3$ channel is \emph{active} on
  this trajectory, carrying the SSR2 idx~18 flag through the odd branch.

\item \textbf{Fourth-order even sign correction.}  Hofmann's Eq.~(41) and its own isotropic
  reduction Eq.~(42) disagree on the sign of the $S^{4}$ block, so one is
  misprinted.  We resolve it with Eq.~(43), whose closed-form roots are
  independent of both and are consistent with only one of the two candidates
  (App.~\ref{app:l4e}).  The roots satisfy one form identically, and only that form's numerator factors over precisely that root set. This resolves the printed inconsistency.  It is, however, no longer the
  only route to that sign: the Vlasov calculation of
  Sec.~\ref{sec:vlasov-full} reproduces Eq.~(41) with the printed
  $-S^{4}$ and rejects the alternative, without using Eq.~(43) at all.  Every
  $\ell\!=\!4_{e}$ number here rests on that agreement between two
  independent routes rather than on either alone.  That includes four of
  the six margin-scan onsets of Sec.~\ref{sec:pip2}.  It does \emph{not}
  include the PIP-II flagged set, where $\ell\!=\!4_{e}$ returns zero growth
  at every in-domain period, nor the $\ell\!=\!3$ correction above,
  which is independent of it.  The supporting census is unaffected too.  Under the adopted sign, every gated
  cell that $\ell\!=\!4_{e}$ flags and $\ell\!=\!2$ does not is \emph{also}
  flagged by $\ell\!=\!3$ --- $0$ exclusive cells out of $13{,}566$ gated, on
  the same $60\!\times\!60$ grid and the same five emittance ratios as
  Table~\ref{tab:census}.  That alone would not settle
  the question --- the \emph{rejected} sign could in principle flag a gated
  cell $\ell\!=\!3$ misses --- so we measure it: recomputing the gated column
  under both signs changes the verdict of \emph{no} cell at any ratio.

  The ungated count does change under the sign replacement.  Ungated, $\ell\!=\!4_{e}$ is the
  sole higher-order flag in $26$ cells of $18{,}000$ ($0.14\%$), all at
  $\varepsilon_{z}/\varepsilon_{x}\!\le\!1.5$.  Recomputing the census under
  the rejected $+S^{4}$ sign raises the ungated higher-order-only count at
  every ratio --- $779$ to $790$ at $1.2$, $778$ to $812$ at $1.5$, $808$ to
  $838$ at $2.5$, $808$ to $875$ at $5.0$, $789$ to $855$ at $8.0$ ---
  shifts of $0.31$, $0.94$, $0.83$, $1.86$, $1.83$ percentage points.
  The sign therefore does move the ungated census count, by up to $1.86$
  points.
  These ungated shifts are production-scan sensitivities: they are computed
  with the same grid-scan root finder that produces the census itself, so what
  they measure is the sensitivity of the \emph{reported} numbers to the sign.
  Re-evaluating them with exact companion-matrix roots moves individual
  entries at the level of a few hundredths of a point; we quote the scan
  values because those are the ones the census is built from.
  The invariance of the gated column is directly verified: recomputing \emph{both} columns under \emph{both} signs on all
  five ratios leaves the gated count identical in every case
  ($148$, $177$, $450$, $621$, $609$ cells), so the exactness above is
  confirmed directly and not merely
  deduced from the adopted-sign result.
  What the sign does move is the split between the ``$\ell\!=\!2$ only'' and
  ``both'' columns, which is why those counts differ from the version of this
  work that carried the opposite sign.  An earlier version of this work adopted the opposite sign, and no
  isotropic-limit check could have distinguished the two: every such check
  compares against Eq.~(42) itself.

\item \textbf{Projection from lattice data to chart coordinates.}  Hofmann's equilibrium has three free parameters
  ($S^{2},\alpha,\hat\eta$) constrained by his Eq.~(5)/(26) closure.  The
  lattice export supplies five independent quantities
  ($k_{x0},k_{z0},\eta_{x},\eta_{z},\varepsilon_{z}/\varepsilon_{x}$), and
  they do not satisfy that closure: the envelope ratio implied by the tune
  depressions disagrees with the one implied by the emittances at every
  period, by a margin that varies along the linac.
  Choosing which subset defines the chart coordinates is therefore a
  projection of an over-determined dataset onto a three-parameter model, and
  different defensible projections give different flagged-period counts.  We
  adopt the pairing fixed by Eq.~(24) --- $\alpha=R_{\rm eff}$ with
  $\hat\eta$ from the same plane assignment (Sec.~\ref{sec:coords}) --- and
  we flag that the PIP-II numbers inherit that choice.  This does not affect
  the census, which places its own points and satisfies the closure by
  construction.

\item \textbf{KV closure throughout.} The dispersion solver and the
  envelope work both assume Kapchinskij--Vladimirskij self-field
  linearity.  Deviations from KV (waterbag, Gaussian) change the
  coherent growth rates, and the associated halo and
  free-energy effects~\cite{Wangler1998,Gluckstern1994Halo,FranchettiHofmannJeon2002}
  lie outside the rms-moment description.  We do not quantify the size
  of the non-KV correction here: the cited halo and free-energy studies
  establish the mechanisms but do not supply a growth-rate correction
  factor for the anisotropic chart, and we are not aware of one in the
  literature for this configuration.  Landau damping from the finite
  incoherent tune spread of a realistic distribution --- absent from KV
  by construction --- would act to attenuate the chart growth rates;
  Ref.~\cite{BurovLebedev2009} treats the analogous damping for a
  coasting beam with chromatic tune spread in a ring, which is not
  directly transferable to a bunched linac, and we do not fold any
  damping term into the dispersion relation.  The reported
  $\gamma/\nu_{0x}$ values should therefore be read as \emph{undamped
  KV-model rates}.  They are not established upper bounds on realistic growth: Landau damping is expected to attenuate them, but the
  non-KV distribution and lattice effects that are absent here are not all
  of one sign, and we have not bounded them.

\item \textbf{Basis of the $S^{2}\!\lesssim\!10$ domain restriction.}
Earlier versions of this work incorrectly described the restriction as a
convergence condition.  Eqs.~(36) and~(41) are \emph{not} truncated power
  series in $S^{2}$, and the gate is not a convergence condition.  As
  Sec.~\ref{sec:vlasov-full} makes explicit, both rows of the defining system
  Eq.~\eqref{eq:appA-vlasov-system} are linear in $S^{2}$, so an
  $n\!\times\!n$ determinant is a polynomial of degree exactly $n$: the
  $S^{4}$ block of Eq.~(36) and the $S^{4},S^{6}$ blocks of Eq.~(41) are
  determinant minors, not successive orders of an expansion.  Within
  Hofmann's model the relations are exact in $S^{2}$ at any intensity, and we
  verify numerically that $D_{3,e}$ is quadratic and $D_{4,e}$ cubic in
  $S^{2}$ to machine precision out to $S^{2}\!=\!40$.

  The truncation is in mode order.  We carry $\ell\!=\!2,3,4$; the
  expansion in $\ell$ is the open one, and there is no reason to expect
  $\ell\!>\!4$ to stay quiet as space charge grows.  That, together with the
  KV and smooth-focusing assumptions above --- both of which degrade at strong
  depression --- is the actual basis for restricting the chart, and it is a
  statement about model adequacy rather than about series convergence.  We
  retain $S^{2}\!\lesssim\!10$ as the operating restriction because it covers
  the moderate-depression regime in which the anisotropic charts have been
  benchmarked against particle-in-cell
  simulation~\cite{HofmannFranchetti2003,HofmannBF2017}, and we adopt it
  \emph{a priori} rather than tuning it to the result.  It is not a limit
  Hofmann himself states: his charts are drawn in $(R,\eta)$ and, at fixed
  emittance ratio, their corners run far beyond it, as the values below show.  At
  $\varepsilon_{z}/\varepsilon_{x}=5$ the strength reaches
  $S^{2}\!\approx\!140$ at $R\!=\!0.3$, $\eta\!=\!0.3$ and
  $S^{2}\!\approx\!330$ at $R\!=\!0.3$, $\eta\!=\!0.2$, while the large-$R$
  corner of the same chart stays well inside
  ($S^{2}\!\approx\!2.4$ at $R\!=\!2.7$, $\eta\!=\!0.3$).  Of the 41
  PIP-II lattice periods (Sec.~\ref{sec:pip2}), two in the
  low-energy SSR1 section fall at $S^{2}\!=\!14$ and $33$; we exclude them
  from quantitative chart claims and defer them to a lattice-periodic
  Floquet treatment, which we identify as future work.  The exclusion should
  be read as ``outside our adopted comparison domain for an
  $\ell\!\le\!4$ KV chart,'' not as ``outside the radius of convergence.''
  This choice does not establish a universal physical-validity boundary at
  $10$ or an error bound on every admitted point; the quantitative findings
  are conditional on this stated domain.

\item \textbf{Fourth-order odd branch and census scope.}  Eq.~(45)'s $S^{4}$ block is not printed by Hofmann; the
  solver formerly carried his isotropic-limit closure Eq.~(46) for it, and on
  that closure the branch was quiet everywhere inside the domain.  The odd quartic basis has two coefficients, compared with three for
$\ell\!=\!4_{e}$, allowing a fully symbolic evaluation of its determinant.
Section~\ref{sec:vlasov-full} gives the anisotropic $S^{4}$ block in all four
variables, reducing to Eq.~(46)
  exactly at isotropy and invariant under Hofmann's interchange.  The solver
  now evaluates that block.

  The anisotropic branch is active on the census grid: the fraction of
  in-domain cells at which the $\ell\!=\!4_{o}$ branch exceeds
  $\gamma_{\rm th}\!=\!10^{-2}$ is $0.47\%$, $3.06\%$, $6.78\%$, $11.64\%$ and
  $15.15\%$ at $\varepsilon_{z}/\varepsilon_{x}\!=\!1.2$, $1.5$, $2.5$, $5.0$
  and $8.0$, with maximum in-domain rates of $0.014$, $0.029$, $0.062$, $0.101$ and
  $0.124$.  That growth was invisible
  to the closure, and it rises with anisotropy in the same way the
  $\ell\!=\!3$ and $\ell\!=\!4_{e}$ channels do.  On the PIP-II trajectory it
  stays below threshold at every one of the $32$ evaluable in-domain periods:
  the only nonzero value is $0.0024$ at SSR2 idx~21, a factor of about $4.2$
  below the cut, so no non-oscillatory PIP-II flag changes.

  The principal census and its flagged-fraction columns nonetheless
  exclude $\ell\!=\!4_{o}$, preserving the original comparison scope:
  the disputed factors concern $\ell\!=\!3$, while the fourth-order odd
  relation is common to the printed and corrected calculations.  A common
  unstable mode can nevertheless change their aggregate disagreement count.
  The activity percentages above include overlap with the retained screen;
  they are not the additional area missed by that screen.  On the identical
  grid, the additional fractions of in-domain cells are $0.47\%$, $1.57\%$,
  $3.19\%$, $4.10\%$ and $3.99\%$, respectively.  Table~\ref{tab:l4oddcoverage}
  gives the disjoint counts and the associated misprint sensitivity.
  Including the branch is therefore a matter of an explicit mode-selection
  flag rather than of missing theory; retaining the principal scope makes its
  numerical comparison reproducible without conflating it with a complete
  fourth-order screen.

\begin{table}[t]
\caption{Fourth-order odd activity and its additional coverage on the principal
$60\!\times\!60$ census grid, inside $S^2\!\le\!10$ at threshold $0.01$.
$N$ is the in-domain denominator, $A$ the odd-active count, $O$ its overlap
with the retained screen, and $\Delta=A-O$ the newly flagged count.
The last column gives printed/corrected verdict flips before and after adding
this common branch; the principal census retains its original mode scope.}
\label{tab:l4oddcoverage}
\begin{ruledtabular}
\begin{tabular}{rrrrrr}
$\varepsilon_z/\varepsilon_x$ & $N$ & $A$ & $O$ & $\Delta$ & flips \\
\hline
1.2 & 2332 & 11 & 0 & 11 & $18\to18$ \\
1.5 & 2419 & 74 & 36 & 38 & $51\to51$ \\
2.5 & 2787 & 189 & 100 & 89 & $47\to44$ \\
5.0 & 2999 & 349 & 226 & 123 & $39\to26$ \\
8.0 & 3029 & 459 & 338 & 121 & $22\to16$ \\
\end{tabular}
\end{ruledtabular}
\end{table}

\item \textbf{Raw-triple training by construction.} The deployed
  calibrated gradient-boosted model is trained on the raw input triple
  $(R,\eta,\varepsilon_{z}/\varepsilon_{x})$ only, with samples
  filtered to $S^{2}\!\le\!10$ at acceptance time; engineered features
  ($\sigma_{0,\rm deg}$,
  $|\varepsilon_{z}/\varepsilon_{x}\!-\!1|$, etc.) and unfiltered
  training are not in the inference path and only appear in the
  historical 9-feature ablation.  A future $\sigma_{0,\rm deg}$ feature based on the
  per-period lattice phase advance extracted from the tracking output
  rather than the heuristic $(R,\varepsilon_{z}/\varepsilon_{x})$
  monotone is left
  as a follow-up.

\item \textbf{Unavailable depressed tunes for seven $\eta\!\ge\!1$ periods.}  Seven of the 41 exported periods carry a depressed
  transverse tune at or above the zero-current value --- idx~9 reaches
  $\eta\!=\!1.50$, half again above $\nu_{0x}$, which no space-charge
  depression can produce.  This is not a choice made in the present analysis:
  the values are in the lattice export itself
  (the exported tune-depression table carries seven entries above unity in its
  $x$ column), so repairing them means regenerating the matched-envelope
  solution upstream, not reprocessing it here.  We therefore quote ``4 of 32''
  against the periods actually evaluated and state the consequence plainly:
  scanning the seven across $\eta\!\in\![0.20,0.99]$, two of them (idx~9 and
  idx~16) would flag somewhere in that range, so a corrected export could give
  5 or 6 flags rather than 4.  The flagged \emph{set} is unchanged under the
  chosen weak-space-charge correction (the $\eta\!=\!0.99$ clamp), which does not establish robustness to the unknown depressed tunes. Regenerating the matched-envelope output is a priority for resolving the PIP-II count.

\item \textbf{Emittance-source sensitivity.}  The chart coordinates
  combine two codes: $(R,\eta)$ from the \textsc{TraceWin}
  transfer-matrix output and $\varepsilon_{z}/\varepsilon_{x}$ from the
  HELIX emittance export (Sec.~\ref{sec:coords}).  The two agree in the
  longitudinal plane but not the transverse one, so the geometric ratio
  falls to ${\approx}1.17$ at the HB650-2 exit under HELIX while
  \textsc{TraceWin}'s own $\varepsilon_{zz'}/\varepsilon_{xx'}$ stays at
  ${\approx}1.67$.  Repeating the full per-period evaluation with
  \textsc{TraceWin}'s ratio in place of HELIX's keeps all four flags and
  adds a fifth (LB650 idx~28, $\gamma/\nu_{0x}\!=\!0.024$, $\ell\!=\!2$),
  and roughly doubles the two SSR2 rates most sensitive to the ratio
  (idx~18: $0.018\!\to\!0.035$; idx~26: $0.031\!\to\!0.057$).  The four HELIX
  flags are therefore a \emph{subset} of the five \textsc{TraceWin} ones ---
  no flag is lost, but one is gained, so the set is nested rather than
  unchanged --- while the individual rates move by up to a factor two.  The
  identity of the $\ell\!=\!3$ flag at idx~18 survives either way.  Resolving the
  transverse-emittance discrepancy between the two codes is required
  before the flagged-period count is quoted as a machine property.

\item \textbf{Failed moment-ODE validation.} An initial 21-dimensional linearized
  second-moment ODE around the smooth-focused matched envelope, intended
  as an independent Floquet check on the dispersion-solver chart, failed
  to recover the $\ell=2$ Hofmann growth rate at a known unstable chart
  point.  Diagnosis and failed-validation protocol are reported in
  the Supplemental Material~\cite{SupplementalMaterial}.  The legacy $\Sigma\!\to\!M\Sigma M^{T}$
  map~\cite{LebedevBogacz2010} is retained as an engineering diagnostic
  rather than a stability-theory result; an independent check on the chart by a lattice-periodic Floquet analysis is left to
  future work.

\item \textbf{Matched-equilibrium theory applied to mismatched periods.}
  The Hofmann chart is a matched-equilibrium KV analysis, and none of the
  flagged PIP-II periods is matched: the export gives transverse mismatch
  factors $M_{x}\!=\!0.51$, $0.32$, $0.07$ and $0.67$ at the four, and
  $M_{x}\!=\!2.86$ at the out-of-domain idx~12 (Sec.~\ref{sec:pip2}).  On the
  PIP-II numbers this is arguably a stronger restriction than the
  $S^{2}$ gate, and unlike the gate it cannot be corrected for: no
  anisotropic mismatched-KV dispersion relation exists to correct \emph{to}.

  This limitation also affects other applications and may be more consequential than the chosen space-charge domain restriction.
  A design-office screen evaluates the chart at whatever $(R,\eta)$ the
  lattice happens to deliver, and a real trajectory is matched only where the
  designer has matched it; the chart, meanwhile, is derived for a matched
  equilibrium and says nothing about what its growth rates mean off that
  state.  In a survey of the chart
  applications we could examine --- the SNS, SPL and ESS studies of
  Ref.~\cite{HofmannFranchetti2003}, the UNILAC and J-PARC examples of
  Ref.~\cite{HofmannBook2017}, and the code-embedded screens that motivated
  this audit --- we found none that reports the mismatch state of the
  trajectory it screens, and the quantity is usually available in the same
  export as the tunes.  The mismatch can be reported as an additional column. Until a corresponding mismatched theory is available, the physical validity of the chart classification at these mismatched points remains unverified.

\item \textbf{No experimental beam.} No PIP-II linac beam data are
  yet available for this validation; the comparison is computational, not experimental.
\end{itemize}

\section{Conclusions}
\label{sec:concl}

\begin{enumerate}
\item \textbf{Changes in stability classification.}  Classifying every cell of the census grid under both forms and
  counting the symmetric difference, the two disagree on $0.73$--$2.11\%$ of
  the cells $S^{2}\!\le\!10$ admits, with no preferred direction.  Under a
  grid-resolution test the direction is stable at four of the five ratios ---
  corrected at $\varepsilon_{z}/\varepsilon_{x}\!=\!1.2$ and $8.0$, printed at
  $2.5$ and $5.0$ --- and does not resolve at $1.5$.  Among the cells that domain
  excludes the disagreement affects $6.65$--$22.27\%$ and the printed relation
  flags more at every ratio, by at least $5$:$1$ and, for
  $\varepsilon_{z}/\varepsilon_{x}\!\ge\!5$, exclusively.  The printed $S^{4}$
  block over-predicts higher-order instability wherever the $S^{4}$ terms are
  large enough for the two forms to part company at all.  The practical
  consequence is about sampling: a comparison
  drawn from inside $S^{2}\!\le\!10$ mostly tests a prediction the two forms
  share, so agreement there is not evidence for the printed $S^{4}$ block.
  Separating them requires either the excluded region, where the disagreement
  is both larger and one-directional, or the single in-domain point identified
  in Sec.~\ref{sec:census}.

\item \textbf{Determinant constraints and the isotropic sign correction.}  Eq.~(35) gives the $\ell\!=\!3$ even potential two
  expansion coefficients, so the dispersion relation is a $2\!\times\!2$
  determinant; that structure alone reproduces six independent features of the
  printed $S^{4}$ block, including which pole pairs are absent.  Requiring the
  $S^{2}$ and $S^{4}$ blocks to describe the same matrix then yields a ratio
  identity, Eq.~\eqref{eq:appA-ratio-identity}, satisfied only by
  $(1\!\mp\!2\hat\eta^{2}/\alpha)$ --- at every $(\alpha,\hat\eta)$, not just
  on the isotropic plane, where the earlier reduction argument stops.  Both
  alternative factor structures are excluded.  Two controls test the method
  rather than the conclusion.  Applied to Eq.~(41), which Hofmann printed
  \emph{with} $\alpha$-coupling factors, the identity holds as printed --- and
  it explains why that equation's factors differ from Eq.~(36)'s by a factor of
  two, since each $S^{4}$ block inherits its own $S^{2}$ factors.  Applied to
  Eq.~(26) it reproduces that relation exactly as printed, a negative control.
  A second erratum follows from Eq.~(43), whose closed-form isotropic roots
  are independent of the two equations in dispute and are satisfied only by
  $+34/(16-\sigma^{2})^{2}-2/(4-\sigma^{2})^{2}$, opposite to the printed
  Eq.~(42); Eq.~(41)'s printed $-S^{4}$ therefore stands.  An earlier version of this work adopted the opposite sign because it reproduced the printed Eq.~(42), and no
  isotropic-limit check could detect the error, since every such check
  compared against Eq.~(42) itself.

\item \textbf{Vlasov--Poisson derivation.}
  Evaluating Hofmann's Eqs.~(19) and~(23) in full, volume term included, with
  the two normalization constants fixed once against his undisputed Eq.~(28),
  reproduces Eq.~(32) as printed, gives Eq.~(36) with the corrected factors and
  the $(3+\hat\eta)$ weight, gives its odd-parity partner exactly as Hofmann's
  interchange predicts, gives Eq.~(41) exactly as printed, and gives the
  $\ell\!=\!4$ odd relation with an anisotropic $S^{4}$ block that Hofmann
  does not print and that reduces to his Eq.~(46) at isotropy
  (Sec.~\ref{sec:vlasov-full}) --- symbolically in all four variables for every
  parity except $\ell\!=\!4_{e}$, and at twelve independent $(\alpha,\hat\eta)$
  for that one.  Both alternative factor structures fail on both $\ell\!=\!3$
  parities.  This
  discharges the completeness assumption the determinant argument required,
  since the pole set is an output rather than an input; it settles four
  $\ell\!=\!4$ pole pairs that the surface term alone predicts and the volume
  term exactly cancels; and it confirms the Eq.~(42) sign by a route that never
  uses Eq.~(43).  The determinant's null vector
contains the mode structure needed to initialize a particle-in-cell
validation, but that vector is not extracted here.

\item \textbf{Higher-order chart coverage and window dependence.}  That these channels go unstable over extended regions is
  Hofmann's result, charted in his Figs.~10 and~13; what the census adds is how
  much of the chart they occupy and how that share is gated.  The fraction of
  operating points flagged by the higher-order channels alone is
  $21.6$--$22.4\%$ over $\varepsilon_{z}/\varepsilon_{x}\!\in\![1.2,8]$ ---
  varying by under $\sim\!10\%$, not exactly flat.  Those points form bands on either side of the
  $\ell\!=\!2$ region rather than surrounding it (Fig.~\ref{fig:census}), so
  a working point held at a comfortable distance from the classical band can
  sit inside a higher-order one.  This is the design-relevant statement, it is
  machine-independent, and it is qualitatively insensitive to the $S^{4}$
  correction below: repeating the census with either alternative factor
  structure moves the gated fraction by $0.17$--$0.77$ percentage points and
  the ungated one by $0.89$--$4.97$, so the qualitative statement survives all
  three while the third significant figure does not.  We do not lead with the
  ungated fraction, because it is a property of the chosen rectangle rather
  than of the physics: the same quantity reads $12.0\%$ under $R\!\le\!2.0$
  and $25.2\%$ under $R\!\ge\!0.5$, and is identically zero on the interior
  window nearest PIP-II.

  Restricted to $S^{2}\!\le\!10$ --- the domain every operating-point claim
  here is made against --- the same fraction reads $6.3\%$ at $1.2$, $16.1\%$
  at $2.5$ and $20.7\%$ at $5.0$, a factor $3.3$.  We report both because the
  gap is itself the useful quantity: the region holds $779$ cells at $1.2$ and
  $808$ at $5.0$ --- within $3.7\%$ --- but the restriction admits $19\%$ of
  the first and $77\%$ of the second.  The higher-order region does not grow with anisotropy, it
  \emph{migrates} into the adopted range for quantitative reporting with
  the $\ell\!\le\!4$ KV chart.  Reading the gated rise as a physical anisotropy effect would be a
  mistake, and an earlier version of this work made it.

\item \textbf{PIP-II flags and local growth.}  The corrected stack flags 4 of 32 evaluable
  in-domain PIP-II periods: three $\ell\!=\!2_o$ tilting/coupling modes and
  one $\ell\!=\!3$ odd-branch flag invisible to an $\ell\!=\!2$ screen.  But
  every unstable period is isolated, and integrating over the actual
  per-period phase advance gives at most $N_{e}\!=\!0.072$ on any single
  period --- a $7\%$ local amplitude scale for the strongest flag, $1\%$ for
  the higher-order one.  These are frozen-period exponents; we do not
  propagate them into a whole-linac gain, which would need mode-resolved
  transfer maps.  PIP-II is a mildly
  anisotropic machine sitting where the gated higher-order-only fraction is
  only $6$--$7\%$ (though $21.6\%$ of the ungated chart space at that
  anisotropy is higher-order-only), so this example tests the framework in a regime of weak predicted growth.  We report this as a quantified negative screening result.

  The single higher-order flag has several important sensitivities.  SSR2 idx~18 is carried by
  the $\ell\!=\!3$ \emph{odd} branch, which we obtain by Hofmann's variable
  interchange --- though that branch is now derived directly on the odd-parity
  basis (Sec.~\ref{sec:vlasov-full}), so the interchange is a confirmed step
  rather than an assumed one; it sits within $0.12\%$ of a band edge in
  $R$, $0.43\%$ in $\eta$ alone and $0.17\%$ in correlated tune-depression
  error; it disappears
  if the threshold is raised to $2\!\times\!10^{-2}$; its rate doubles if the
  emittance ratio is taken from \textsc{TraceWin} rather than from
  HELIX; and it is evaluated at a transverse mismatch factor of
  $0.32$ in a theory written for matched equilibria.  Each of those is
  established in its own place (Secs.~\ref{sec:pip2} and~\ref{sec:limits});
  together, they limit the flag to an indication for further investigation.

\item \textbf{Root verification and the excluded oscillatory channel.}  At every audited point, the production
  scan's real-root growth rate agrees with an exact-numerator companion-matrix
  solve to $5.2\!\times\!10^{-17}$, so the reported values there are not
  limited by root finding.  That comparison alone would not be enough --- it
  covers ten PIP-II points and compares the \emph{maximum} growth per branch,
  so a missed subdominant root, or one of even multiplicity, would leave it
  unchanged --- so we also compare root \emph{sets} over the census domain
  (Sec.~\ref{sec:track1}): at the $1524$ sampled points inside
  $S^{2}\!\le\!10$ every one of $678$ exact roots is recovered with no
  spurious root.  At the $476$ sampled points outside it, all $523$ exact
  roots are also recovered with no missed or spurious root.  The earlier
  uniform-only scan lost a root pair sharing one grid cell; the hybrid scan
  now resolves it, and Sec.~\ref{sec:track1} retains that failure as a
  historical regression example.  That is a sampled statement about the
  adopted domain, not a proof of chart-wide completeness, and the failure mode
  it exposes is grid resolution at large $S^{2}$ rather than anything in the
  dispersion relations.  The ten-point audit also shows the
  $\ell\!=\!4_{e}$ and $\ell\!=\!3_{e}$ blocks carry oscillatory roots at $5$
  of the $10$ PIP-II periods examined, at periods where the same branch
  returns exactly zero non-oscillatory growth.  Their formal magnitudes
  are comparable to the non-oscillatory rates we do report, and one exceeds
  the bound Hofmann quotes for such modes; Sec.~\ref{sec:track1validation}
  gives the full accounting and explains why we record it as an unresolved
  discrepancy rather than a third erratum.  We do not propagate them, but their
  existence means the scope restriction is a stronger one than the literature
  phrasing suggests.

\item \textbf{Mismatch reporting in chart applications.}  The chart is a
  matched-equilibrium KV analysis, and none of the four flagged PIP-II periods
  is matched ($M_{x}\!=\!0.51$, $0.32$, $0.07$, $0.67$).  That is not special
  to PIP-II: surveying the applications we could examine --- the SNS, SPL and
  ESS studies of Ref.~\cite{HofmannFranchetti2003}, the UNILAC and J-PARC
  examples of Ref.~\cite{HofmannBook2017}, and the code-embedded screens that
  motivated this audit --- we found none that states it, though the quantity
  usually sits in the same export as the tunes.  Unlike the $S^{2}$ gate this
  one cannot be corrected for, since no anisotropic mismatched-KV dispersion
  relation exists to correct \emph{to}; but it can be reported, at the cost of
  a column, and on the PIP-II numbers it is arguably the stronger restriction
  of the two.

\item \textbf{Computational tests that distinguish the two forms.}
  Anyone maintaining a code that carries these relations can check it in a
  single evaluation.  At
  $(R,\eta,\varepsilon_{z}/\varepsilon_{x})\!=\!(1.454237,0.297288,1.5)$ ---
  $S^{2}\!=\!9.68$, inside the adopted model domain --- the printed $\ell=3$
  relation returns a non-oscillatory $\gamma/\nu_{0x}\!=\!0.1503$ and the
  corrected one returns exactly zero.  For a code audit this is a
  presence/absence question rather than a comparison of magnitudes, so it
  is robust at the stated reporting threshold, although numerical roots
  still require residual and pole checks.  That point is not a ready-made
  \emph{experimental} test: its corrected full implemented spectrum has
  maximum $0.1160$ against the printed value $0.1503$, including fourth-order
  odd oscillatory growth.  The retained-channel value $0.0997$ is therefore
  not its full-spectrum maximum.  The second target of Sec.~\ref{sec:census},
  $(0.971186,0.426102,5.0)$ at $S^{2}\!=\!6.89$, remains farther inside the
  domain and has a larger spectral ratio: $0.0472$ against $0.1176$, or
  $2.49$.  Its originally quoted $0.009$ corrected value excludes fourth-order
  odd growth.  Both targets remain useful computational discriminators;
  observable experimental discrimination requires a specified excitation and
  diagnostic, and is not established by a spectral maximum alone.
\end{enumerate}

\begin{acknowledgments}
This work was produced by FermiForward Discovery Group, LLC
under Contract No.~89243024CSC000002 with the U.S. Department of
Energy, Office of Science, Office of High Energy Physics.
Large language model tools were used to assist with computational implementation, checks of mathematical and numerical consistency, and manuscript editing. The author takes responsibility for the scientific content and conclusions.
\end{acknowledgments}

\appendix

\section{The anisotropic $\ell=3_{e}$ and $\ell=4_{e}$ blocks: isotropic-limit audit, determinant structure, and derivation from Eqs.~(19)/(23)}
\label{app:anisotropic}

This appendix gives the explicit anisotropic $S^{4}$ blocks of the
$\ell=3_{e}$ and $\ell=4_{e}$ dispersion functions used in this work,
shows their analytic reduction to Hofmann's isotropic-limit
forms Eqs.~(37) and~(42), and establishes the
$(1\mp 2\hat\eta^{2}/\alpha)$ factors on the $\ell=3_{e}$
$\alpha$-coupling residues.

The appendix develops complementary checks with different
assumptions and domains of applicability.  Sec.~\ref{app:l3e} shows the factors are \emph{required} by the
isotropic-limit reduction; Sec.~\ref{app:l3e-derivation} shows they are
fixed at every $(\alpha,\hat\eta)$ by the determinant
structure;
and Sec.~\ref{sec:vlasov-full} \emph{derives} them, together with the
$\ell=4_{e}$ relation exactly as printed, from Hofmann's Vlasov--Poisson
equations~(19) and~(23) --- symbolically in all four variables for
$\ell=3$ (both parities) and $\ell=4_{o}$, and, for $\ell=4_{e}$, exactly in
$(\sigma,S^{2})$ at each of twelve independent $(\alpha,\hat\eta)$ rather
than symbolically in $(\alpha,\hat\eta)$.  The full kinetic
derivation is presented in Sec.~\ref{sec:vlasov-full}.

The numerical implementation was checked with a 516-test regression
suite, including 16 symbolic checks of the Appendix~\ref{app:anisotropic}
algebra, 25 checks of the Vlasov--Poisson derivation of
Sec.~\ref{sec:vlasov-full}, and 81 checks of Hofmann's Eq.~(24)/(25)
coordinate identities.  Numerical work uses \textsc{NumPy}~\cite{NumPy2020},
\textsc{SciPy}~\cite{SciPy2020}, \textsc{scikit-learn}~\cite{Sklearn2011},
and \textsc{Matplotlib}~\cite{Matplotlib2007}.  All $20{,}000$
surrogate-training labels and the metrics on the $4{,}000$ held-out
samples were revalidated against the current solver.  The failed-validation
result for the second-moment ODE closure is documented in the Supplemental
Material~\cite{SupplementalMaterial}.

\subsection{$\ell=3_{e}$ block (Eq.~(36))}
\label{app:l3e}

We write the dispersion function $D_{3,e}(s;\alpha,\hat\eta,S^{2})$ in
the form
\begin{equation}
  D_{3,e}(s) = (1+\hat\eta)^{3} + \frac{S^{2}}{8}\,\mathcal{N}_{2}(s)
  + \frac{S^{4}}{8}\,\mathcal{N}_{4}(s) = 0,
  \label{eq:appA-l3struct}
\end{equation}
with the $S^{2}$ single-pole block
\begin{multline}
  \mathcal{N}_{2}(s) =
    \frac{(1-5\hat\eta)}{1-s}
    + \frac{(9+27\hat\eta+24\hat\eta^{2})}{9-s} \\
    + \frac{(1-2\alpha)(1-2\hat\eta^{2}/\alpha)(3+\hat\eta)}{(1-2\alpha)^{2}-s} \\
    + \frac{(1+2\alpha)(1+2\hat\eta^{2}/\alpha)(3+\hat\eta)}{(1+2\alpha)^{2}-s}
  \label{eq:appA-l3sp2}
\end{multline}
reproducing Hofmann's printed expression term-by-term.  The $S^{4}$
block, however, must be written as
\begin{multline}
  \mathcal{N}_{4}(s) =
    -\frac{1}{(1-s)^{2}}
    + \frac{3}{(1-s)(9-s)} \\
    + \frac{3(1-2\alpha)\,(1-2\hat\eta^{2}/\alpha)}{(9-s)\bigl[(1-2\alpha)^{2}-s\bigr]} \\
    + \frac{3(1+2\alpha)\,(1+2\hat\eta^{2}/\alpha)}{(9-s)\bigl[(1+2\alpha)^{2}-s\bigr]}.
  \label{eq:appA-l3sp4-corrected}
\end{multline}
Hofmann's printed Eq.~(36) lists the third and fourth terms of
$\mathcal{N}_{4}$ without the $(1\mp2\hat\eta^{2}/\alpha)$
factors that we have included here.  We show below that those
factors are required by the internal consistency of Hofmann's
analysis: with the factors absent, Eq.~(36) does not reduce to
its own isotropic-limit form Eq.~(37).

\paragraph*{Isotropic-limit reduction at $\alpha=\hat\eta=1$.}
At isotropy the four poles of $\mathcal{N}_{4}$ collapse: the
$\alpha$-dependent pole pair $(1\mp2\alpha)^{2}$ at $\alpha=1$ becomes
$\{1, 9\}$, coinciding with the existing two anisotropic poles.
Term-by-term, with the factors of Eq.~\eqref{eq:appA-l3sp4-corrected}:
\begin{align}
  \text{T1:} \quad &-\frac{1}{(1-s)^{2}}, \nonumber\\
  \text{T2:} \quad &\frac{3}{(1-s)(9-s)}, \nonumber\\
  \text{T3:} \quad &\frac{3(-1)\,(-1)}{(9-s)\,(1-s)} \;=\; \frac{3}{(9-s)(1-s)}, \nonumber\\
  \text{T4:} \quad &\frac{3(3)\,(3)}{(9-s)\,(9-s)} \;=\; \frac{27}{(9-s)^{2}}. \nonumber
\end{align}
The two cross-pole terms T2 and T3 are identical (combine to
$6/[(1-s)(9-s)]$), giving the iso-limit
\begin{equation}
  \mathcal{N}_{4}\bigl|_{\alpha=\hat\eta=1}
    = -\frac{1}{(1-s)^{2}}
      + \frac{6}{(1-s)(9-s)}
      + \frac{27}{(9-s)^{2}}.
  \label{eq:appA-iso-block}
\end{equation}
Bringing onto the common denominator $(1-s)^{2}(9-s)^{2}$:
\begin{align}
  -(9-s)^{2} &+ 6(1-s)(9-s) + 27(1-s)^{2} \nonumber\\
  &= -[81-18s+s^{2}] + 6[9-10s+s^{2}] \nonumber\\
  &\qquad {}+ 27[1-2s+s^{2}] \nonumber\\
  &= (-81+54+27) + (18-60-54)s \nonumber\\
  &\qquad {}+ (-1+6+27)s^{2} \nonumber\\
  &= 0 - 96s + 32s^{2}
  \;=\; -32\,s\,(3-s). \label{eq:appA-key-id}
\end{align}
Hence
\begin{equation}
  \frac{S^{4}}{8}\,\mathcal{N}_{4}\bigl|_{\alpha=\hat\eta=1}
  = -\,\frac{4\,S^{4}\,s\,(3-s)}{(9-s)^{2}\,(1-s)^{2}},
  \label{eq:appA-iso-target}
\end{equation}
which is exactly the $S^{4}$ term of Hofmann's Eq.~(37).  Dropping
the $(1\mp2\hat\eta^{2}/\alpha)$ factors from T3 and T4 changes the
residue at the $(9-s)^{-2}$ pole from $27/8$ to $9/8$ and the
identity~\eqref{eq:appA-key-id} fails: the printed Eq.~(36) cannot
reduce to Eq.~(37) without those factors.

\subsubsection{Determinant-structure derivation of the $S^{4}$ factors}
\label{app:l3e-derivation}

The iso-limit identity of the previous subsection fixes the
$\alpha$-coupling factors only \emph{on} the isotropic plane.  Their
off-isotropic form, however, is not free: it is determined by the algebraic
structure of Hofmann's own Eq.~(35), together with the set of terms his
Eq.~(36) does and does not contain.

Eq.~(35) writes the $\ell\!=\!3$ even perturbed potential as
$\Phi^{\rm(in)}_{3e} = a_{0}x^{3} + a_{2}xy^{2}$ --- \emph{two} expansion
coefficients.  The dispersion relation is therefore the vanishing of a
$2\!\times\!2$ determinant,
\begin{equation}
  D_{3e} = \det\Bigl[\,\mathbf{M}_{0}
    + S^{2}\!\sum_{k}\frac{\mathbf{A}_{k}}{d_{k}-s}\Bigr],
  \label{eq:appA-det}
\end{equation}
and for $2\!\times\!2$ matrices
$\det(\mathbf B+\mathbf C)=\det\mathbf B+\det\mathbf C+m(\mathbf B,\mathbf C)$
with the polarization
$m(\mathbf B,\mathbf C)=B_{11}C_{22}+C_{11}B_{22}-B_{12}C_{21}-C_{12}B_{21}$.
Expanding Eq.~\eqref{eq:appA-det} in $S^{2}$ therefore gives, order by order,
\begin{equation}
  \begin{aligned}
    \det\mathbf{M}_{0} &= (1+\hat\eta)^{3},\\
    m(\mathbf{M}_{0},\mathbf{A}_{k}) &= \text{$S^{2}$ residues},\\
    \left.\begin{array}{l}
      \det\mathbf{A}_{k}\\ m(\mathbf{A}_{k},\mathbf{A}_{l})
    \end{array}\right\} &= \text{$S^{4}$ terms}.
  \end{aligned}
  \label{eq:appA-orders}
\end{equation}

The harmonic content of the basis fixes which coefficient drives which
resonance: $x^{3}$ carries $\cos(3\varphi_{x})$ and $\cos\varphi_{x}$, hence
the poles $s\!=\!9$ and $s\!=\!1$; $xy^{2}$ carries $\cos\varphi_{x}$ and
$\cos(\varphi_{x}\!\pm\!2\varphi_{y})$, hence $s\!=\!1$ and
$s\!=\!(1\!\mp\!2\alpha)^{2}$.  Consequently $\mathbf{A}_{2}$ (at $s\!=\!9$)
and $\mathbf{A}_{3},\mathbf{A}_{4}$ (at the $\alpha$-dependent poles) are
rank one, while $\mathbf{A}_{1}$ (at $s\!=\!1$, driven by both coefficients)
is rank two.  This accounts for six independent features of the printed $S^{4}$ block without any adjustable input: it explains why a
$(1-s)^{-2}$ term is present, why $(9-s)^{-2}$, $g_{3}^{-2}$ and
$g_{4}^{-2}$ terms are absent, and why there is no $(g_{3},g_{4})$
cross-term (both $\mathbf{A}_{3}$ and $\mathbf{A}_{4}$ project onto the same
coefficient, so $m(\mathbf{A}_{3},\mathbf{A}_{4})\!=\!0$ identically).

The printed Eq.~(36) also omits the $(1,g_{3})$ and $(1,g_{4})$ cross-terms,
i.e.\ it asserts
$m(\mathbf{A}_{1},\mathbf{A}_{3})=m(\mathbf{A}_{1},\mathbf{A}_{4})=0$.
Writing $\mathbf{A}_{k}=\mathbf{u}_{k}\otimes\mathbf{e}_{2}^{\mathsf T}$ for
$k=3,4$ gives
$m(\mathbf{A}_{1},\mathbf{A}_{k})=\det[\,\mathrm{col}_{1}\mathbf{A}_{1},
\mathbf{u}_{k}]$, so both vanishing forces
$\mathbf{u}_{3}\parallel\mathbf{u}_{4}$: a single scalar $r$ satisfies
$\mathbf{u}_{4}=r\,\mathbf{u}_{3}$.  That same $r$ then governs both
$\alpha$-coupling ratios,
\begin{equation}
  \frac{m(\mathbf{M}_{0},\mathbf{A}_{4})}{m(\mathbf{M}_{0},\mathbf{A}_{3})}
  = r =
  \frac{m(\mathbf{A}_{2},\mathbf{A}_{4})}{m(\mathbf{A}_{2},\mathbf{A}_{3})},
  \label{eq:appA-ratio-identity}
\end{equation}
because both are determinants against the same direction
$\mathbf{u}_{3}$.  Eq.~\eqref{eq:appA-ratio-identity} is a hard constraint
linking the $S^{2}$ and $S^{4}$ blocks at \emph{every} $(\alpha,\hat\eta)$.

Evaluating it settles the factor structure.  The $S^{2}$ block of Eq.~(36)
gives
$r=(1\!+\!2\alpha)(1\!+\!2\hat\eta^{2}/\alpha)/
[(1\!-\!2\alpha)(1\!-\!2\hat\eta^{2}/\alpha)]$.  The printed $S^{4}$
numerators $3(1\!\mp\!2\alpha)$ give $r=(1\!+\!2\alpha)/(1\!-\!2\alpha)$,
which disagrees unless $\hat\eta\!=\!0$; the $\ell\!=\!4$-style factors
$(1\!\mp\!\hat\eta^{2}/\alpha)$ likewise disagree.  Only
$(1\!\mp\!2\hat\eta^{2}/\alpha)$ --- the same factors the printed $S^{2}$
block already carries --- satisfy
Eq.~\eqref{eq:appA-ratio-identity} identically.  Both statements are verified
by symbolic regression tests.

The common $2\!\times\!2$ matrix underlying the $S^{2}$
and $S^{4}$ blocks therefore fixes the correction away from isotropy, using
Hofmann's Eq.~(35) basis and the term set of Eq.~(36).  The
iso-limit identity~\eqref{eq:appA-key-id} then fixes the remaining overall
normalization.  This argument assumes that the printed
Eq.~(36) is complete in \emph{which} pole pairs it contains, so that the
absent $(1,g_{3})$ and $(1,g_{4})$ terms are genuinely zero rather than a
second omission.  The direct evaluation
of Hofmann's Eq.~(19)/(23) integrals in Sec.~\ref{sec:vlasov-full} removes
this assumption. The corresponding coefficient matrix also permits
extraction of the mode eigenvectors, which are not obtained from the scalar
determinant alone.

\subsubsection{Direct derivation of the $\alpha$-coupling residues}
\label{app:l3e-vlasov}

The determinant argument above establishes the factor structure from
Eq.~(36)'s own term set.  It can be replaced by a direct derivation from
Hofmann's Eqs.~(19) and~(23), which also removes the assumption about that
term set.  The step that makes it tractable is his choice of momentum
variables: Eq.~(19) introduces polar coordinates $p_{x}=P\cos\Theta$,
$T^{1/2}p_{y}=P\sin\Theta$, so at the surface term's evaluation point
$P^{2}\!=\!0$ \emph{both momenta vanish}.  The unperturbed orbits Eq.~(13)
then collapse to
\begin{equation}
  \begin{aligned}
    x' &= x\cos u, & y' &= y\cos\alpha u,\\
    p_{x}' &= -x\nu_{x}\sin u, & p_{y}' &= -y\nu_{y}\sin\alpha u,
  \end{aligned}
  \label{eq:appA-orbits-P0}
\end{equation}
with $u=\varphi'-\varphi$, and the integrand
$p_{x}'\partial\Phi/\partial x' + T p_{y}'\partial\Phi/\partial y'$ becomes
elementary.

The anisotropy enters through the weight $T$ on the $y$-term, and Hofmann's
Eqs.~(8) and~(9) identify it:
\begin{equation}
  T\frac{\nu_{y}}{\nu_{x}}
  = \frac{a^{2}\nu_{x}^{2}}{b^{2}\nu_{y}^{2}}\frac{\nu_{y}}{\nu_{x}}
  = \frac{a^{2}\nu_{x}}{b^{2}\nu_{y}}
  = \frac{\varepsilon_{x}}{\varepsilon_{y}}
  \;\equiv\; \mathcal{E}.
  \label{eq:appA-T-emittance}
\end{equation}
The $T$-weighting \emph{is} the emittance ratio.  This is the origin of the
$\alpha$-coupling factors.

Finally, for rational $\alpha$ the periodic-orbit integral with Hofmann's
$(e^{-2\pi i m\omega/\nu_{x}}-1)^{-1}$ prefactor collapses to a simple kernel,
\begin{equation}
  \frac{1}{e^{-i\sigma L}-1}\int_{0}^{L}\!\!e^{-i\sigma u}\sin\beta u\,du
  = \frac{-\beta}{\beta^{2}-\sigma^{2}},
  \qquad L=2\pi m,
  \label{eq:appA-kernel}
\end{equation}
so each harmonic $\beta$ present in the integrand produces a pole at
$\sigma^{2}\!=\!\beta^{2}$ with numerator $\beta$ times its coefficient.

\paragraph*{$\ell=2$ odd, as a check.}  Here $\Phi=a_{1}xy$ is harmonic, so
Eq.~(19) is trivial and the whole relation comes from Eq.~(23).  The integrand
reduces to $-xy\nu_{x}[\sin u\cos\alpha u + \mathcal{E}\sin\alpha u\cos u]$,
whose harmonics are $\beta=1\pm\alpha$ with coefficients
$(1\pm\mathcal{E})/2$.  Eq.~\eqref{eq:appA-kernel} then gives numerators
$(1\pm\alpha)(1\pm\hat\eta^{2}/\alpha)$ on poles $(1\pm\alpha)^{2}$ ---
Eq.~(32) exactly as printed.

\paragraph*{$\ell=3$ even.}  With $\Phi=a_{0}x^{3}+a_{2}xy^{2}$ one has
$\partial\Phi/\partial x'=3a_{0}x'^{2}+a_{2}y'^{2}$ and
$\partial\Phi/\partial y'=2a_{2}x'y'$.  Decomposing column by column:
\begin{equation}
  \begin{array}{lll}
    a_{0}\ (x^{3}): & \beta=1:\ \tfrac{3}{4}, & \beta=3:\ \tfrac{3}{4},\\[2pt]
    a_{2}\ (xy^{2}): & \beta=1:\ \tfrac{1}{2},
      & \beta=1\pm2\alpha:\ \tfrac{1}{4}(1\pm2\mathcal{E}).
  \end{array}
  \label{eq:appA-l3-harmonics}
\end{equation}
Three things follow at once.  The pole $s\!=\!1$ is driven by \emph{both}
coefficients and $s\!=\!9$ by $a_{0}$ alone, while $s\!=\!(1\!\mp\!2\alpha)^{2}$
are driven by $a_{2}$ alone --- precisely the column support the determinant
argument assumed, here derived rather than inferred.  The kernel then gives
$\alpha$-coupling numerators $(1\!\mp\!2\alpha)(1\!\mp\!2\hat\eta^{2}/\alpha)$
--- the full $\alpha$ dependence of Eq.~(36)'s printed $S^{2}$ residues
$\mathcal{R}^{(2)}_{\pm}$, Eq.~\eqref{eq:appA-S2-residues}.  And since the
$S^{4}$ block is $\det\mathbf{R}$ over the same residue matrices, it inherits
the same factors: the correction of
Eq.~\eqref{eq:appA-l3sp4-corrected} follows without any appeal to the isotropic
limit or to the completeness of Eq.~(36).

\paragraph*{What the surface term alone does \emph{not} give.}  The surface
term does not reproduce $\mathcal{R}^{(2)}_{\pm}$ in full: the scalar weight
$(3+\hat\eta)$ is absent.  That is expected.  Unlike the $\ell\!=\!2$ odd case
above, the $\ell\!=\!3$ potential is not harmonic ---
$\nabla^{2}(a_{0}x^{3}+a_{2}xy^{2}) = (6a_{0}+2a_{2})x \neq 0$ --- so Eq.~(19)
contributes a volume term, and that term is what supplies the
$\hat\eta$-dependent normalization.  The same is true of $\ell\!=\!4_{e}$
below, whose printed common weight is $3(1+4\hat\eta+\hat\eta^{2})/8$.  Taken
by itself the surface term therefore settles the pole locations, the column
support, and the $\alpha$ dependence, but not the overall $\hat\eta$ scale.

That is enough for the $S^{4}$ correction --- the disputed quantity is an
$\alpha$-coupling factor, and the missing weight multiplies the $+$ and $-$
residues \emph{identically}, so it cancels from every ratio used here,
including Eq.~\eqref{eq:appA-iteration-ratio} and the $T_{+}/T_{-}$ test of
Eq.~\eqref{eq:appA-l4-ratio}.  But it is not the whole calculation, and in
Sec.~\ref{sec:vlasov-full} we carry it out: evaluating the volume term as well
closes the derivation, supplies $(3+\hat\eta)$, and removes the completeness
assumption entirely.

\paragraph*{Where the factor of two comes from.}  The same computation for
$\ell\!=\!4_{e}$ ($\Phi=a_{0}x^{4}+a_{2}x^{2}y^{2}+a_{4}y^{4}$) gives, on the
$a_{2}$ column, harmonics $\beta=2,\,2\alpha,\,2\pm2\alpha$ with the
$\alpha$-coupling coefficients $\tfrac{1}{4}(1\pm\mathcal{E})$ --- factors
$(1\!\mp\!\hat\eta^{2}/\alpha)$, exactly as Eq.~(41) prints them.  The
difference from $\ell\!=\!3$ is structural: there $\partial(xy^{2})/\partial y'
= 2x'y'$ supplies a factor two \emph{and} the product
$\sin\alpha u\cos\alpha u=\tfrac{1}{2}\sin2\alpha u$ shifts the resonance from
$1\!\pm\!\alpha$ to $1\!\pm\!2\alpha$.  Pole and factor move together, which is
why the emittance weight doubles exactly when the resonance order does.  The
factor of two between Eqs.~(36) and~(41) is therefore required, not anomalous.

\subsubsection{Two independent controls}
\label{app:l3e-controls}

We test the derivation against two published relations that do not require the proposed correction.

\paragraph*{Positive control: Eq.~(41).}  Hofmann's $\ell\!=\!4_{e}$ relation
\emph{was} printed with $\alpha$-coupling factors --- $(1\!\mp\!\hat\eta^{2}
/\alpha)$ --- on both its $S^{2}$ and $S^{4}$ blocks.  Its basis, Eq.~(40),
has three coefficients ($a_{0}x^{4}+a_{2}x^{2}y^{2}+a_{4}y^{4}$), hence a
$3\!\times\!3$ determinant and a highest power $S^{6}$, which is exactly what
Eq.~(41) contains.  Applying the same ratio test to its $\alpha$-coupling
residues gives
\begin{equation}
  \frac{T_{+}}{T_{-}} = \frac{M_{10}}{M_{9}} = \frac{M_{12}}{M_{11}}
  = \frac{(1+\alpha)(1+\hat\eta^{2}/\alpha)}
         {(1-\alpha)(1-\hat\eta^{2}/\alpha)},
  \label{eq:appA-l4-ratio}
\end{equation}
i.e.\ Eq.~(41) satisfies the identity \emph{as printed}.  The same test that
flags Eq.~(36) passes Eq.~(41) untouched.

The sign correction
of App.~\ref{app:l4e} leaves this control unchanged.  Changing the overall sign of the $S^{4}$
block --- $M_{9}$ through $M_{12}$ together --- multiplies every term by $-1$.  Both ratios in
Eq.~\eqref{eq:appA-l4-ratio} are therefore unchanged, since a common factor
cancels between numerator and denominator.  The control tests the
$\alpha$-\emph{dependence} of the residues; the misprint is in their overall
sign.  The two are independent, and the positive control stands either way.

The same constraint explains the factor-of-two difference between the equations.  The constraint is that
each relation's $S^{4}$ $\alpha$-coupling residues carry \emph{its own}
$S^{2}$ factors: Eq.~(36)'s $S^{2}$ block carries
$(1\!\mp\!2\hat\eta^{2}/\alpha)$ and Eq.~(41)'s carries
$(1\!\mp\!\hat\eta^{2}/\alpha)$, so the corrected $S^{4}$ blocks inherit those
respectively.  The factor of two therefore follows from the residues of the respective lower-order blocks.

\paragraph*{A pattern across all five printed relations.}  The
$\alpha$-coupling residues Hofmann prints are not arbitrary: throughout the
paper, a pole at $(m\!\mp\!k\alpha)^{2}$ carries the residue factor
$(m\!\mp\!k\hat\eta^{2}/\alpha)$.  Collecting every instance:

\begin{center}
\begin{tabular}{llll}
\hline\hline
Eq. & mode & pole & printed factor \\
\hline
(32) & $\ell\!=\!2_{o}$ & $(1\!\mp\!\alpha)^{2}$  & $(1\!\mp\!\hat\eta^{2}/\alpha)$ \\
(41) & $\ell\!=\!4_{e}$ & $4(1\!\mp\!\alpha)^{2}$ & $(1\!\mp\!\hat\eta^{2}/\alpha)$ \\
(45) & $\ell\!=\!4_{o}$ & $(1\!\mp\!\alpha)^{2}$  & $(1\!\mp\!\hat\eta^{2}/\alpha)$ \\
(45) & $\ell\!=\!4_{o}$ & $(1\!\mp\!3\alpha)^{2}$ & $(1\!\mp\!3\hat\eta^{2}/\alpha)$ \\
(45) & $\ell\!=\!4_{o}$ & $(3\!\mp\!\alpha)^{2}$  & $(3\!\mp\!\hat\eta^{2}/\alpha)$ \\
(36) & $\ell\!=\!3_{e}$, $S^{2}$ & $(1\!\mp\!2\alpha)^{2}$ & $(1\!\mp\!2\hat\eta^{2}/\alpha)$ \\
\hline
(36) & $\ell\!=\!3_{e}$, $S^{4}$ & $(1\!\mp\!2\alpha)^{2}$ & \emph{none} \\
\hline\hline
\end{tabular}
\end{center}

Every printed $\alpha$-coupling residue in the paper obeys the rule.  The
$S^{4}$ block of Eq.~(36) is the single exception --- on the same poles for
which its own $S^{2}$ block obeys it.  Restoring
$(1\!\mp\!2\hat\eta^{2}/\alpha)$ is precisely what the pattern demands, and
the factor of two follows from the pole being $(1\!\mp\!2\alpha)^{2}$ rather
than $(1\!\mp\!\alpha)^{2}$.  This is a third argument, independent of both the
isotropic-limit identity and the ratio identity, and it is verified against the
transcriptions using symbolic regression tests.

\paragraph*{Physical interpretation of the factors.}  Hofmann states below Eq.~(24) that the ratio of emittances is
$\hat\eta^{2}/\alpha$, so every $\alpha$-coupling factor in the paper can be
rewritten as an emittance imbalance:
\begin{equation}
  \Bigl(m \mp k\,\frac{\hat\eta^{2}}{\alpha}\Bigr)
  \;=\;
  \Bigl(m \mp k\,\frac{\varepsilon_{x}}{\varepsilon_{y}}\Bigr).
  \label{eq:appA-factor-physical}
\end{equation}
The rule of the previous paragraph then reads: \emph{a resonance
$m\nu_{x}\!\mp\!k\nu_{y}$ carries a residue weighted by
$m\!\mp\!k\,\varepsilon_{x}/\varepsilon_{y}$.}  This is what one expects of a
coupled resonance that exchanges $m$ quanta of $x$-action against $k$ quanta
of $y$-action: the driving term weights the two planes' emittances by exactly
those integers, and vanishes when the weighted emittances balance.  It is also
Hofmann's own physical reading of the $\ell\!=\!2$ odd branch, which he
describes as a difference resonance whose instability ``is an exchange of
emittance between $x$ and $y$'' (his p.~4718).

Read this way the correction requires no reference to the isotropic limit at
all.  Eq.~(36)'s $\alpha$-coupling poles are $(1\!\mp\!2\alpha)^{2}$, so
$m\!=\!1$, $k\!=\!2$, and both its blocks must carry
$(1\!\mp\!2\varepsilon_{x}/\varepsilon_{y})$.  Its $S^{2}$ block does; its
printed $S^{4}$ block does not.  This interpretation follows the pattern of the printed residues; the derivation from Hofmann's Eq.~(19)/(23) integrals is a separate calculation.  It does,
however, explain \emph{why} the pattern holds, and it independently reproduces
the factor of two from the pole structure alone.

\paragraph*{A structural question in Eq.~(41), and its resolution.}  Applying
the same pole-pair audit to the $\ell\!=\!4_{e}$ relation accounts for 17 of
the 21 possible pole pairs of its $3\!\times\!3$ structure: the twelve printed
$S^{4}$ terms are all predicted, and five absences follow from column support
(a pole driven by a single expansion coefficient cannot pair with itself or
with another pole driven by the same one).  Four absences are \emph{not}
explained by that audit --- the pairs coupling $s\!=\!4$ or $s\!=\!4\alpha^{2}$
to the sum/difference poles $4(1\!\mp\!\alpha)^{2}$.

On the surface term alone those four pairs \emph{should} be present.  The
$\ell\!=\!4_{e}$ residue at $s\!=\!4$ is driven by $a_{0}$ (in the $x^{4}$ row)
and by $a_{2}$ (in the $x^{2}y^{2}$ row), while the residues at
$4(1\!\mp\!\alpha)^{2}$ are driven by $a_{2}$ alone; the corresponding
$3\!\times\!3$ minor is a determinant of three vectors lying in three different
rows, generically nonzero.  So either the Eq.~(19) volume term cancels them
exactly, or Eq.~(41) carries omissions of the same kind as Eq.~(36).

Evaluation of the volume term gives exact cancellation.  The
full calculation of Sec.~\ref{sec:vlasov-full} reproduces Eq.~(41) with a ratio
that is independent of $\sigma$, which means the derived relation has precisely
the pole set Eq.~(41) prints --- no more and no fewer.  The four pairs are
present in the surface term and removed by the volume term.  This resolves the four missing-pair question and confirms the pole set of Eq.~(41) in the stated verification domain.

\paragraph*{A negative control: Eq.~(26) is correct as printed.}  Hofmann's
Eq.~(26) relates the depressed frequency ratio to its zero-intensity value.
Deriving it directly from his Eq.~(5) pair,
$\nu_{x}^{2}=\nu_{0x}^{2}-\omega_{p}^{2}/(1+\hat\eta)$ and
$\nu_{y}^{2}=\nu_{0y}^{2}-\omega_{p}^{2}\hat\eta/(1+\hat\eta)$, and using
$\nu_{0x}^{2}/\nu_{x}^{2}=1+S^{2}/(1+\hat\eta)$, gives
\begin{equation}
  \alpha^{2} = \alpha_{0}^{2}
    + \frac{S^{2}}{1+\hat\eta}\bigl(\alpha_{0}^{\,2}-\hat\eta\bigr),
  \label{eq:appA-eq26-corrected}
\end{equation}
an exact identity --- the residual is symbolically zero --- and this is
Eq.~(26) exactly as Hofmann prints it.

This provides a negative control. An earlier version of this work incorrectly reported Eq.~(26) as a second erratum.  It arose because the scanned journal's math fonts carry no
\texttt{ToUnicode} map, so the equations cannot be extracted mechanically, and
the discrepancy was inferred rather than read.  Working from a full \LaTeX{}
transcription of pp.~4714--4721 shows the exponent is present in the original.
Eq.~(26) follows from Eq.~(5),
which is a check on our reading of Hofmann's normalization, and the variant
with $\alpha_{0}$ in place of $\alpha_{0}^{2}$ --- which is what one obtains by
misreading the exponent --- leaves a residual proportional to
$(\nu_{0y}-\nu_{0x})$ that vanishes at $\alpha_{0}\!=\!1$.  Both statements are verified symbolically.  Eq.~(26) does not enter our
solver in any case.

The corrected interpretation of Eq.~(26) is retained in the verification
record, alongside the source transcription and the Eq.~(42) sign check below.  The two errata we
do report --- Eq.~(36)'s $S^{4}$ factors and Eq.~(42)'s sign --- rest on
internal inconsistencies within Hofmann's own printed equations, not on
reading a scan.

\subsubsection{Isotropic and residue-pattern consistency checks}
\label{app:l3e-selfconsistency}

The isotropic reduction and residue-pattern checks originally identified the
candidate factors. They provide simple checks independent of the determinant
argument and the Vlasov calculation, with the limits stated below.
\label{app:l3e-iteration}

The closing identity~\eqref{eq:appA-key-id} establishes that the
$(1\!\mp\!2\hat\eta^{2}/\alpha)$ factors are \emph{required} on the
$\alpha$-coupling residues of the $S^{4}$ block at the isotropic point
$\alpha\!=\!\hat\eta\!=\!1$ in order to recover Hofmann's stated
Eq.~(37); restoring them lifts the residue at the $(9-s)^{-2}$ pole
from the printed $9/8$ to the required $27/8$.  This is a genuine,
non-circular constraint on the iso-limit magnitudes.

Without the determinant argument of
App.~\ref{app:l3e-derivation} or the Vlasov calculation of
Sec.~\ref{sec:vlasov-full}, matching the off-isotropic factors to their
$S^{2}$ partners on the same poles is a motivated hypothesis: in the printed $S^{2}$ block of
Eq.~(36) the
$\alpha$-coupling residues are
\begin{equation}
  \mathcal{R}^{(2)}_{\pm} \;=\; (1\!\mp\!2\alpha)\,
                                (1\!\mp\!2\hat\eta^{2}/\alpha)\,
                                (3+\hat\eta),
  \label{eq:appA-S2-residues}
\end{equation}
with the upper sign for the $(1\!-\!2\alpha)^{2}$ pole and the lower
sign for the $(1\!+\!2\alpha)^{2}$ pole.  The corrected $S^{4}$
residues at the cross-products with the $(9{-}s)$ pole are taken to
share the same $(1\!\mp\!2\alpha)(1\!\mp\!2\hat\eta^{2}/\alpha)$
structure:
\begin{equation}
  \mathcal{R}^{(4)}_{\pm} \;=\; 3\,(1\!\mp\!2\alpha)\,
                                  (1\!\mp\!2\hat\eta^{2}/\alpha),
  \label{eq:appA-S4-residues}
\end{equation}
i.e.\ the form proposed in
Eq.~\eqref{eq:appA-l3sp4-corrected}.  As a self-consistency check, the ratios of
Eq.~\eqref{eq:appA-S4-residues} to Eq.~\eqref{eq:appA-S2-residues}
satisfy
\begin{equation}
  \frac{\mathcal{R}^{(4)}_{+}}{\mathcal{R}^{(2)}_{+}}
  \;=\;
  \frac{\mathcal{R}^{(4)}_{-}}{\mathcal{R}^{(2)}_{-}}
  \;=\;
  \frac{3}{3+\hat\eta}.
  \label{eq:appA-iteration-ratio}
\end{equation}
The shared universal ratio confirms that the two $\alpha$-coupling
residues of the proposed $S^{4}$ block sit in an internally consistent
relation to those of the $S^{2}$ block, as one expects from a
self-consistent determinant expansion on a fixed pole set; both ratios
in Eq.~\eqref{eq:appA-iteration-ratio} are properties of the proposed
form and do not by themselves prove the correctness of that form.
As a sensitivity check on the factor structure, we have verified symbolically that two alternative factor
structures: (a) the unfactored printed form, and (b) an
$\ell\!=\!4$-style $(1\!\mp\!\hat\eta^{2}/\alpha)$ (without the factor
of two), both fail to satisfy the iso-limit
identity~\eqref{eq:appA-key-id}: the unfactored form gives residue
$9/8$ at the $(9{-}s)^{-2}$ pole, and the $\ell\!=\!4$-style form
gives $18/8$, both incompatible with the $27/8$ demanded by Eq.~(37).
The $(1\!\mp\!2\hat\eta^{2}/\alpha)$ structure is therefore the unique
single-factor form in this family consistent with the iso-limit closing
identity --- which is what pointed to it, and which
Sec.~\ref{sec:vlasov-full} later confirms by direct derivation.

The case for the correction therefore rests on four considerations of
increasing strength: (i) the iso-limit closing identity
\eqref{eq:appA-key-id}, which constrains only the
$\alpha\!=\!\hat\eta\!=\!1$ plane; (ii) the self-consistent
mirroring~\eqref{eq:appA-iteration-ratio} between the $S^{2}$ and
$S^{4}$ $\alpha$-coupling residues, a structural plausibility
check; (iii) the asymmetry with Hofmann's Eq.~(41), whose printed
$S^{4}$ block does carry analogous $\alpha$-coupling factors of the
form $(1\!\mp\!\hat\eta^{2}/\alpha)$, indicating that the omission in
Eq.~(36) is a localized print artifact rather than a convention
difference; and (iv) the direct evaluation of Hofmann's Eqs.~(19)
and~(23), Sec.~\ref{sec:vlasov-full}, which derives the factors outright.

Checks (i)--(iii) originally identified the correction and remain useful
because they are simple to evaluate and independent of the two normalization
constants fixed against Eq.~(28) in (iv). The full calculation (iv)
establishes the correction by direct derivation.
The symbolic verification of
Eq.~\eqref{eq:appA-iteration-ratio} and of the
identity~\eqref{eq:appA-key-id} was performed using internal
\textsc{SymPy} regression tests.

\subsection{Relation to the full kinetic derivation}

The routes above establish the corrected $S^{4}$ factors by internal
consistency.  The self-contained evaluation of Hofmann's Eqs.~(19) and~(23) ---
volume term included, which is what closes the remaining freedom --- is given
in the main text, Sec.~\ref{sec:vlasov-full}.

\subsection{$\ell=4_{e}$ block (Eq.~(41))}
\label{app:l4e}

The implemented $\ell=4_{e}$ dispersion function contains all 12~anisotropic
$S^{4}$ terms and 6~anisotropic $S^{6}$ terms of Hofmann's Eq.~(41).
\paragraph*{The $S^{4}$ sign, and a second misprint (Eq.~(42)).}
Hofmann's Eq.~(41) carries $-S^{4}$ on its second-order block, while his
isotropic reduction Eq.~(42) prints
$S^{4}[-34/(16-\sigma^{2})^{2}+2/(4-\sigma^{2})^{2}]$.  These two are mutually
inconsistent: reducing Eq.~(41) at $\alpha=\hat\eta=1$ with the printed
$-S^{4}$ does not give the printed Eq.~(42).  One of them is misprinted, and
the choice is not cosmetic --- the two resolutions give materially different
$\ell\!=\!4_{e}$ growth rates.

Hofmann's Eq.~(43) supplies five closed-form isotropic roots for this check.  They are printed immediately below Eq.~(42), so the independence we
rely on is not editorial but numerical: a sign corrupted while typesetting
Eq.~(42) cannot propagate into a separately typeset root set.  Substituting them into
Eq.~(42) shows that it is satisfied identically only by
\begin{equation}
  S^{4}\Bigl[\,\frac{34}{(16-\sigma^{2})^{2}}-\frac{2}{(4-\sigma^{2})^{2}}\Bigr],
  \label{eq:appA-eq42-corrected}
\end{equation}
and not by the printed form.  Clearing denominators gives an equivalent polynomial identity:
only Eq.~\eqref{eq:appA-eq42-corrected} yields a numerator equal to
$-16\prod_{i}(\sigma^{2}-\sigma_{i}^{2})$ over exactly the Eq.~(43) roots and
no others.  So \emph{Eq.~(42) carries the typo}, Eq.~(41)'s printed $-S^{4}$
is correct, and we adopt it.

We record that an earlier version of this work adopted $+S^{4}$ in Eq.~(41), on
the reasoning that only that sign reproduces the printed Eq.~(42).  That
reasoning was sound in form but validated against the misprinted equation; the
error was invisible to every isotropic-limit check we had, because those checks
compared against Eq.~(42) itself.  It was found only by testing against
Eq.~(43), which the checks had not used.  This is the second misprint the
present internal-consistency audit locates, after the Eq.~(36) $S^{4}$
$\alpha$-coupling factors, and it is missed by a reduction test that uses the printed isotropic equation as its reference.  The resolution is verified symbolically.

With that sign, the $S^{6}$ block (unchanged between the two candidate forms)
and the $S^{4}$ block both reduce to Eq.~\eqref{eq:appA-eq42-corrected}
symbolically at the level of every pole residue and numerically to
floating-point zero across a dense $(s,S^{2})$ grid.

The reduction at $\alpha=\hat\eta=1$ collapses the 12~anisotropic
$S^{4}$ terms (labeled $M_{1}\!\dots\!M_{12}$ in this calculation) onto the two independent iso poles
$(4-s)^{-2}$ and $(16-s)^{-2}$.  Grouping the terms by their iso-limit
contribution (verified symbolically; see below):
the self-coupling triple $M_{1}\!+\!M_{6}\!+\!M_{7}$ sums to
$+2/(4-s)^{2}$; the group $M_{3}\!+\!M_{10}\!+\!M_{12}$ sums to
$-34/(16-s)^{2}$; and the remaining cross-terms
$M_{2}\!+\!M_{4}\!+\!M_{5}\!+\!M_{8}$ sum to zero.  The four
$\alpha$-coupling terms do \emph{not} cancel among themselves:
$M_{9}\!+\!M_{10}\!+\!M_{11}\!+\!M_{12}=-15/(16-s)^{2}$ at iso (with
$M_{9}\!=\!M_{11}\!=\!0$ there), and it is the pairing of $M_{10},M_{12}$
with the $\alpha$-independent $M_{3}$ that produces the $-34$ numerator.
The bare subgroup sum is
\begin{equation}
  \bigl[\,M_{1}+\cdots+M_{12}\bigr]_{\alpha=\hat\eta=1}
  = \tfrac{2}{(4-s)^{2}} - \tfrac{34}{(16-s)^{2}},
  \label{eq:appA-l4-Msum}
\end{equation}
but that is \emph{not} the contribution to $D_{4,e}$.  Eq.~(41) carries
$-S^{4}$ on this block, so the contribution is
$-S^{4}$ times Eq.~\eqref{eq:appA-l4-Msum}, i.e.\
$S^{4}[\,34/(16-s)^{2} - 2/(4-s)^{2}\,]$ --- which is the
\emph{sign-corrected} Eq.~\eqref{eq:appA-eq42-corrected}, not the printed
Eq.~(42).  The distinction is easy to lose because the bare sum happens to
equal the printed form exactly; that coincidence is the whole content of the
Eq.~(42) misprint.  These subgroup sums are
checked symbolically.  The $S^{6}$ block
reduces analogously to the three terms of Eq.~(42).  In our reference
implementation the per-grid reduction to Eq.~(42) at iso evaluates
to floating-point zero (no rounding tolerance).

\subsection{Verification protocol}
\label{app:verify}

The corrected anisotropic blocks are verified by three checks:
(i)~per-mode iso-limit reduction
($\bigl|D_{\ell}^{\rm aniso}(s,1,1) - D_{\ell}^{\rm iso}(s)\bigr| < 10^{-12}$ for
$\ell=3_{e}$, exactly $0$ for $\ell=4_{e}$);
(ii)~required divergence from the legacy iso-limit prototype at
$\alpha\!\ne\!1$, $\hat\eta\!\ne\!1$ (the corrected anisotropic forms
must differ from the legacy iso-limit closure where they are no
longer mathematically identical);
(iii)~symmetry of the $\ell=3$ odd branch under the variable
interchange $(\alpha,\hat\eta)\!\to\!(1/\alpha,1/\hat\eta)$ per
Ref.~\cite{Hofmann1998} p.~4719.
All three checks pass at their stated tolerances.

\subsection{Minor-truncation sensitivity of the $\ell=4_{e}$ block}
\label{app:convergence}

The $S^{2}\!\lesssim\!10$ bound used throughout the manuscript is an
\emph{a~priori} domain restriction, and --- as Sec.~\ref{sec:limits} now
states --- it is a model-adequacy restriction rather than a convergence
condition, because Eq.~(41) is an exact cubic in $S^{2}$ rather than a
truncated series.  The diagnostic below measures sensitivity to omitted minors: dropping the $S^{4}$ and $S^{6}$
terms removes required minors of a finite determinant, so what it measures is
how much of the $\ell\!=\!4_{e}$ rate those minors carry, i.e.\ a
sensitivity rather than a remainder.  With that reading we evaluate the
$\ell\!=\!4_{e}$ growth rate with the determinant retained to
$S^{2}$-only,
$S^{2}\!+\!S^{4}$, and full $S^{2}\!+\!S^{4}\!+\!S^{6}$, at every PIP-II
period flagged above $\gamma_{\rm th}\!=\!10^{-2}$ by the combined
solver (so the set includes periods whose flag is carried by
$\ell\!=\!2$ or $\ell\!=\!3$, at which the $\ell\!=\!4_{e}$ block is
quiet), and report, for the $\ell\!=\!4_{e}$ block at each, the $S^{6}$
contribution
$|\gamma_{S^{2}\!+\!S^{4}} - \gamma_{\rm full}|/\gamma_{\rm full}$.

\emph{Result.}\quad On the corrected PIP-II trajectory the
$\ell\!=\!4_{e}$ block is quiet at every flagged period, in-domain and
out: all six flagged periods return
$\gamma_{S^{2}}\!=\!\gamma_{S^{2}+S^{4}}\!=\!\gamma_{\rm full}\!=\!0$,
so the ratio $|\gamma_{S^{2}+S^{4}}-\gamma_{\rm full}|/\gamma_{\rm full}$
is $0/0$ rather than zero; the numerical implementation regularizes the denominator and prints
$0$.  The diagnostic therefore provides no relative-rate information on this trajectory --- there is no
$\ell\!=\!4_{e}$ rate to attribute to any minor.  The flagged set is carried by the
$\ell\!=\!2_o$ tilting/coupling mode and the $\ell\!=\!3$ odd branch
(Sec.~\ref{sec:pip2}), neither of which involves an $S^{6}$ block --- the
$\ell\!=\!3$ determinant is $2\!\times\!2$ and so terminates exactly at
$S^{4}$, with no $S^{6}$ term to omit.    We note that the iso-limit
regression tests are \emph{not} a substitute for it: they verify algebraic
reduction to Hofmann's closed forms, which is a transcription check, and say
nothing about the size of any individual minor.  We retain the diagnostic as a
guard: were a future lattice or emittance revision to activate
$\ell\!=\!4_{e}$, it would report the truncation sensitivity at that
point.  This is a weaker \emph{ex~post} check than the previous version
of this appendix claimed, and it is one reason the $S^{2}\!\lesssim\!10$
bound is stated as an \emph{a~priori} domain restriction from the
physical scaling rather than as an empirically calibrated threshold.

The $\ell\!=\!3_{e}$ block of Eq.~(36) contains only $S^{2}$ and $S^{4}$
terms, and that is not an omission to be repaired: its $2\!\times\!2$
determinant terminates at $S^{4}$ exactly, so there is no $S^{6}$ minor and
none is missing from Ref.~\cite{Hofmann1998}.  The analogous diagnostic at
$\ell\!=\!3_{e}$ is therefore the $S^{2}$-only versus $S^{2}\!+\!S^{4}$
comparison, which is also implemented in the truncation diagnostic.  The $S^{2}\!\lesssim\!10$ restriction is inherited for
$\ell\!=\!3$ on the same model-adequacy grounds as for $\ell\!=\!4$
(Sec.~\ref{sec:limits}) --- the mode-order truncation and the KV assumptions
--- not from any $\ell\!=\!3$-internal convergence property, of which there
is none to test.

\subsection{Search for published correction notices}
\label{app:correction-search}

We searched for a published correction to Ref.~\cite{Hofmann1998} and found none
in the records examined, most recently on 5~September~2026.  The search
combines the journal's own index, the INSPIRE-HEP record, and Crossref title
queries.  APS publishes an erratum as a separate article whose title names
its source, for example ``Erratum: $\ldots$
[\textit{Phys.\ Rev.\ Accel.\ Beams} \textbf{20}, 014202 (2017)]''; we therefore
check whether a returned notice actually identifies the cited work.  The
archived preliminary scan covered 18 journal articles and located that
erratum to Ref.~\cite{HofmannBF2017}, which is cited with its original
article.  That preliminary scan does not cover every entry in the present
bibliography and is not a complete count of publisher correction notices.
Crossref's \texttt{relation} object
cannot by itself establish whether an erratum exists: its
\texttt{update-to}/\texttt{updated-by} entries are empty for
Ref.~\cite{HofmannBF2017}, despite the published erratum.  Our searches found no notice addressing
the two errors examined in this paper; this negative search result does not prove
that no correction exists.  The conclusion is limited to the records and queries examined.

\section{A fast solver-labeled surrogate}
\label{app:surrogate}
\label{sec:track3}

Sections~\ref{sec:vlasov-full} and~\ref{sec:census} show that the corrected
$\ell\!=\!3$ relation changes individual stability classifications even
when the aggregate fraction changes little. Here we test whether a surrogate
trained on solver labels retains that spatial structure. Despite near-unity
in-distribution ranking and $98.1\%$ cell agreement, the classifier misses an
entire narrow $\ell\!=\!3$ odd band and half the flagged PIP-II periods.
These failures limit its use as a standalone screen and support direct
evaluation of the corrected relations, which are finite polynomials in
$S^{2}$ and inexpensive to compute.

A practical question for operations is whether the deterministic
Hofmann chart can be evaluated cheaply enough to serve as a
working-point screen, e.g.\ for online use during commissioning,
or for embedding inside a fault-recovery decision layer.  The
dispersion solver of Sec.~\ref{sec:track1} runs in $\mathcal{O}(\rm
ms)$ per call; a classifier surrogate is $\mathcal{O}(\rm \mu s)$
per call after one-time training providing a potential speed advantage.  We do \emph{not} propose the surrogate as a
physics result; it is a screening tool that does not replace
the solver.  The deployed surrogate is trained on the raw input triple
$(R,\eta,\varepsilon_{z}/\varepsilon_{x})$ only; a prior unfiltered
comparison added the
legacy generator's engineered non-linear features and found no
measurable ranking gain, so we did not carry them into the deployed,
validity-gated model (Sec.~\ref{sec:track3-raw}).  We also confirm by a
cross-distribution test
(Sec.~\ref{sec:track3-cd}) that a representative synthetic-rule
labeling scheme (a hand-tuned summed-Gaussian-band score in
the zero-current phase advance $\sigma_{0}$ --- the undepressed phase
advance per focusing cell, in degrees --- and
$\varepsilon_{z}/\varepsilon_{x}$) fails to generalize; we do not claim, and the test does not establish, that
\emph{every} synthetic-rule labeling scheme would fail.

\subsection{Five-baseline comparison}

We train five classifiers (logistic regression, RBF-SVM, random
forest, gradient boost, kNN) on $N_{s}\!=\!20{,}000$ points drawn
uniformly in $R\in[0.1,3.0]$, $\eta\in[0.2,0.99]$,
$\varepsilon_{z}/\varepsilon_{x}\in[0.9,9.0]$ and labeled by the
iso-limit-validated solver of Sec.~\ref{sec:track1} as ``unstable'' when
$\gamma > 0.01\,\nu_{0x}$.
The training distribution is intentionally restricted to the
adopted model domain $S^{2}\!\le\!10$ (Sec.~\ref{sec:track1}):
samples that fall outside this gate, computed pointwise from
$(R,\eta,\varepsilon_{z}/\varepsilon_{x})$ via the
$S^{2}(R,\eta)$ map of Eq.~\eqref{eq:S2_def}, are dropped before
training and are flagged as out-of-scope at inference time rather
than scored.  In addition to the $S^{2}\!\le\!10$ physics gate, the
deployed surrogate enforces a hard \emph{domain gate}: it refuses to
score points that fall outside its training hyperrectangle
$R\!\in\![0.1,3.0]$, $\eta\!\in\![0.2,0.99]$,
$\varepsilon_{z}/\varepsilon_{x}\!\in\![0.9,9.0]$, including periods
in the $\eta\!>\!0.99$ no-current-depression region where the solver
returns trivial stability and points at $R\!>\!3$ where the
trajectory has left the surrogate's sampled support.  Both gates are
applied jointly at inference time, and the PIP-II results of
Sec.~\ref{sec:pip2} report agreement only on the periods that pass
both.
The deployed surrogate consumes the raw input triple
$(R,\eta,\varepsilon_{z}/\varepsilon_{x})$ only; engineered
non-linear features ($\sigma_{0,\rm deg}$,
$|\varepsilon_{z}/\varepsilon_{x}-1|$, etc.) appear only in the
ablation study of Sec.~\ref{sec:track3-raw} and are not used in any
of the calibrated gradient-boosting metrics quoted below or in the PIP-II
cross-check of Sec.~\ref{sec:pip2}.  This keeps the surrogate
aligned with the scope of the underlying chart: it predicts what the chart predicts on the gated subset, and refuses to
predict elsewhere.  All models are wrapped in
\texttt{StandardScaler} pipelines (\texttt{scikit-learn}~\cite{Sklearn2011})
and evaluated with grouped
4-fold CV (groups are 4$\times$4 buckets in $(\varepsilon_{z}/\varepsilon_{x},\eta)$)
to reduce near-neighbor leakage between adjacent operating points; we
note the grouping is on $(\varepsilon_{z}/\varepsilon_{x},\eta)$ only, so
$R$ is shared across folds and the reported CV metrics are mildly
optimistic in $R$.  All ML metrics reported
in this section, and the PIP-II surrogate-vs-solver agreement
quoted in Sec.~\ref{sec:pip2}, are computed on the gated subset;
the two PIP-II periods at $S^{2}\!>\!10$
(Table~\ref{tab:pip2_flagged}) and the seven periods outside the
training hyperrectangle are not scored by the surrogate.

Table~\ref{tab:ml} reports cross-validated accuracy, ROC--AUC,
Brier loss, and expected calibration error
(ECE~\cite{Guo2017}, computed with 10 equal-width probability bins) per
classifier.  The gradient-boosted trees~\cite{Friedman2001} achieve the highest ranking score
(AUC=0.994), the random forest~\cite{Breiman2001} is close behind
(AUC=0.991), and the linear logistic-regression baseline gives AUC=0.867.

\begin{table}[t]
\caption{Five-baseline machine-learning surrogate trained on
$N_{s}\!=\!20{,}000$ solver-labeled, validity-gated points
($S^{2}\!\le\!10$ filter applied at sample-acceptance; raw-triple
features $(R,\eta,\varepsilon_{z}/\varepsilon_{x})$).  The five baseline
rows above the rule report grouped 4-fold cross-validation means.  The
``GB~$+$~isotonic'' row reports the \emph{held-out} metrics
($N\!=\!4{,}000$ fresh gated samples at an independent seed) of the
isotonic-calibrated gradient-boosted model used for probability reporting
(Brier $0.012$, ECE $0.008$); the calibrated and uncalibrated held-out
metrics were evaluated separately on the same independent sample.
Gradient boosting has the highest cross-validated ROC--AUC; isotonic calibration lowers the
held-out ECE to $0.008$, an $\sim\!6\times$ reduction relative to the
uncalibrated model on the same held-out set (ECE $0.047$), at the cost of
$0.001$ in accuracy.
Training on the synthetic-rule labels gives cross-distribution ROC--AUC $0.634$.  The MLP baseline is omitted as a wall-time choice
(sklearn's single-threaded MLP is the bottleneck at
$N_{s}\!=\!20{,}000$); we do not claim it would have ranked above the
gradient-boosted model.}
\label{tab:ml}
\begin{ruledtabular}
\begin{tabular}{lcccc}
classifier        & acc    & ROC--AUC & Brier   & ECE \\
\hline
logreg            & 0.836 & 0.867    & 0.114   & 0.048 \\
RBF-SVM           & 0.862 & 0.918    & 0.097   & 0.048 \\
random forest     & 0.950 & 0.991    & 0.048   & 0.100 \\
\textbf{gradient boost} & \textbf{0.967} & \textbf{0.994} & \textbf{0.027} & 0.035 \\
kNN               & 0.921 & 0.965    & 0.059   & 0.031 \\
\hline
GB $+$ isotonic$^\ast$ & 0.983 & 0.999 & 0.012 & \textbf{0.008} \\
\hline\hline
\multicolumn{5}{l}{\footnotesize $^\ast$~held-out, not CV-comparable to rows above.} \\
\multicolumn{5}{l}{\textit{Cross-distribution (train$\to$test):}} \\
\multicolumn{5}{l}{synth $\to$ solver  : acc=0.476, AUC=0.634} \\
\multicolumn{5}{l}{solver $\to$ synth  : acc=0.640, AUC=0.719} \\
\end{tabular}
\end{ruledtabular}
\end{table}

\subsection{Calibration and ablation}
\label{sec:track3-raw}

\begin{figure*}[t]
  \centering
  \includegraphics[width=0.98\textwidth]{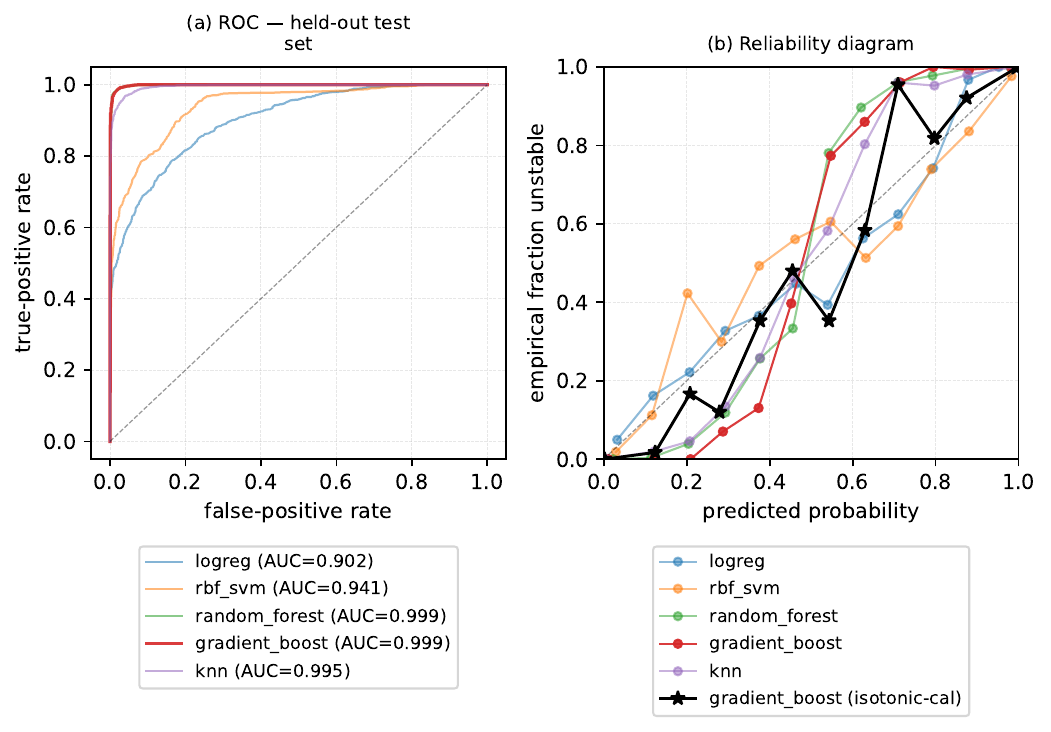}
  \caption{ROC and reliability diagram for the five baselines on a
  held-out test set ($N\!=\!4{,}000$ fresh gated samples); the AUC
  values shown in the legend are computed on this held-out set and are
  correspondingly higher than the more conservative cross-validated
  values reported in Table~\ref{tab:ml}, by which the gradient-boosted model
  was selected.  This held-out set is drawn from the same uniform
  gated distribution as the training set (different seed), so its near-unity
  AUC is an in-distribution \emph{interpolation} metric, not a test of
  generalization to a different operating-point distribution; the
  $n\!=\!32$ PIP-II cross-check (Sec.~\ref{sec:pip2}) and the
  cross-distribution test (Sec.~\ref{sec:track3-cd}) are the
  out-of-distribution evidence.  The calibrated curve in the reliability panel is the
  isotonic-wrapped gradient-boosted model.}
  \label{fig:fig7}
\end{figure*}

\begin{figure*}[tbp]
  \centering
  \includegraphics[width=0.98\textwidth]{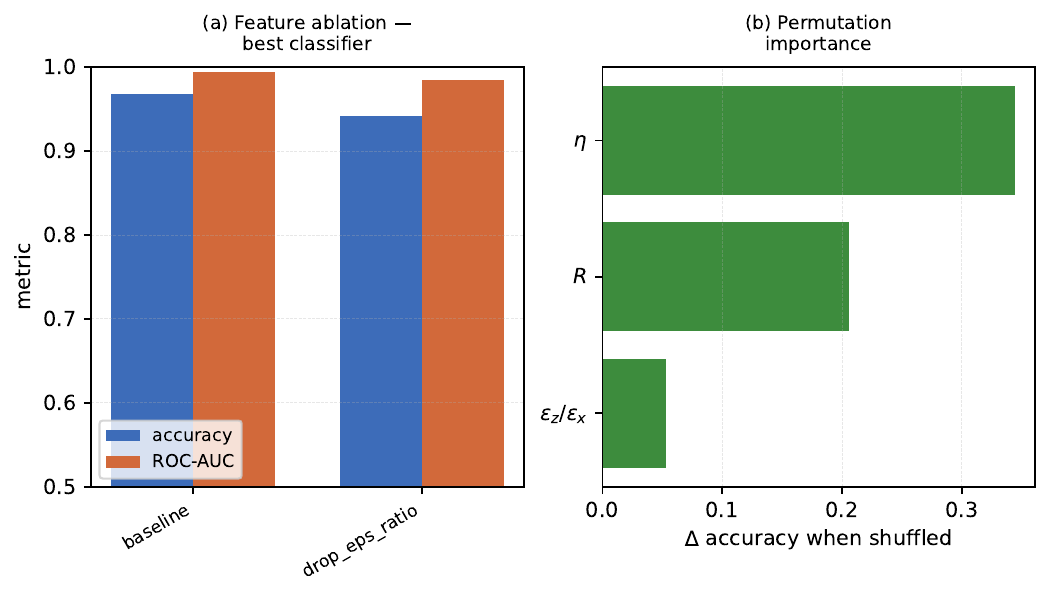}
  \caption{Feature ablation (left) and permutation importance (right)
  for the deployed gradient-boosted classifier on the $N_{s}\!=\!20{,}000$
  gated training subset.  The deployed surrogate already uses only the
  raw triple $(R,\eta,\varepsilon_{z}/\varepsilon_{x})$, so the
  ablation panel shows only the baseline (ROC--AUC $0.994$) and the
  \texttt{drop\_eps\_ratio} subset (dropping
  $\varepsilon_{z}/\varepsilon_{x}$, AUC drops to $0.985$).
  Permutation importance is led by $\eta$
  ($\Delta_{\rm acc}\!\simeq\!0.34$) ahead of $R$ ($0.21$), with
  $\varepsilon_{z}/\varepsilon_{x}$ contributing $0.05$ on the gated
  domain.}
  \label{fig:fig8}
\end{figure*}

Figure~\ref{fig:fig7} shows the ROC curves and the reliability diagram
including the isotonic-calibrated
wrapper~\cite{NiculescuMizilCaruana2005}.  The calibrated model is a
\texttt{CalibratedClassifierCV} wrapper that refits the base gradient-boosted
model on internal 3-fold splits and fits the isotonic map on the
held-out portion of each; it is therefore a refit of the selected
architecture, not literally the CV-scored estimator object, and its
metrics (Table~\ref{tab:ml}, final row) are reported on the independent
held-out set rather than under CV.
Figure~\ref{fig:fig8} reports the feature ablation and permutation
importance for the deployed gradient-boosted model on the gated training
subset.  The deployed surrogate already consumes only the raw triple
$(R,\eta,\varepsilon_{z}/\varepsilon_{x})$; the single ablation
beyond the trivial baseline is therefore \texttt{drop\_eps\_ratio}
(dropping $\varepsilon_{z}/\varepsilon_{x}$), which costs
$\Delta\text{ROC--AUC}\!\approx\!0.010$ relative to the baseline
(0.994~$\to$~0.985).  Permutation importance is led by the tune
depression ($\Delta_{\rm acc}\!=\!0.344$ for $\eta$ and $0.206$ for
$R$), with $\varepsilon_{z}/\varepsilon_{x}$ contributing only
$\Delta_{\rm acc}\!=\!0.053$.  The emittance ratio is therefore the
weakest of the three coordinates for this classifier on this sampling
domain --- consistent with the small ablation cost above, and a caveat
on reading the surrogate as a probe of anisotropy sensitivity.  We note
the importance is computed on the training set and is therefore
optimistic.  Note also that these are gated-domain statistics: the
$S^{2}\!\le\!10$ filter removes much of the strongly anisotropic corner,
so the low $\varepsilon_{z}/\varepsilon_{x}$ weight is partly a property
of the sampled subspace rather than of the chart as a whole.  The previous-version unfiltered
9-feature comparison (with $\sigma_{0,\rm deg}$ and its derived
products) was evaluated separately; the
present deployed model is gated to the raw triple by construction.

\subsection{Spatial distribution of surrogate errors}
\label{app:surrogate-chart}

\begin{figure*}[t]
  \centering
  \includegraphics[width=0.98\textwidth]{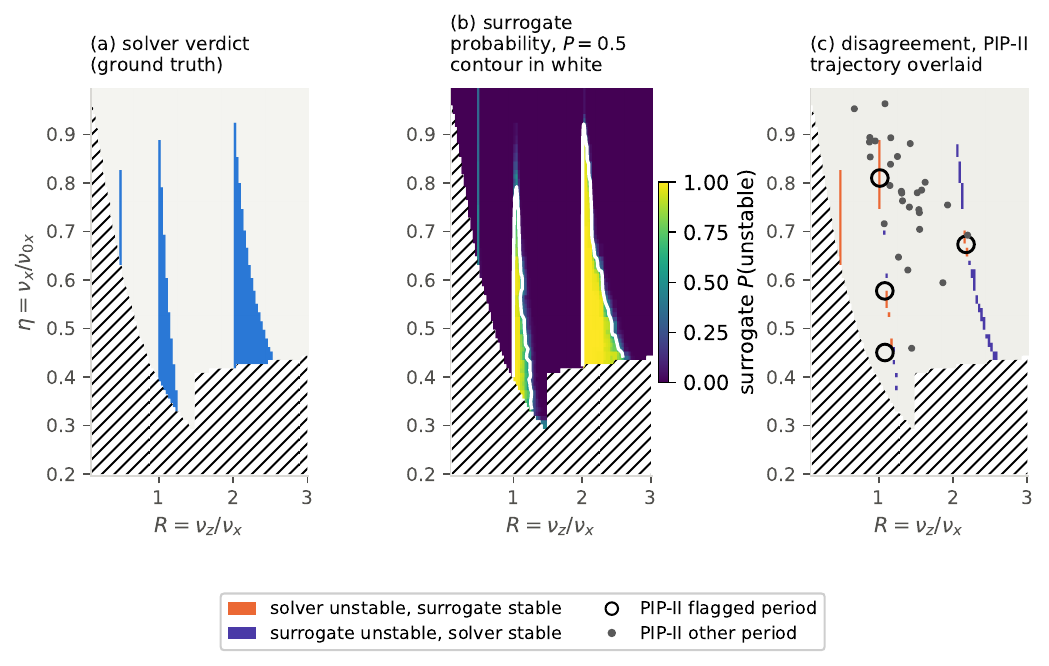}
  \caption{Surrogate versus solver in the gated chart space at
  $\varepsilon_{z}/\varepsilon_{x}=1.5$.  (a)~Solver verdict.  (b)~Calibrated
  surrogate probability with its $P\!=\!0.5$ contour in white.  (c)~Disagreement
  map with the in-domain PIP-II periods overlaid; open circles are the
  flagged periods.  Diagonal hatching marks the $S^{2}\!>\!10$ region, which is
  neither trained on nor scored.  The surrogate captures the two wide bands and
  misses the narrow $\ell\!=\!3$ odd band at $R\!\approx\!0.49$ completely.}
  \label{fig:surrogate_chart}
\end{figure*}

The aggregate metrics do not locate the surrogate's errors, whose spatial distribution determines its usefulness as a screen.  Figure~\ref{fig:surrogate_chart} renders the comparison
spatially at $\varepsilon_{z}/\varepsilon_{x}\!=\!1.5$.

Over the gated chart space ($5{,}272$ cells) the surrogate reproduces the
solver verdict on $98.1\%$, with chart-space recall $0.916$ and precision
$0.919$ --- comfortably better than its $0.50$ recall on the PIP-II subset,
which is a small sample sitting near boundaries.  The spatial distribution of the errors is nonuniform.

The solver produces three unstable bands at this emittance ratio, $593$ cells
in total.  The surrogate reproduces the two wide ones with errors confined to
their edges: a mixed band at $R\!\approx\!1.0$--$1.2$ ($215$ cells, $147$
$\ell\!=\!2$-dominated and $68$ $\ell\!=\!4_{e}$-dominated, $11\%$ missed) and
an $\ell\!=\!3$ odd band at $R\!\approx\!2.0$--$2.5$ ($356$ cells, $1\%$
missed).  Notably it misses none of the $68$ $\ell\!=\!4_{e}$ cells.

The surrogate misses the third band entirely: a single-column $\ell\!=\!3$ odd
feature at $R\!\approx\!0.49$, $22$ cells, all $22$ classified stable.  There
the solver returns $\gamma/\nu_{0x}\!=\!0.011$--$0.021$ and the surrogate
returns $P\!=\!0.38$--$0.45$.  The probability field retains the feature: the model puts the band well above its background
probability and yet below $0.5$ everywhere across it.  The structure is
present in the calibrated probability field and destroyed by thresholding.

This comparison shows a limitation of aggregate ranking metrics: a model with
ROC--AUC $0.994$ and $98\%$ cell agreement misses $100\%$ of a band a design
screen would need to see.  Since the band survives in the probability field, a surrogate used as a \emph{pre-filter} should be read at a threshold
set by the cost of a missed flag --- here anything below $P\!\approx\!0.35$
would retain it --- rather than at the nominal $0.5$; that is an argument for
reporting calibrated probabilities, as we do in
the Supplemental Material~\cite{SupplementalMaterial}, rather than verdicts.  The missed band is also on the same channel as the flag it misses on the PIP-II trajectory
(Sec.~\ref{sec:pip2}, SSR2 idx~18, also $\ell\!=\!3$ odd), which suggests the
PIP-II failure is an instance of this rather than an accident of the small
sample.  This supports retaining the solver for verification: the surrogate is a fast pre-filter, and its failures fall
precisely on the higher-order structure that motivates the method.

\subsection{Transfer between synthetic and solver labels}
\label{sec:track3-cd}

To verify the surrogate reflects physics rather than memorized
labeling rules, we evaluate it across the distribution boundary.
The gradient-boosted surrogate trained on synthetic-rule labels and
tested on solver-labeled held-out data achieves ROC--AUC $0.634$,
a substantial drop from the solver-trained-and-tested baseline of
$0.994$.  Here the legacy generator labels a point ``unstable'' by
thresholding (at $0.6$) a score formed as the \emph{sum} of four
Gaussian ``bands'' in a heuristic zero-current phase advance
$\sigma_{0}$, each gated by $\eta$, plus a linear
$0.4\,|\varepsilon_{z}/\varepsilon_{x}-1|$ anisotropy penalty, followed
by a $1\%$ random label flip that caps the attainable
cross-distribution ranking score; $R$ is
reconstructed from $\sigma_{0}$ and
$\varepsilon_{z}/\varepsilon_{x}$ afterward and is not itself a
labeling input.  For this tested rule, \emph{this specific
legacy labeling rule does not generalize to the validated
physics distribution}.  We do not claim that every synthetic
labeling scheme fails, only that this representative one does; on the
strength of this single negative result we prefer solver-generated
labels over heuristic rules of this form for any working-point-screen
deployment.  The
inverse direction (solver-trained, synth-tested) recovers AUC
$0.719$, consistent with partial signal in the synthetic rule but
well below the in-distribution baseline.  This cross-distribution gap
conflates two effects: the label-rule mismatch and a covariate shift
between the two generators (the synthetic set spans
$\varepsilon_{z}/\varepsilon_{x}\!\in\![0.5,3.0]$ and is concentrated
at low $R$, $\bar R\!\approx\!0.5$, whereas the solver set spans
$[0.9,9.0]$ with $\bar R\!\approx\!1.3$ and $\sim\!65\%$ of its points
at $\varepsilon_{z}/\varepsilon_{x}\!>\!3$, outside the synthetic
support).  We do not separate the two contributions quantitatively;
the label-rule mismatch alone is sufficient to motivate
solver-generated labels.

\section{Per-period PIP-II validity table}
\label{app:perperiod}

The full per-period table (all 41 rows) is provided in the Supplemental
Material~\cite{SupplementalMaterial}; the
compact subset of periods with solver-flagged growth above
$\gamma_{\rm th}\!=\!10^{-2}$ is reproduced in the main text
(Table~\ref{tab:pip2_flagged}).  Each row lists the chart coordinates
$(R,\eta,\varepsilon_{z}/\varepsilon_{x})$, the dimensionless
space-charge parameter $S^{2}$, the solver growth rate in each retained
mode order, the combined growth rate, and a model-domain flag
(\textbf{N}~$=$~$S^{2}\!>\!10$, outside the adopted model domain;
\textbf{Y} otherwise).  Seven of the 39 in-domain records carry an
exported $\eta\!\ge\!1$ and are marked $\ddagger$ there: they are stable
by convention rather than physically evaluated, which is why the
flagged count in Sec.~\ref{sec:pip2} is quoted against the remaining
32.  The final listed energy ($\sim\!715$~MeV at
HB650-2 idx~40) is below the nominal PIP-II extraction energy of
800~MeV: the 41 entries are the \emph{cavity-period boundaries} of the
physics-design lattice's accelerating section, and the final period is
the last cavity-period sampling point before the HB650-2 exit and
beam-transfer line~\cite{PIP2BTL}.  Two periods are flagged at $S^{2}\!>\!10$ (SSR1
idx~8, $S^{2}=14$; SSR1 idx~12, $S^{2}=33$);
their chart growth-rate predictions should be cross-validated against a treatment valid outside the adopted model
domain before being read as instability claims.  Of the 41 periods, six exhibit
solver-flagged growth above $\gamma_{\rm th}\!=\!10^{-2}$; four of those
six sit inside the in-domain subset.

\end{document}


\title{Supplemental Material:\\
Higher-order space-charge stability in anisotropic beams: Vlasov--Poisson derivation, refined dispersion relations, and stability charts}

\author{Abhishek Pathak}
\thanks{Contact author: abhishek@fnal.gov}
\affiliation{Fermi National Accelerator Laboratory, Batavia, IL 60510, USA}
\date{September 6, 2026}

\maketitle

This supplemental material contains four items: the probabilistic-margin
analysis under an engineering jitter model, with its ensemble Monte~Carlo
uncertainty; the second-moment ODE closure, retained as a failed-validation
record; section-level worst-case growth rates by mode order; and the full
per-period table of the PIP-II application, which is described first.  That
table gives the
chart coordinates $(R,\eta,\varepsilon_{z}/\varepsilon_{x})$, the
dimensionless space-charge parameter $S^{2}$, the per-mode-order
growth rates $\gamma_{\ell}/\nu_{0x}$ at $\ell\!=\!2,3,4_{e}$, the
combined growth rate
$\max_{\ell\in\{2,3,4_{e}\}}\gamma_{\ell}/\nu_{0x}$, and the
domain flag (\textbf{N}~$=$~$S^{2}\!>\!10$, outside the adopted chart
domain; \textbf{Y} otherwise --- a model-adequacy restriction, not a
convergence condition; see the main text) for each of the 41
lattice periods of the PIP-II physics-design trajectory.  The geometric
emittance ratio $\varepsilon_{z}/\varepsilon_{x}$ is taken from the
HELIX normalized-emittance export, de-normalized by $\beta\gamma$ in
all three planes (consistent $(z,z')$ convention).

The compact subset of periods showing solver-flagged coherent growth
above $\gamma_{\rm th}\!=\!10^{-2}$ is reproduced in the compact
flagged-period table of the main manuscript.  Table~\ref{tab:perperiod_supp}
below carries the per-period chart coordinates, growth rates and model-domain status
flag.  Local exponents in the main text are calculated from the zero-current
phase advance $\sigma_{0x}$ and period length, from which the accumulated
$N_{e}$ is formed.
Note also that the $\gamma_{\ell=3}$ column below is the maximum over the even
and odd branches, so it does not by itself identify SSR2 idx~18 as an
odd-branch flag; the flagged-period table of the main manuscript lists the two
parities separately. The calculation resolves both parities for all 41 periods.

The table was calculated using the coordinate definitions and dispersion
relations given in the main text.

\onecolumngrid
{\scriptsize
\begin{longtable}{rlrrrrrrrrrrc}
\caption{Per-period PIP-II solver output and model-domain validity.
``comb.'' is $\max_{\ell\in\{2,3,4_{e}\}}\gamma_{\ell}/\nu_{0x}$;
``in domain'' $=$ \textbf{N} indicates $S^{2}\!>\!10$, i.e.\ outside the
adopted model domain.  Thirty-nine records satisfy that gate, but seven of
them (idx~9, 11, 13, 14, 16, 24, 25) carry an exported $\eta\!\ge\!1$ --- a
depressed tune above the zero-current value, which is unphysical --- and are
assigned $S^{2}\!=\!0$.  Those are stable \emph{by convention} rather than
physically evaluated, and are marked $\ddagger$.  The main text therefore
quotes four flags among the remaining $32$ evaluable periods.
$\varepsilon\!\equiv\!\varepsilon_{z}/\varepsilon_{x}$ (geometric).}
\label{tab:perperiod_supp} \\
\hline\hline
idx & section & $z$\,[m] & $E$\,[MeV] & $R$ & $\eta$ & $\varepsilon$ & $S^{2}$ & $\gamma_{\ell\!=\!2}$ & $\gamma_{\ell\!=\!3}$ & $\gamma_{\ell\!=\!4_{e}}$ & comb. & in dom. \\
\hline\hline
\endfirsthead
\hline\hline
idx & section & $z$\,[m] & $E$\,[MeV] & $R$ & $\eta$ & $\varepsilon$ & $S^{2}$ & $\gamma_{\ell\!=\!2}$ & $\gamma_{\ell\!=\!3}$ & $\gamma_{\ell\!=\!4_{e}}$ & comb. & in dom. \\
\hline\hline
\endhead
0 & HWR & 1.1 & 2.3 & 1.31 & 0.78 & 1.68 & 0.7 & 0.000 & 0.000 & 0.000 & 0.000 & Y \\
1 & HWR & 1.8 & 2.6 & 1.31 & 0.78 & 1.69 & 0.7 & 0.000 & 0.000 & 0.000 & 0.000 & Y \\
2 & HWR & 2.5 & 3.0 & 1.41 & 0.75 & 1.71 & 0.7 & 0.000 & 0.000 & 0.000 & 0.000 & Y \\
3 & HWR & 3.2 & 3.6 & 0.95 & 0.89 & 1.73 & 0.5 & 0.000 & 0.000 & 0.000 & 0.000 & Y \\
4 & HWR & 3.8 & 4.6 & 1.32 & 0.76 & 1.77 & 0.8 & 0.000 & 0.000 & 0.000 & 0.000 & Y \\
5 & HWR & 4.5 & 6.5 & 1.15 & 0.79 & 1.77 & 0.8 & 0.000 & 0.000 & 0.000 & 0.000 & Y \\
6 & HWR & 5.2 & 8.3 & 1.93 & 0.75 & 1.78 & 1.5 & 0.000 & 0.000 & 0.000 & 0.000 & Y \\
7 & HWR & 5.9 & 10.1 & 1.44 & 0.46 & 1.78 & 3.4 & 0.000 & 0.000 & 0.000 & 0.000 & Y \\
8 & SSR1 & 7.9 & 11.6 & 2.04 & 0.36 & 1.78 & 13.9 & 0.000 & 0.070 & 0.000 & 0.070 & \bfseries N \\
9 & SSR1 & 9.2 & 13.4 & 1.52 & 1.50 & 1.77 & 0.0 & 0.000 & 0.000 & 0.000 & 0.000 & Y$^{\ddagger}$ \\
10 & SSR1 & 10.5 & 16.0 & 1.08 & 0.58 & 1.77 & 3.0 & 0.049 & 0.000 & 0.000 & 0.049 & Y \\
11 & SSR1 & 11.8 & 20.0 & 0.71 & 1.06 & 1.76 & 0.0 & 0.000 & 0.000 & 0.000 & 0.000 & Y$^{\ddagger}$ \\
12 & SSR1 & 14.1 & 21.9 & 2.80 & 0.25 & 1.67 & 33.2 & 0.000 & 0.186 & 0.000 & 0.186 & \bfseries N \\
13 & SSR1 & 15.4 & 25.0 & 5.97 & 1.14 & 1.65 & 0.0 & 0.000 & 0.000 & 0.000 & 0.000 & Y$^{\ddagger}$ \\
14 & SSR1 & 16.7 & 27.7 & 3.48 & 1.13 & 1.61 & 0.0 & 0.000 & 0.000 & 0.000 & 0.000 & Y$^{\ddagger}$ \\
15 & SSR1 & 18.0 & 30.3 & 1.42 & 0.88 & 1.61 & 0.3 & 0.000 & 0.000 & 0.000 & 0.000 & Y \\
16 & SSR2 & 21.0 & 36.0 & 1.14 & 1.05 & 1.53 & 0.0 & 0.000 & 0.000 & 0.000 & 0.000 & Y$^{\ddagger}$ \\
17 & SSR2 & 23.2 & 41.5 & 0.67 & 0.95 & 1.51 & 0.4 & 0.000 & 0.000 & 0.000 & 0.000 & Y \\
18 & SSR2 & 27.8 & 50.6 & 2.18 & 0.67 & 1.50 & 2.7 & 0.000 & 0.018 & 0.000 & 0.018 & Y \\
19 & SSR2 & 30.1 & 58.1 & 1.63 & 0.80 & 1.46 & 1.1 & 0.000 & 0.000 & 0.000 & 0.000 & Y \\
20 & SSR2 & 34.7 & 70.1 & 1.25 & 0.85 & 1.47 & 0.5 & 0.000 & 0.000 & 0.000 & 0.000 & Y \\
21 & SSR2 & 36.9 & 78.8 & 1.02 & 0.81 & 1.46 & 0.9 & 0.016 & 0.000 & 0.000 & 0.016 & Y \\
22 & SSR2 & 41.6 & 90.8 & 2.20 & 0.69 & 1.45 & 2.4 & 0.000 & 0.000 & 0.000 & 0.000 & Y \\
23 & SSR2 & 43.8 & 100.4 & 1.39 & 0.62 & 1.44 & 1.6 & 0.000 & 0.000 & 0.000 & 0.000 & Y \\
24 & SSR2 & 48.4 & 112.5 & 0.92 & 1.23 & 1.39 & 0.0 & 0.000 & 0.000 & 0.000 & 0.000 & Y$^{\ddagger}$ \\
25 & SSR2 & 50.7 & 121.9 & 1.44 & 1.09 & 1.41 & 0.0 & 0.000 & 0.000 & 0.000 & 0.000 & Y$^{\ddagger}$ \\
26 & SSR2 & 55.3 & 134.6 & 1.08 & 0.45 & 1.33 & 6.4 & 0.031 & 0.000 & 0.000 & 0.031 & Y \\
27 & SSR2 & 57.5 & 144.5 & 0.88 & 0.89 & 1.34 & 0.6 & 0.000 & 0.000 & 0.000 & 0.000 & Y \\
28 & LB650 & 73.4 & 158.6 & 1.07 & 0.72 & 1.26 & 1.6 & 0.000 & 0.000 & 0.000 & 0.000 & Y \\
29 & LB650 & 80.2 & 183.6 & 1.08 & 0.96 & 1.22 & 0.1 & 0.000 & 0.000 & 0.000 & 0.000 & Y \\
30 & LB650 & 87.0 & 221.6 & 1.52 & 0.78 & 1.19 & 1.4 & 0.000 & 0.000 & 0.000 & 0.000 & Y \\
31 & LB650 & 93.9 & 265.0 & 1.55 & 0.74 & 1.19 & 1.8 & 0.000 & 0.000 & 0.000 & 0.000 & Y \\
32 & LB650 & 100.7 & 312.2 & 1.58 & 0.79 & 1.22 & 1.3 & 0.000 & 0.000 & 0.000 & 0.000 & Y \\
33 & LB650 & 107.5 & 356.7 & 1.86 & 0.59 & 1.19 & 4.1 & 0.000 & 0.000 & 0.000 & 0.000 & Y \\
34 & LB650 & 114.3 & 406.2 & 1.16 & 0.84 & 1.20 & 0.6 & 0.000 & 0.000 & 0.000 & 0.000 & Y \\
35 & LB650 & 121.2 & 454.8 & 1.16 & 0.89 & 1.22 & 0.4 & 0.000 & 0.000 & 0.000 & 0.000 & Y \\
36 & LB650 & 128.0 & 454.8 & 1.54 & 0.74 & 1.23 & 1.7 & 0.000 & 0.000 & 0.000 & 0.000 & Y \\
37 & HB650-1 & 138.8 & 502.0 & 1.55 & 0.70 & 1.23 & 2.2 & 0.000 & 0.000 & 0.000 & 0.000 & Y \\
38 & HB650-2 & 150.1 & 562.2 & 0.89 & 0.85 & 1.23 & 0.9 & 0.000 & 0.000 & 0.000 & 0.000 & Y \\
39 & HB650-2 & 161.3 & 635.5 & 0.88 & 0.88 & 1.20 & 0.7 & 0.000 & 0.000 & 0.000 & 0.000 & Y \\
40 & HB650-2 & 172.5 & 715.2 & 1.27 & 0.65 & 1.17 & 2.8 & 0.000 & 0.000 & 0.000 & 0.000 & Y \\
\hline\hline
\end{longtable}}
\twocolumngrid

\clearpage

\begin{table}[htbp]
\caption{Worst-case coherent growth rates $\gamma/\nu_{0x}$ along the
PIP-II trajectory by mode order and section.  ``comb.\,all'' is the
maximum across $\ell\!\in\!\{2,3,4_{e}\}$ over every period in the
section, including those at $S^{2}\!>\!10$ (marked ``$\dagger$'');
``comb.\,in'' restricts the maximum to the in-domain
subset $S^{2}\!\le\!10$ and is the conclusion-relevant figure.  The SSR1 and SSR2
``comb.\,in'' maxima are set by the $\ell\!=\!2_o$ tilting/coupling
mode of Eq.~(32), while the even-envelope branch of Eq.~(28) has no
negative real root at the flagged periods; SSR1's larger ``comb.\,all'' value ($0.186$) comes from the
out-of-domain $\ell\!=\!3$ odd rate at idx~12, which is excluded from
quantitative claims.  SSR2's $\ell\!=\!3$ column ($0.018$, idx~18) is an
in-domain higher-order flag: it is below the section's $\ell\!=\!2$
maximum but is the only flag at that period.  Values from the corrected anisotropic solver of
Sec.~III of the main text applied to the PIP-II physics-design trajectory~\cite{PIP2CDR,PIP2FDR2024}.}
\label{tab:pip2_sections}
\begin{ruledtabular}
\begin{tabular}{lccccccc}
section  & $N_{p}$ & $\eta_{\min}$ & $\ell{=}2$ & $\ell{=}3$ & $\ell{=}4_{e}$ & comb.\,all & comb.\,in \\
\hline
HWR      & 8  & 0.46 & 0.00 & 0.00 & 0.00 & 0.00 & 0.00 \\
SSR1$^\dagger$    & 8  & 0.25 & 0.049 & 0.186 & 0.00 & 0.186 & 0.049 \\
SSR2     & 12 & 0.45 & 0.031 & 0.018 & 0.00 & 0.031 & 0.031 \\
LB650    & 9  & 0.59 & 0.00 & 0.00 & 0.00 & 0.00 & 0.00 \\
HB650-1  & 1  & 0.70 & 0.00 & 0.00 & 0.00 & 0.00 & 0.00 \\
HB650-2  & 3  & 0.65 & 0.00 & 0.00 & 0.00 & 0.00 & 0.00 \\
\end{tabular}
\end{ruledtabular}
\end{table}

\clearpage

\section{Second-moment ODE closure: Failed-validation diagnostic}
\label{app:momentode}

A 21-dimensional linearized second-moment ODE around the
smooth-focused matched envelope was constructed in the spirit of
Sacherer~\cite{Sacherer1971}, Lapostolle~\cite{Lapostolle1971}, and
the intense-beam envelope-stability studies of
Refs.~\cite{LundBukh2004,Qiang2018Envelope}, with the intent of
producing a lattice-periodic Floquet stability indicator
$\max_{p}\ln|\lambda_{\max}(S_{p})|$ as an independent check on the
dispersion-solver chart of Sec.~III of the main text.  A first-cut
implementation using \emph{uncoupled} per-plane KV self-fields
$K_{i}(\Sigma)\!=\!Q/\Sigma_{ii}$ returns growth at machine zero at
$(R,\eta,\varepsilon_{z}/\varepsilon_{x})\!=\!(1.70,0.28,5.73)$, a point
well inside the domain ($S^{2}\!=\!6.3$) where the Hofmann $\ell\!=\!2$
analysis gives $\gamma/\nu_{0x}\!=\!0.169$: a clean qualitative failure.
The comparison is made at that chart point and not merely near it: because
$R$ is defined on the \emph{depressed} tunes (Sec.~II of the main text), the
matched envelope is built by imposing $\nu_{z}\!=\!R\,\nu_{x}$ and inverting
the matched-envelope relation $\nu_{i}\!=\!\varepsilon_{i}/\sigma_{i}^{2}$ for
$k_{z}$, giving $k_{z}\!=\!(Q/\sigma_{z}^{2}+\nu_{z}^{2})^{1/2}$.  Setting
$k_{z}\!=\!R\,k_{x}$ instead would fix the \emph{zero-current} ratio and
silently realize $\nu_{z}/\nu_{x}\!=\!5.13$; the test asserts the realized
$(R,\eta)$ before integrating.  The failure reflects the closure structure: with uncoupled self-fields the one-period moment map has
$|\lambda_{\max}|\!=\!1$ to machine precision at every period length tried,
i.e.\ the closure is exactly neutrally stable and cannot produce growth of
either sign, rather than producing growth that is merely too small.

We attribute the failure to two missing physics elements: the
cross-plane KV coupling
$K_{x}\!=\!2Q/[\sigma_{x}(\sigma_{x}+\sigma_{y})]$, and the
derivative coupling
$\partial K_{i}/\partial\Sigma_{jj}$ in the envelope-perturbation
matrix.  The full three-plane Sacherer derivation~\cite{Sacherer1971,Qiang2018Envelope} is beyond the present scope. The initial implementation and its failing validation test were retained
in the internal analysis record. The failure is described here so that
future revisits inherit the correct boundary condition rather than the
misleading zero-growth return value.

\clearpage

\section{Probabilistic margins under an engineering jitter model}
\label{sec:track2}

A nominally stable working point close to a chart boundary can become
unstable under small changes in current, matching or tune. We characterize
its response to a specified engineering jitter model by a probability map.
The resulting margins describe this perturbation model and supplement the
deterministic classification.
For a nominal operating point $(R_{0},\eta_{0})$ we define
\begin{equation}
  P(\text{unstable} \mid R_{0},\eta_{0})
  = \frac{1}{N_{\rm mc}} \sum_{i=1}^{N_{\rm mc}}
    \Theta\bigl(\gamma_{i} - \gamma_{\rm th}\bigr),
  \label{eq:punstable}
\end{equation}
where $\Theta$ is the Heaviside step and each $\gamma_{i}$ is the
higher-order solver output (Sec.~III of the main text) evaluated at a
perturbed point drawn from an engineering jitter budget.  Here the rate is
the maximum non-oscillatory growth in the retained principal mode set:
both second-order parities, both third-order parities, and fourth-order
even.  The separate fourth-order odd and oscillatory sensitivities in the
main paper are excluded from this auxiliary probability model.  The
budget amplitudes are author-chosen to bracket the dominant operational
error sources in a CW H$^{-}$ linac at the PIP-II design level (they are
representative engineering allocations, not values transcribed from a
single specification document):
beam-current jitter from the ion-source/RFQ chain
($\sigma_{I}\!=\!3\%$); residual envelope
mismatch from imperfect inter-section matching
($\sigma_{\rm mm}\!=\!10\%$, a conservative envelope-mismatch
amplitude relative to the matched RMS size, well above what the
inter-cryomodule matching sections are designed to deliver);
emittance-ratio scatter from measurement and end-to-end tracking
uncertainty ($\sigma_{\varepsilon}\!=\!5\%$); and tune jitter from
quadrupole and solenoid power-supply regulation
($\sigma_{\nu}\!=\!2\%$).  The sweep below measures the mismatch
sensitivity of selected contour positions and widths.  It does not establish
weak dependence throughout the chart or for the budget as a whole: the
current, tune, and emittance amplitudes are not swept here.  Each perturbation
acts directly on
the chart coordinates $(R,\eta,\varepsilon_{z}/\varepsilon_{x})$ that the
solver consumes.
Explicitly, the $i$-th sample is drawn from the nominal point
$(R_{0},\eta_{0},\varepsilon_{0})$ as
\begin{align}
  R_{i} &= R_{0} + \delta_{R,i}, \qquad
        \delta_{R,i}\sim\mathcal{N}\!\bigl(0,(\sigma_{\nu}R_{0})^{2}\bigr),
        \notag\\
  \eta_{i} &= \eta_{0}\,(1-\delta_{I,i})\,(1+\tfrac{1}{2}\delta_{M,i})
              + \delta_{\eta,i}, \notag\\
  \varepsilon_{i} &= \varepsilon_{0}\,(1+\delta_{\varepsilon,i}),
\end{align}
with $\delta_{I}\sim\mathcal{N}(0,\sigma_{I}^{2})$ (beam-current jitter;
higher current lowers the depressed tune, hence the minus sign),
$\delta_{M}\sim\mathcal{N}(0,\sigma_{\rm mm}^{2})$ (envelope mismatch),
$\delta_{\eta}\sim\mathcal{N}(0,(\sigma_{\nu}\eta_{0})^{2})$ (tune jitter),
and $\delta_{\varepsilon}\sim\mathcal{N}(0,\sigma_{\varepsilon}^{2})$, all
independent across the four sources and across $i=1,\dots,N_{\rm mc}$.
The space-charge strength is recomputed per sample from
the $S^{2}$ definition of Sec.~II of the main text,
\begin{equation}
  S_{i}^{2} = S^{2}(R_{i},\eta_{i},\varepsilon_{i}),
\end{equation}
and $\gamma_{i}$ follows from the solver described in Sec.~III of the main text, evaluated at
$(R_{i},\eta_{i},\varepsilon_{i})$ (the implementation additionally
applies physical guards $\varepsilon_{i}\!\ge\!0.01$, $R_{i}\!\ge\!10^{-6}$, and
$\eta_{i}\!\in\![0.2,2.0]$, which essentially never bind on the chart
interior).  The envelope-mismatch term enters only
through $\eta$, with the $\tfrac{1}{2}$ prefactor chosen so that a
$\sigma_{\rm mm}=10\%$ RMS mismatch acts as an effective $5\%$
tune-depression perturbation; this is a deliberately simple proxy for a
full mismatch-to-coordinate map.  The parameter sweep tests the sensitivity of the
$P\!=\!0.5$ contour in \emph{position} as well as in width: a contour
can remain sharply determined while translating.  Throughout this summary,
a column's position is the first crossing encountered as $\eta$ increases,
with linear interpolation between grid points.  This is a reproducible
selection rule; it does not follow a named physical branch when a column
contains several crossings.

Across a four-fold change in mismatch amplitude, from
$\sigma_{\rm mm}=5\%$ to $20\%$, the formal whole-chart median
contour position moves by $\Delta\eta\!=\!1.4\times10^{-3}$ over
$34$ columns with a median-map crossing in both configurations.  The
whole-chart median envelope width grows from
$6.8\times10^{-3}$ through $7.1\times10^{-3}$ to
$1.0\times10^{-2}$
(a factor of $1.45$, at the sweep fidelity $B\!=\!40$,
$N_{\rm mc}\!=\!100$).  These width medians use $34$, $34$, and $32$
columns, respectively, with crossings in both percentile surfaces; an
absent crossing is not assigned zero width.  The displacement is not
uniform: it reaches $6.4\times10^{-2}$ at the right chart boundary
$R\!=\!2.0$.  The median first crossings in that column have
$S^{2}>10$, so this maximum is a formal result outside the adopted domain,
not an in-domain operating-point uncertainty.

The domain restriction changes what those aggregate statistics describe.
Only $6$ of the $34$ median first crossings in each sweep lie at nominal
points with $0<\eta<1$ and $S^{2}\le10$.  Comparing those same six
columns between $5\%$ and $20\%$ mismatch gives a median displacement
$1.2\times10^{-3}$ and a maximum $1.1\times10^{-2}$.  Five columns have
all three percentile-surface crossings inside the nominal domain at every
sweep amplitude.  Their median envelope widths are $0.0166$, $0.0163$,
and $0.0210$, respectively; that sequence is not monotonic.  Thus the small
median position change survives this restriction, while the whole-chart
widths understate the typical widths of the admitted envelopes.  The
position change is smaller than these median sampling widths, but neither
comparison establishes a uniform physical margin over the chart.
Table~\ref{tab:contourdomain} reports the admitted counts and both interval
definitions.  The current, tune, and emittance budget amplitudes have not
been swept, so the result concerns these three mismatch choices within the
stated engineering proxy.
We adopt $\gamma_{\rm th}\!=\!0.01\,\nu_{0x}$ as the \emph{reporting}
threshold --- in the idealized model any $\gamma\!>\!0$ is unstable, so the cut
selects what we consider operationally worth flagging rather than marking a
mathematical boundary.  The deterministic PIP-II conclusions of
the PIP-II section of the main text (its in-domain flagged-period
count, and the section-level deterministic growth rates tabulated below) do \emph{not} depend on
this jitter budget; the budget enters only the probabilistic
contours.  The sensitivity test above compares three mismatch amplitudes;
it does not establish continuous smoothness under changes to every budget term.

Two additional assumptions affect the interpretation of the probability map.

\emph{Samples that leave the model domain.}  A jitter draw moves the sample
point, and near the $S^{2}\!=\!10$ boundary it can move it out of the adopted
domain.  At $R=1$, $\eta=0.40128345$, and
$\varepsilon_z/\varepsilon_x=1.475$, where $S^{2}=9.5$, an independent
perturbation-only check with $10^{6}$ draws (seed 19) gives
$P(S_i^2>10)=0.3733$ under the nominal budget, with Monte~Carlo standard
error $0.0005$.  This is a point-specific example; $S^{2}$ alone does not
determine the leakage probability.  We do \emph{not} reject or redraw those samples, so
$P$ is taken over all draws and, within roughly one jitter width of the
boundary, includes formal solver evaluations outside the domain in which this
paper claims quantitative validity.  Rejection would condition the probability on domain membership. To quantify leakage in the unconditioned estimate, the grid routine optionally returns
$P_{\rm out}\!=\!P(S_{i}^{2}\!>\!10)$ alongside $P$.  The two questions ---
is the point unstable within the model, and does jitter push it outside the
model --- can therefore be reported separately.  The whole-chart crossing
statistics below include locations outside the adopted domain; we retain
those formal statistics and separately identify the crossings inside it.
Even a nominal crossing inside the domain can have appreciable jitter
leakage.  Independent perturbation-only checks quantify the leakage at the reported
median crossings; the nominal-domain restriction is not a conditioning or rejection of
individual draws.

\emph{Correlation between the two chart coordinates.}  We draw $\delta R$ and
$\delta\eta$ independently, but $R\!=\!\nu_{z}/\nu_{x}$ and
$\eta\!=\!\nu_{x}/\nu_{0x}$ share $\nu_{x}$, so a magnet-induced change in
$\nu_{x}$ moves both.  The independent draws are therefore a
diagonal-covariance engineering proxy, not a lattice-derived tune-response
model; correlations arising from the shared $\nu_{x}$ are not included.  A lattice-based extension could perturb $(\nu_{x},\nu_{z},\nu_{0x})$ directly, or use a
covariance matrix from magnet-error tracking.  The present proxy is auxiliary; no deterministic conclusion in the main text depends on the jitter budget.

\subsection{Ensemble Monte~Carlo uncertainty (parametric bootstrap)}

\begin{figure*}[t]
  \centering
  \includegraphics[width=0.98\textwidth]{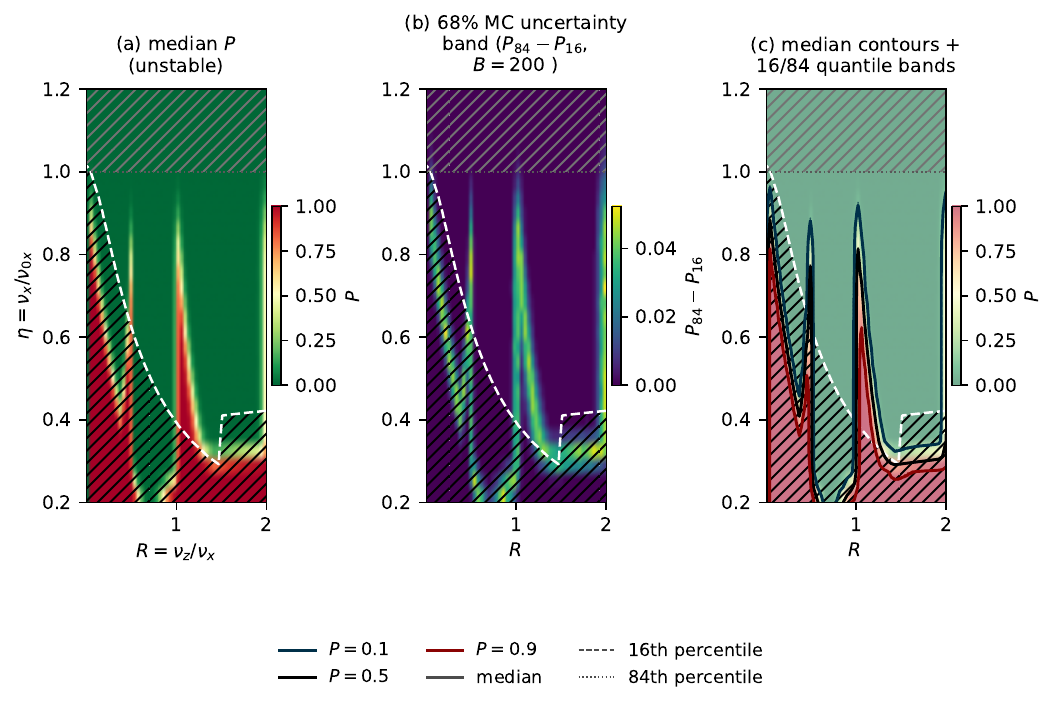}
  \caption{Probabilistic Hofmann chart at
  $\varepsilon_{z}/\varepsilon_{x}=1.475$ with ensemble Monte~Carlo
  uncertainty intervals ($B=200$ resamples, $N_{\rm mc}=500$
  perturbations per resample).  The value
  $\varepsilon_{z}/\varepsilon_{x}\!=\!1.475$ is the representative
  PIP-II mid-linac point (the SSR2 period idx~20; the SSR2 section
  mean is $\approx\!1.44$), illustrating a mildly anisotropic,
  representative PIP-II emittance ratio.  Higher-order stop bands elsewhere
  in this chart slice also depend on $R$ and $\eta$.
  Panel (a) median $P$. Panel (b) 68\% MC uncertainty band
  $P_{84}-P_{16}$ across $B=200$ ensemble resamples.
  Panel (c) median contours at
  $P=0.1,\,0.5,\,0.9$ overlaid on the median surface.
  The dashed white contour marks the nominal $S^{2}\!=\!10$ boundary,
  and the dotted gray line marks $\eta=1$; diagonal hatching identifies
  $S^{2}\!>\!10$ and $\eta\!\ge\!1$ outside the adopted comparison
  domain (the Limitations section and the $\ell\!=\!4$ minor-truncation
  appendix of the main text).  Probabilities in the hatched regions are
  formal only.  The boundaries apply to nominal coordinates; jitter draws
  can leave the domain even from a nominal point inside it.}
  \label{fig:fig6}
\end{figure*}

The probability map of Eq.~\eqref{eq:punstable} is itself a
Monte~Carlo estimate and therefore carries sampling noise; a contour
drawn through it has a finite positional uncertainty.  To quantify
that uncertainty we generate $B$ independent realizations of the
full $N_{\rm mc}$-sample block, each at an independent seed
(re-drawing fresh Gaussian perturbations rather than resampling a
fixed dataset; the procedure is therefore a \emph{parametric bootstrap}
in the sense of a Monte~Carlo replication of the sampling
model~\cite{Efron1979,EfronTibshirani1993}, not a nonparametric
resampling of a fixed dataset).  We recompute $P$ for every
realization and report the pointwise $16/50/84$ percentile maps
$P_{16}, P_{50}, P_{84}$.  The median map
$P_{50}$ is our central estimate of the probability surface; the
$68\%$ band $P_{84}-P_{16}$ measures how sharply the surface (and
hence each iso-probability contour) is determined.
Figure~\ref{fig:fig6} shows the median surface, the
ensemble Monte~Carlo uncertainty band, and the median contours at
$P=0.1,\,0.5,\,0.9$ with their $16/84$ quantile envelopes.

The pointwise uncertainty band is narrow where the sampled maps saturate
near $P=0$ or $P=1$, and is concentrated around probability transitions.
This sampling band describes the precision of the probability estimate;
it does not by itself measure the physical displacement of a stability
boundary.  The probabilistic contours supplement the deterministic
verdict by describing the response to the specified jitter distribution.
At the production scale of Fig.~\ref{fig:fig6} ($B=200$ realizations and
$N_{\rm mc}=500$ samples per realization), the formal whole-chart envelope
between the $16$th- and $84$th-percentile probability surfaces has median
width $\Delta\eta\!\approx\!3.7\!\times\!10^{-3}$ at $P=0.5$.
There are $42$ columns with both percentile-surface crossings; exactly
$1$ has a width above $5\!\times\!10^{-2}$.  At the right chart boundary,
$R\!=\!2.0$, the envelope has the $0.384$ maximum; the next widest is
$1.2\!\times\!10^{-2}$.  The median crossing in the widest column has
$S^{2}=14.02$, and its envelope spans both sides of the adopted domain
boundary.  It is therefore a formal out-of-domain result, not evidence
of a large in-domain operating-point uncertainty.  The pointwise band
on the probability surface itself peaks at $P_{84}-P_{16}=0.052$.
Both the median and maximum widths are retained to expose this
nonuniformity; neither is a chart-wide physical margin.

The domain audit finds $43$ median-map first crossings, of which $9$
lie at nominal points satisfying $0<\eta<1$ and $S^{2}\le10$.
Only $8$ columns have all three percentile-surface crossings inside that
domain.  Their median and maximum envelope widths are $0.0080$ and
$0.0117$, respectively (Table~\ref{tab:contourdomain}).  Independent
perturbation-only checks at the nine admitted median crossings give
out-of-domain draw fractions ranging from $0$ to $0.7315$, using
$10^{5}$ draws per point.  These are separate diagnostic draws, not
recovered flags from the production ensemble; the stored record specifies
each point and seed.  The nominal-domain selection consequently does not
establish small leakage of individual perturbations.

The envelope widths are calculated as the separation between selected
$P=0.5$ crossings of the $16$th- and $84$th-percentile probability maps.
An envelope between contours of percentile surfaces is not, in general,
the $16$--$84$ percentile interval of the contour \emph{position} across
realizations.  We also calculate the latter statistic directly:
locate the first increasing-$\eta$ crossing
$\eta^{(b)}_{0.5}(R)$ in every realization, then take the
$16/50/84$ percentiles of those positions.  An interval is reported only
where all realizations have a crossing; missing crossings are recorded
rather than dropped from the ensemble or assigned zero width.  In the
production run, $41$ columns meet that requirement.  Their formal median
and maximum position-interval widths are $0.0031$ and $0.0506$.
Nine intervals have all three position quantiles inside the nominal
domain, with median and maximum widths $0.0062$ and $0.0085$.
The admitted columns need not coincide with those selected from the
percentile-surface envelopes.  No identity between these two interval
definitions is assumed, and neither procedure matches named physical
branches between realizations.

All four ensembles use the same $28$ vertical grid points over
$0.2\le\eta\le1.2$, with spacing $\Delta\eta_{\rm grid}=1/27$;
the production and sweep grids have $50$ and $40$ horizontal points,
respectively.  Linear interpolation can yield sampling widths smaller
than that spacing, but the reported interval does not include spatial
discretization error.  No grid-convergence error bound for these
probabilistic contours is established here.  The intervals also exclude
uncertainty in the engineering allocations, their omitted correlations,
and the physical model.  The contour position is thus a useful diagnostic
of this specified proxy, with the stated first-crossing convention and
sampling uncertainty, rather than a validated quantitative margin for
machine operation.

\begin{table*}[t]
\caption{Nominal-domain audit of the $P=0.5$ first crossings.
The median column gives admitted median-map crossings divided by all
median-map crossings.  An admitted envelope requires all three
percentile-surface crossings to satisfy $0<\eta<1$ and $S^{2}\le10$.
Direct position intervals require a crossing in every realization;
all three position quantiles must lie inside the same nominal domain
for the interval to be admitted.  Width pairs give the median and maximum
in $\eta$ over the admitted columns.  Neither selection conditions the
jitter draws on remaining inside the domain.  The production configuration
uses $B=200$, $N_{\rm mc}=500$; the mismatch sweeps use $B=40$,
$N_{\rm mc}=100$.}
\label{tab:contourdomain}
\begin{ruledtabular}
\begin{tabular}{lccccc}
Configuration & Median & Envelopes & Envelope widths & Position intervals & Position widths \\
 & in/all & admitted & median / max & admitted & median / max \\
Production & $9/43$ & $8$ & $0.0080$ / $0.0117$ & $9$ & $0.0062$ / $0.0085$ \\
$5\%$ sweep & $6/34$ & $5$ & $0.0166$ / $0.0186$ & $6$ & $0.0126$ / $0.0179$ \\
$10\%$ sweep & $6/34$ & $5$ & $0.0163$ / $0.0228$ & $6$ & $0.0138$ / $0.0209$ \\
$20\%$ sweep & $6/34$ & $5$ & $0.0210$ / $0.0275$ & $5$ & $0.0191$ / $0.0220$ \\
\end{tabular}
\end{ruledtabular}
\end{table*}

\bibliographystyle{apsrev4-2}